\documentclass[twocolumn,pra,showpacs,superscriptaddress,amssymb,amsmath,amsmath,floatfix]{revtex4-1}
\usepackage{graphicx}
\usepackage{epstopdf}
\usepackage{bm}
\usepackage[colorlinks=true,linkcolor=blue,urlcolor=blue,citecolor=blue]{hyperref}
\usepackage{comment}
\usepackage{float}
\usepackage{cancel}
\usepackage{times}
\usepackage{xcolor}
\usepackage{wrapfig}
\usepackage{longtable}
\usepackage{derivative}

\newcommand{\be}{\begin{equation}}
\newcommand{\ee}{\end{equation}}

\usepackage{xcolor}

\begin{document}
\title{Diatomic molecular anions of alkali-metal and alkaline-earth-metal atoms}

\author{Sana Akkari}
\affiliation{Laboratory of Interfaces and Advanced Materials, Faculty of Science, University of Monastir, 5019 Monastir, Tunisia}
\author{Hela Ladjimi}
\affiliation{Faculty of Physics, University of Warsaw, Pasteura 5, 02-093 Warsaw, Poland}
\author{Wissem Zrafi}
\affiliation{Laboratory of Interfaces and Advanced Materials, Faculty of Science, University of Monastir, 5019 Monastir, Tunisia}
\affiliation{University of Sousse, Electronic Department, Higher Institute of Applied Sciences and Technology of Sousse, 4003 Sousse, Tunisia}
\author{Hamid Berriche}
\affiliation{Laboratory of Interfaces and Advanced Materials, Faculty of Science, University of Monastir, 5019 Monastir, Tunisia}
\affiliation{Department of Mathematics and Physics, School of Arts and Sciences, American University of Ras Al Khaimah, P.O. Box 10021, RAK, UAE}
\author{Marcin Gronowski}
\affiliation{Faculty of Physics, University of Warsaw, Pasteura 5, 02-093 Warsaw, Poland}
\author{Micha{\l} Tomza}
\email{michal.tomza@fuw.edu.pl}
\affiliation{Faculty of Physics, University of Warsaw, Pasteura 5, 02-093 Warsaw, Poland}

\date{\today}

\begin{abstract}

Studies of anions are inherently more challenging than investigations of neutrals and cations because of the diffuse and weakly bound character of an anionic electron. Here, we present a comprehensive computational examination of ground-state diatomic molecular anions composed of alkali-metal (Li, Na, K, Rb, Cs, Fr) and alkaline-earth-metal (Be, Mg, Ca, Sr, Ba, Ra) atoms. We study 21 alkali-metal diatomic anions in the X$^{2}\Sigma^{+}$ electronic state and 36 alkali-metal--alkaline-earth-metal diatomic anions in the X$^{1}\Sigma^{+}$ electronic state.  The calculations employ a hierarchy of the coupled cluster methods, combined with large Gaussian basis sets and small-core relativistic energy-consistent pseudopotentials for heavier elements. We compute potential energy curves, permanent electric dipole moments, and static polarizabilities, and we assess convergence and uncertainties of our results. Additionally, using the multireference configuration interaction and equation-of-motion electron-attachment coupled cluster methods, we investigate excited electronic states of alkali-metal molecular anions, including valence-bound and dipole-bound states. We predict crossings between ground neutral and excited anionic states, which may enhance resonant electron attachment and subsequent anion dissociation. This finding may be relevant for experiments with mixtures of ultracold ground-state alkali-metal molecules and Rydberg atoms.

\end{abstract}

\maketitle

\section{Introduction}

Ultracold atoms and molecules provide unprecedented opportunities to explore quantum phenomena in physics and chemistry. Their applications range from studying controlled chemical reactions~\cite{BohnScience17,KarmanNP24} and high-precision spectroscopy probing fundamental laws of nature~\cite{DemilleScience17,DeMilleNP24} to investigating exotic phases of quantum many-body systems~\cite{GrossScience17} and quantum computing~\cite{CornishNP24}. Charged atomic and molecular ions further expand these possibilities in quantum science and technology due to their stronger interparticle interactions, extended trapping lifetimes, and easier single-particle manipulation, which enable full control over their quantum states~\cite{TomzaRMP19,DeissNP24}. Despite substantial progress in systems based on cold molecular cations, research on cold molecular anions remains comparatively limited.

Trapped alkaline-earth-metal atomic cations have been successfully laser-cooled and employed in a plethora of quantum experiments because of their favorable electronic structure~\cite{EschnerJOSAB03}. They have also been combined with ultracold gases of alkali-metal and alkaline-earth-metal atoms in cold hybrid ion-atom systems~\cite{TomzaRMP19}. Experimentally realized mixtures include Yb$^+$+Yb~\cite{GrierPRL09}, Yb$^+$+Rb~\cite{ZipkesNature10}, Ba$^+$+Rb~\cite{SchmidPRL10,SchmidtPRL20}, Ca$^+$+Rb~\cite{HallPRL11}, Ca$^+$+Li~\cite{HazePRA13}, Ca$^+$+Na~\cite{SmithAPB14}, Sr$^+$+Rb~\cite{MeirPRL16}, Cs$^+$+Rb~\cite{DuttaPRL17}, Yb$^+$+Li~\cite{JogerPRA17}, and Ba$^+$+Li~\cite{WeckesserNature21}. These systems have been used to observe charge-transfer chemical reactions~\cite{ZipkesNature10,HallMP13}, spin dynamics~\cite{SikorskyNC18}, buffer-gas cooling~\cite{FeldkerNP20}, and magnetic Feshbach resonances~\cite{WeckesserNature21}. A larger number of diatomic combinations have also been investigated theoretically (see, e.g., Refs.~\cite{TomzaRMP19,SmialkowskiPRA20} and references therein). 

Cold molecular cations can be produced from cold ion-atom mixtures \textit{via} spontaneous or stimulated charge-transfer radiative association~\cite{HallMP13,daSilvaNJP2015,ZrafiNJP20} or magnetoassociation~\cite{IdziaszekPRA09,TomzaPRA15}, followed by optical stabilization. They can also be generated through chemical reactions with trapped atomic ions~\cite{MolhavePRA00,HansenAC12,PatelJPCL26}. Alternatively, they can be obtained directly \textit{via} photoionization of neutral molecules~\cite{SullivanPCCP11,JyothiPRL16} or by sympathetic cooling of molecular ions from higher temperatures~\cite{RellergertN13,Hansen14}.

Atomic and molecular anions remain relatively unexplored at ultralow temperatures. Their study is inherently more challenging due to the diffuse and weakly bound nature of the excess electron~\cite{SimonsJPCA08, SimonsJPCA23}. This characteristic, on one hand, restricts the existence of excited electronic states necessary for laser cooling or optical manipulation and, on the other hand,  increases the computational complexity of theoretical predictions, which are highly sensitive to the details of electron correlation treatment~\cite{SimonsARPC11}. Searches for laser-coolable anions are motivated by their potential applications in precision spectroscopy and in sympathetic cooling of antiprotons~\cite{KellerbauerNJP06}, a key step toward the subsequent formation of antihydrogen~\cite{CerchiariPRL18}.

Bound excited electronic states exist for some transition-metal and f-block atoms~\cite{PeggRPP04}. However, only a small number of elements support multiple bound excited states~\cite{BilodeauPRL00,WalterPRL07}, including states of opposite parity that are essential for optical cycling and laser cooling~\cite{OMalleyPRA10,WalterPRL14,TangPRL19}. Laser-cooling schemes have been predicted for a range of molecular anions isovalent with C$_2^-$~\cite{YzombardPRL15}, and the first spectroscopic measurements of the relevant transitions in C$_2^-$ were reported recently~\cite{NotzoldPRA22}.

In many cases, molecular anions are more stable (i.e., bound) than their atomic counterparts due to the delocalization of the excess electron and its stronger interaction with the dipole and higher electric multipole moments of the polar neutral core. As a result, a large number of stable diatomic and polyatomic anions have already been investigated experimentally and theoretically~\cite{RienstraCR02,SimonsJPCA08,SimonsJPCA23}. Polar molecules with sufficiently large electric dipole moments ($\gtrsim$2.5~debye) can form anionic dipole-bound states (DBSs)~\cite{FranccoisIJMPB96,JordanARPC03} in addition to conventional valence states. DBSs are characterized by highly diffuse orbitals, analogous to Rydberg states, and are experimentally difficult to observe and theoretically challenging to model. Nevertheless, two dipole-bound states of $\sigma$ and $\pi$ symmetry were recently observed experimentally in the diatomic KI$^{-}$ anion~\cite{LuJPCL21}. 

Evaporative, buffer-gas, and sympathetic cooling of molecular anions have been considered as alternatives to laser cooling. Laser-induced forced evaporative cooling of OH$^-$ anions~\cite{TauchNP23} and helium buffer-gas cooling of OH$^-$ anions~\cite{HauserNP15} were demonstrated experimentally. In addition, OH$^-$ and O$^-$ anions were immersed in an ultracold gas of Rb atoms~\cite{DeiglmayrPRA12,HassanNC22,HassanJCP22}, while several similar molecular anions mixed with alkali-metal and alkaline-earth-metal atoms were investigated theoretically~\cite{GonzalezCP15,KasJCP17,TomzaPCCP17,KasPRA19}. However, associative electronic detachment leads to significant reactive losses, which hinder sympathetic cooling of OH$^-$ anions with Rb atoms~\cite{ByrdPRA13,HassanJCP22}.

Recent experiments on mixtures of ultracold Rydberg atoms and ground-state alkali-metal molecules~\cite{GuttridgePRL23,ZhuPRL25} open new possibilities for molecular spectroscopy~\cite{PatschJPCL22,ZouPRL26}, collisional cooling~\cite{HuberPRL12,ZhaoPRL12,ZhangPRL24}, quantum computing~\cite{ZhangPRXQ22,WangPRXQ22,Bai2026,Zhang2026}, and quantum simulation~\cite{DobrzynieckiPRA23}. These systems are governed by interactions between low-energy Rydberg electrons and ground-state polar molecules, where the transient or resonant formation of anionic states can play an important role. However, our knowledge of alkali-metal molecular anions remains limited. Only a few anions such as Li$_2^-$~\cite{SarkasZPD94}, Na$_2^-$~\cite{MchughJCP89,EatonCPL92}, K$_2^-$~\cite{MchughJCP89,EatonCPL92}, Rb$_2^-$~\cite{MchughJCP89,EatonCPL92},  Cs$_2^-$~\cite{MchughJCP89}, NaK$^-$~\cite{EatonCPL92}, KRb$^-$~\cite{EatonCPL92}, RbCs$^-$~\cite{EatonCPL92}, and KCs$^-$~\cite{EatonCPL92} were experimentally studied using laser photoelectron spectroscopy. Likewise, a small number of species, including Li$_2^-$~\cite{AndersenJCP76,DixonJCP77,ShepardJCP78,SunilCPL84,KonowalowCPL84,MichelsCPL85,BoldyrevJCP93,PetchJCS95,HogreveEPJD00,NasiriCTC17,DunningJCC24}, 
Na$_2^-$~\cite{ShepardJCP78,PartridgeJCP83,SunilCPL84}, K$_2^-$~\cite{PartridgeJCP83}, Rb$_2^-$~\cite{PartridgeJCP83,KraussJCP90},  Cs$_2^-$~\cite{KraussJCP90}, and LiNa$^{-}$~\cite{ShepardJCP78,BoldyrevJCP93} were theoretically investigated. Alkali-metal--alkaline-earth-metal anions remain even less explored, with only a few theoretical works on 
LiBe$^-$~\cite{BauschlicherJCP92,BoldyrevJCP93}, NaBe$^-$~\cite{BauschlicherJCP92}, LiMg$^-$~\cite{BauschlicherJCP92,BoldyrevJCP93}, NaMg$^-$~\cite{BauschlicherJCP92}, and KCa$^-$~\cite{MoussaNJP21}.

Here, we fill this gap by presenting a theoretical investigation of the ground-state electronic properties of 57 homo- and heteronuclear diatomic anions formed from alkali-metal (Li, Na, K, Rb, Cs, Fr) and alkaline-earth-metal (Be, Mg, Ca, Sr, Ba, Ra) atoms. We employ \textit{ab initio} quantum-chemical methods to characterize the electronic structure of these molecular anions, including computations of potential energy curves, permanent electric dipole moments, static electric dipole polarizabilities, and spectroscopic constants. The calculations are performed using coupled-cluster methods, including full triple excitations, in conjunction with small-core relativistic energy-consistent pseudopotentials for heavier elements and large Gaussian basis sets. The convergence and accuracy of the results are assessed by systematically varying the orbital basis-set size and the wave-function model for selected anions. Additionally, excited valence and dipole-bound states of alkali-metal molecular anions are investigated using the multireference configuration interaction and equation-of-motion electron-attachment coupled cluster methods with single and double excitations, for which crossings between ground neutral and excited anionic states are predicted and discussed. 

The structure of the paper is as follows. In Section~\ref{sec:theory}, we describe the theoretical methods employed. In Section~\ref{sec:results}, we present and discuss the results. In Section~\ref{sec:summary}, we provide a summary and outlook.

\section{Theoretical methods}
\label{sec:theory}

The interaction of a closed-shell alkali-metal anion in its ground $^1S$ state with an open-shell alkali-metal atom in its ground $^2S$ state leads to a molecular electronic ground state of doublet $X^2\Sigma^+$ symmetry. In contrast, the interaction between a ground-state alkali-metal anion and a ground-state alkaline-earth-metal atom, both in their ground $^1S$ states, results in a closed-shell molecular electronic ground state of $X^{1}\Sigma^+$ symmetry. In both cases, the resulting molecular anionic states are adequately described by single-reference electronic-structure methods over the entire range of internuclear separations.

We use the computational scheme based on the composite approach to calculate potential energy curves within the Born-Oppenheimer approximation for the relevant molecular electronic states. This approach has been successfully applied and validated in our previous studies of molecules containing alkali-metal and alkaline-earth-metal atoms~\cite{GronowskiPRA20,SmialkowskiPRA20,KarmanPRA23,LadjimiPRA23,LadjimiPRA24}. The final interaction energies, $V_{\text{int}}(R)$, as functions of the internuclear distance $R$, are obtained as the sum of two contributions
\begin{equation}
 V_{\text{int}}(R) = V_{\text{CCSD(T)}}^{\text{apwCV5Z+bf}}(R) + \delta V_{\text{CCSDT}}^{\text{apwCVTZ}}(R),
 \label{eq:Vint}
\end{equation}
where the leading term, $V_{\text{CCSD(T)}}^{\text{apwCV5Z+bf}}(R)$, is the interaction energy calculated using the spin-restricted open-shell (for the $X^2\Sigma^+$ states) or closed-shell (for the $X^1\Sigma^+$ states) coupled-cluster method including single, double, and perturbative triple excitations [CCSD(T)]~\cite{BartlettRMP07} together with the augmented correlation consistent polarized weighted core-valence quintuple-$\zeta$ quality basis sets (aug-cc-pwCV5Z)~\cite{PrascherTCA11,HillJCP17}. The atomic basis sets are further augmented in these calculations by a set of (4$s$4$p$3$d$3$f$1$g$) bond functions (bf)~\cite{Bond2024} to accelerate the convergence towards the complete basis set limit~\cite{TaoJCP92}.

The second term in Eq.~\eqref{eq:Vint}, $\delta V_{\text{CCSDT}}^{\text{apwCVTZ}}(R)$, is the electron-correlation correction to the interaction energy to account for the contribution of the iterative full triple excitations in the coupled cluster method given by
\begin{equation}
\delta V_{\text{CCSDT}}^{\text{apwCVTZ}}(R) = V_{\text{CCSDT}}^{\text{apwCVTZ}}(R) - V_{\text{CCSD(T)}}^{\text{apwCVTZ}}(R),  
\label{eq:VSDT}
\end{equation}
where $V_{\text{CCSDT}}^{\text{apwCVTZ}}(R)$ is the interaction energy computed using the coupled cluster method restricted to single, double, and triple excitations (CCSDT), while $V_{\text{CCSD(T)}}^{\text{apwCVTZ}}(R)$ is obtained with the CCSD(T) method, both employing the same augmented correlation-consistent polarized weighted core-valence triple-$\zeta$ quality basis sets (aug-cc-pwCVTZ)~\cite{PrascherTCA11,HillJCP17}.

The interaction energies, $V_{method}^{basis}(R)$, in Eqs.~\eqref{eq:Vint} and \eqref{eq:VSDT} are obtained using the super-molecule method with the basis-set superposition error corrected by the Boys-Bernardi counterpoise scheme~\cite{BoysMP70} 
\begin{equation}
V_{method}^{basis}(R)={E}_{\text{anion+atom}}(R)-{E}_{\text{anion}}(R)-{E}_{\text{atom}}(R),
\end{equation}
where $E_{\text{anion+atom}}(R)$ is the total energy of the molecular anion, while $E_{\text{anion}}(R)$ and $E_{\text{atom}}(R)$ are the total energies of the isolated atomic anion and atom, respectively, each evaluated in the full dimer basis set at the same separation $R$, using the specified electronic-structure \textit{method} and \textit{basis} set.

All electrons in the Li, Be, Na, and Mg atoms as well as the outer-shell electrons in heavier atoms are described explicitly using the augmented correlation-consistent weighted core-valence Gaussian basis sets (aug-cc-pwCV$n$Z)~\cite{PrascherTCA11,HillJCP17}. Scalar relativistic effects in heavier atoms are taken into account by replacing the inner-shell electrons with small-core relativistic energy-consistent pseudopotentials (ECPs) from the Stuttgart library~\cite{DolgCR12}. Specifically, the K, Ca, Rb, Sr, Cs, Ba, Fr, and Ra atoms are described using the ECP10MDF, ECP10MDF, ECP28MDF, ECP28MDF, ECP46MDF, ECP46MDF, ECP78MDF, and ECP78MDF pseudopotentials, respectively~\cite{LimJCP05,LimJCP06}. These pseudopotentials are employed with the corresponding aug-cc-pwCV$n$Z-PP basis sets~\cite{HillJCP17}. The electrons of two outermost shells are correlated, i.e., $(n-1)s^{2}$ $(n-1)p^{6}$ $ns^{1}$ for alkali-metal and $(n-1)s^{2}$ $(n-1)p^{6}$ $ns^{2}$ for alkaline-earth-metal atoms.

Additional convergence tests are performed for the exemplary KRb$^{-}$ and RbSr$^{-}$ molecular anions. A family of aug-cc-pwCV$n$Z basis sets with $n$=D, T, Q, and 5  is employed. The complete basis set (CBS) limit is extrapolated using the two-point formula~\cite{HelgakerJCP97}. Furthermore, potential energy curves are also obtained using several  lower-level electronic structure methods, including the restricted Hartree-Fock (RHF), the coupled cluster method restricted to single and double excitations (CCSD), the configuration interaction method restricted to single and double excitations (CISD) and its variant with the Davidson correction (CISD+Q), and also the multireference configuration interaction method including single and double excitations with and without the Davidson correction (MRCISD and MRCISD+Q)~\cite{Helgaker00}.

We investigate excited electronic states, including valence and dipole-bound states, using the multireference configuration interaction and equation-of-motion electron-attachment coupled cluster methods restricted to single and double excitations (MRCISD and EOM-EA-CCSD)~\cite{NooijenJCP95} with the aug-cc-pwCV5Z and aug-cc-pCVQZ basis sets, respectively. We calculate potential energy curves for the lowest excited states of Li$_2^-$, RbCs$^-$, and NaCs$^-$. Additionally, we compute electron binding energies with EOM-EA-CCSD for dipole-bound states of six alkali-metal diatomic anions (NaK$^{-}$, LiK$^{-}$, NaRb$^{-}$, LiRb$^{-}$, NaCs$^{-}$, and LiCs$^{-}$), whose neutral counterparts possess permanent electric dipole moments sufficiently large to support dipole-bound states. These energies are obtained at the equilibrium geometries of the corresponding neutral molecules~\cite{LadjimiPRA24}. 

The dipole-bound electron is highly diffuse, which imposes special requirements on the basis set. To address this, we augment the basis set by adding diffuse Gaussian basis functions centered at the midpoint between the atoms. We initially tested basis functions proposed for describing continuum and Rydberg electrons~\cite{KaufmannJPB89}; however, their inclusion led to poor convergence behavior. Instead, we constructed and employed a new sequence of even-tempered basis sets of increasing size, namely (2$s$2$p$1$d$), (3$s$3$p$2$d$1$f$), (4$s$4$p$3$d$2$f$1$g$),  (5$s$5$p$4$d$3$f$2$g$1$h$),  (6$s$6$p$5$d$4$f$3$g$2$h$1$i$), (7$s$7$p$6$d$5$f$4$g$3$h$2$i$), and (8$s$8$p$7$d$6$f$5$g$4$h$3$i$). The most compact (largest) exponents were chosen to match those corresponding to double augmentation in the d-aug-cc-pwCV5Z-PP basis set for the Cs atom~\cite{HillJCP17}. These basis sets in CFOUR format are available in the Supplemental Material~\cite{SM}.

To obtain the molecular spectroscopic constants, we use the cubic spline method to interpolate the potential energy curves. The equilibrium internuclear distance $R_{e}$ is defined by $\frac{dV_{\text{int}}(R)}{dR}|_{R_{e}} = 0$, while the depth of the potential energy well $D_{e}$ is determined by $D_{e}=-V_{\text{int}}(R_{e})$. The harmonic constant, $\omega_{e}$, is calculated at the equilibrium distance defined as
$\omega_{e} = {\sqrt{\frac{1}{\mu} \frac{d^{2}V_{\text{int}}(R)}{dR^{2}}|_{R_{e}}}}$, where $\mu$ is the reduced mass of the molecule. The anharmonicity constant, $\omega_{e}x_{e}\approx-Y_{20}$, is obtained by fitting the Dunham expansion to the lowest vibrational energy levels~\cite{LadjimiPRA24}.
The equilibrium rotational constant, $B_e$, is identified as $B_e = \frac{\hbar^{2}}{2 \mu R_e^{2}}$.

Permanent electric dipole moments and static electric dipole polarizabilities are calculated using the finite field approach with the CCSD(T) method and the aug-cc-pwCV5Z basis sets augmented by bond functions. The strength of the used external field perturbation is $\pm0.0001~e/a_0^2$. The $z$ axis is oriented from the atom with a larger ionization potential to the atom with a smaller ionization potential, and the origin is at the center of mass.

The EOM-EA-CCSD calculations are carried out using the CFOUR package~\cite{MatthewsJCP20}, while the CCSDT energies are obtained with the MRCC program~\cite{KallayJCP20} interfaced with MOLPRO. All remaining electronic structure calculations are performed using the MOLPRO package of \textit{ab initio} programs~\cite{MOLPRO_brief,WernerWIRCMS12}.

%******************************************TABI******************************************
\begin{table*}[ht]
 \begin{ruledtabular}
\caption{Characteristics of alkali-metal and alkaline-earth-metal atoms: ionization potential $E_\text{IP}$, electron affinity $E_\text{EA}$, the lowest S-P excitation energy $E_\text{S-P}$ ($^2$S-$^2$P for alkali-metal atoms and $^1$S-$^3$P for alkaline-earth-metal atoms), and the static electric dipole polarizability of neutral atom $\alpha_X$, cation $\alpha_{X^+}$, and anion $\alpha_{X^-}$. Present theoretical values are compared with the most accurate available experimental or theoretical data. Experimental electron affinities and excitation energies are averaged on spin-orbit manifolds.} 
\label{table:atoms}
\begin{tabular*}{\textwidth}{@{\extracolsep{\fill}}lllllll}
 Atom & $E_\text{IP}$ (cm$^{-1}$) & $E_\text{EA}$ (cm$^{-1}$)    & $E_\text{S-P}$ (cm$^{-1}$)  & $\alpha_{X}$ ($a^3_0$) &  $\alpha_{X^{+}}$ ($a^3_0$)   & $\alpha_{X^{-}}$ ($a^3_0$)  \\
\hline
Li      
& 43 481                        & 4977              &14 902            &164.2                   &0.191   &806 \\
& 43 487~\cite{NIST_ASD}$^{a}$ & 4985~\cite{HaefflerPRA96}$^{a}$  &14 904~\cite{NIST_ASD}$^{a}$      &164.2~\cite{MiffreEPJD06}$^{a}$ &0.192~\cite{McneillJCP20}$^{b}$&794~\cite{SahooPRA20}$^{b}$\\                
Na     
& 41 384          &4416         &16 914     &162.9        &1.013  &1021\\ 
& 41 449~\cite{NIST_ASD}$^{a}$&4419~\cite{AndersenJPC99}$^{a}$ & 16 968~\cite{NIST_ASD}$^{a}$&162.7~\cite{EkstromPRA95}$^{a}$  &0.997~\cite{McneillJCP20}$^{b}$ &953~\cite{SahooPRA20}$^{b}$\\              
K       
& 35 053         &4036          & 13 062     &278.8         &5.481   &1456  \\
& 35 010~\cite{NIST_ASD}$^{a}$ &4044~\cite{AnderssonAPS00}$^{a}$    &13 024~\cite{NIST_ASD}$^{a}$    & 290.0~\cite{GregoirePRA15}$^{a}$   &5.481~\cite{McneillJCP20}$^{b}$&1354~\cite{SahooPRA20}$^{b}$\\                    
Rb      
& 33 649          &3910         &12 731           & 322.6        &9.187   &1665\\      
& 33 691~\cite{NIST_ASD}$^{a}$       &3919~\cite{FreyJPB78}$^{a}$           &12 737~\cite{NIST_ASD}$^{a}$       & 320.1~\cite{GregoirePRA15}$^{a}$   &9.187~\cite{McneillJCP20}$^{b}$ &1508~\cite{SahooPRA20}$^{b}$\\                                  
Cs    
& 31 428      &3791          &11 647           & 391.1           &15.98  &1988 \\      
& 31 406~\cite{NIST_ASD}$^{a}$     &3804~\cite{AndersenJPC99}$^{a}$ &11 548~\cite{NIST_ASD}$^{a}$       & 401.2~\cite{GregoirePRA15}$^{a}$   &15.98~\cite{McneillJCP20}$^{b}$ & 1804~\cite{SahooPRA20}$^{b}$\\                                 
Fr    
&32 428              &3836               &13 062           & 325.0         &19.72    &1827\\      
&32 849~\cite{NIST_ASD}$^{a}$&3920~\cite{LandauJCP01}$^{b}$&13 362~\cite{NIST_ASD}$^{a}$& 317.8~\cite{DereviankoPRL99}$^{b}$ &19.72~\cite{McneillJCP20}$^{b}$ &1620~\cite{SahooPRA20}$^{b}$\\         
Be    
& 75 171                        &  $< 0$           &21 966              & 37.7       &24.51   & - \\      
& 75 193~\cite{NIST_ASD}$^{a}$  &  $\lesssim 0$      &21 980~\cite{NIST_ASD}$^{a}$  & 37.7~\cite{MitroyPRA03}$^{a}$    & 24.50~\cite{JiangADNDT15}$^{b}$ & -\\    
Mg    
& 61 569           & $< 0$         &21 784          & 71.5        &35.30     & -\\      
& 61 671~\cite{NIST_ASD}$^{a}$      & $\lesssim 0$~\cite{AndersenJPC99}$^{a}$ &21 891~\cite{NIST_ASD}$^{a}$       & 71.3~\cite{PorsevJOTP06}$^{b}$     & 34.99~\cite{JiangADNDT15}$^{b}$  & -\\      
Ca    
& 49 378           & $<0$       &15 283       &156.1       &73.87    & -\\       
& 49 306~\cite{NIST_ASD}$^{a}$       & 172~\cite{PetruninPRL96}$^{a}$        &15 263~\cite{NIST_ASD}$^{a}$       & 157.1~\cite{PorsevJOTP06}$^{b}$     &75.46~\cite{JiangADNDT15}$^{b}$&-\\
Sr    
& 45 876             & 44       &14 680              & 198.6       &90.07 & -\\     
& 45 932~\cite{NIST_ASD}$^{a}$      & 313~\cite{AndersenPRA97}$^{a}$          &14 705~\cite{NIST_ASD}$^{a}$      & 197.2~\cite{PorsevJOTP06}$^{b}$     &90.18~\cite{JiangADNDT15}$^{b}$&-\\    
Ba    
& 41 915           & 585          &13 044              & 274.2             &121.0    & -\\      
& 42 035~\cite{NIST_ASD}$^{a}$       &870~\cite{PetruninPRL95}$^{a}$            &13 083~\cite{NIST_ASD}$^{a}$       & 273.5~\cite{PorsevJOTP06}$^{b}$ &121.2~\cite{JiangADNDT15}$^{b}$&-\\      
Ra   
& 42 208           & $<0$           &15 261              & 250.5            &106.4    & - \\       
& 42 573~\cite{NIST_ASD}$^{a}$      &807~\cite{AndersenPR04}$^{b}$           &15 391~\cite{NIST_ASD}$^{a}$       & 248.6~\cite{LimPRA04}$^{b}$        &107.4~\cite{LimJCP06}$^{b}$&-\\
\end{tabular*}    
\end{ruledtabular}    
    \begin{flushleft}
    \footnotesize{$^a$ Experimental values. \\
    $^b$ Theoretical values.  }
    \end{flushleft}
\end{table*}             
%****************************************************************************

\section{Results and discussion}
\label{sec:results}
\subsection{Atomic properties}

We begin by analyzing the electronic properties of alkali-metal and alkaline-earth-metal atoms to benchmark the accuracy of the employed \textit{ab initio} methods. Table~\ref{table:atoms} presents the ionization potentials, electron affinities, lowest S-P excitation energies, and static electric dipole polarizabilities for all considered atoms. In addition, static electric dipole polarizabilities are calculated for the corresponding cations and anions. The present theoretical results are compared with the most accurate available experimental and theoretical data.

The calculated ionization potentials and lowest $S$-$P$ excitation energies are in close agreement with experimental data, with deviations of 6--421~cm$^{-1}$ and 2--300~cm$^{-1}$, corresponding to relative errors of 0.01--1.3\% and 0.01--2.2\%, respectively. The static electric dipole polarizabilities of the neutral atoms, corresponding cations, and alkali-metal anions also agree well with available reference values. The absolute differences range from 0 to 11.2~$a_0^3$ for neutral atoms, from 0 to 1.59~$a_0^3$ for cations, and from 12 to 207~$a_0^3$ for anions, corresponding to relative errors of 0--3.8\%, 0--2.1\%, and 1.5--12.7\%, respectively. The calculated electron affinities of alkali-metal atoms agree with experimental data within 3--84~cm$^{-1}$ (0.07--2.1\%). The electron affinities of alkaline-earth-metal atoms are less accurate because of the weakly bound and often metastable nature of their anions. This limitation is not expected to affect the present molecular results significantly, since all relevant dissociation limits involve an alkali-metal anion and a neutral atom.
 
The overall agreement between the calculated atomic properties and the most accurate available experimental and theoretical data is satisfactory. This level of agreement indicates that the chosen methods, basis sets, and energy-consistent pseudopotentials adequately capture scalar relativistic effects and electron correlation, and that the calculations are close to the basis-set convergence limit. Consequently, the present methodology should provide a reliable description of the interatomic interactions and molecular anion properties discussed in the following sections.

\subsection{Molecular electronic structure}

%************FigI************************
\begin{figure} [!tb]
\centering
\includegraphics[width=\columnwidth]{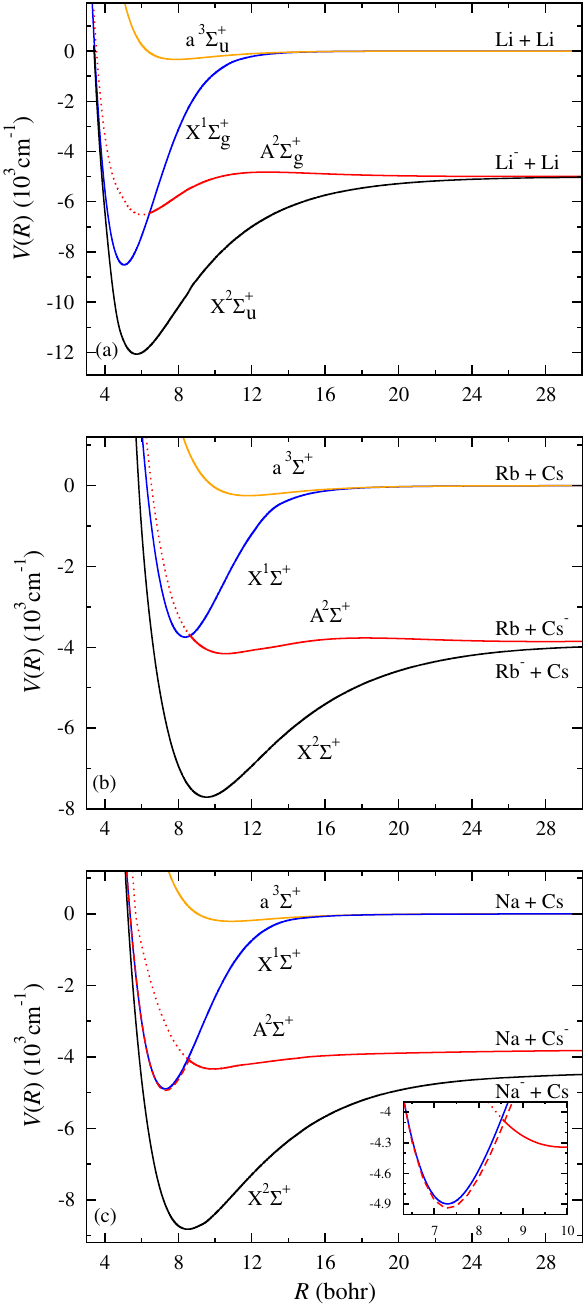}
\caption{Potential energy curves for the lowest X$^2\Sigma^+$ and A$^2\Sigma^{+}$ electronic states of the (a) Li$_2^-$, (b) RbCs$^-$, and (c) NaCs$^-$ alkali-metal molecular anions, together with the X$^{1}\Sigma^{+}$ and a$^{3}\Sigma^{+}$ states of the corresponding neutral molecules. Solid lines denote the valence-bound electronic states, dotted lines indicate unstable anionic states that can spontaneously detach an electron, and the dashed line in panel (c) represents a dipole-bound state.}
\label{fig:excited-states}
\end{figure}
%************************************

To identify the general electronic-structure features of alkali-metal molecular anions, we begin by considering three representative systems: Li$_2^{-}$, RbCs$^{-}$, and NaCs$^{-}$. Figure~\ref{fig:excited-states} presents the potential energy curves (PECs) of their ground X$^2\Sigma^+$ and lowest excited A$^2\Sigma^+$ states obtained with the MRCISD method. For comparison, the X$^1\Sigma^+$ and a$^3\Sigma^+$ states of the corresponding neutral molecules, taken from Ref.~\cite{LadjimiPRA24}, are also shown. The neutral-atom dissociation limit is chosen as the zero of energy, and the anionic PECs are shifted vertically so that their asymptotic energies reproduce the experimental atomic electron affinities. For the homonuclear Li$_2^{-}$ system, the gerade and ungerade symmetry labels are included explicitly. The dipole-bound state in panel (c) is obtained by subtracting the electron binding energy calculated with the EOM-EA-CCSD method from the accurate neutral X$^1\Sigma^+$ curve.

In all three systems, the ground X$^2\Sigma^+$ anionic state lies lowest in energy over the chemically relevant range of internuclear distances and exhibits a pronounced potential minimum. For Li$_2^-$, the X$^2\Sigma_u^+$ and A$^2\Sigma_g^+$ states correlate at large internuclear distances with the same, symmetry-degenerate Li$^-+$Li dissociation limit. In the heteronuclear anions, the two states instead correlate with distinct charge-localized asymptotes: the X$^2\Sigma^+$ state approaches the lower Rb$^-+$Cs and Na$^-+$Cs limits, because Rb and Na have larger electron affinities than Cs, whereas the A$^2\Sigma^+$ state approaches the higher Rb$+$Cs$^-$ and Na$+$Cs$^-$ limits, respectively. For both Li$_2^-$ and RbCs$^-$, the A$^2\Sigma^+$ state exhibits a potential barrier at intermediate internuclear separations, associated with exchange and charge-transfer interactions between configurations in which the excess electron is localized on different atoms.

The neutral X$^1\Sigma^+$ state and the valence-excited anionic A$^2\Sigma^+$ state are close in energy around their potential minima and cross as functions of the internuclear distance. On the short-range side of each crossing, the anionic A$^2\Sigma^+$ state lies above the neutral X$^1\Sigma^+$ state and is therefore embedded in the electron-detachment continuum. These portions of the anionic potentials, shown by dotted lines, correspond to temporary anions that can decay by spontaneous electron detachment. Because they are obtained using a standard bound-state treatment, their energies should be regarded as approximate diabatic continuations of the valence-excited bound states. At larger internuclear distances, the A$^2\Sigma^+$ state falls below the neutral detachment threshold and becomes electronically bound. The resulting near-threshold temporary states may provide resonant pathways for the enhanced capture of low-energy electrons and could therefore play a role in collisions between ultracold ground-state molecules and Rydberg atoms~\cite{GuttridgePRL23,ZhuPRL25}. A detailed treatment of these collision processes and resonance properties is, however, beyond the scope of the present work.

The nature of the near-threshold excited anionic states also depends strongly on the permanent electric dipole moment of the corresponding neutral molecule, which determines whether an additional dipole-bound state can occur. Li$_2$ cannot support such a state because its dipole moment vanishes by inversion symmetry. The dipole moment of RbCs, 1.21~D at equilibrium~\cite{LadjimiPRA24}, is also too small to bind an electron. By contrast, the much larger dipole moment of NaCs, 4.98~D~\cite{LadjimiPRA24}, supports a dipole-bound anionic state. Its PEC, represented by the dashed curve in Fig.~\ref{fig:excited-states}(c), closely follows the neutral X$^1\Sigma^+$ potential while remaining slightly below the electron-detachment threshold, as expected because the diffuse excess electron only weakly perturbs the neutral molecular core. Near $R\approx8.5$~bohr, the dipole-bound and valence-bound states undergo an avoided crossing and exchange their dominant electronic character. Consequently, the lowest excited state of NaCs$^-$ has predominantly dipole-bound character at shorter distances and predominantly valence-bound character at larger distances.

%************FigI************************
\begin{figure*}[!htb]
\centering
\includegraphics [width=\textwidth] {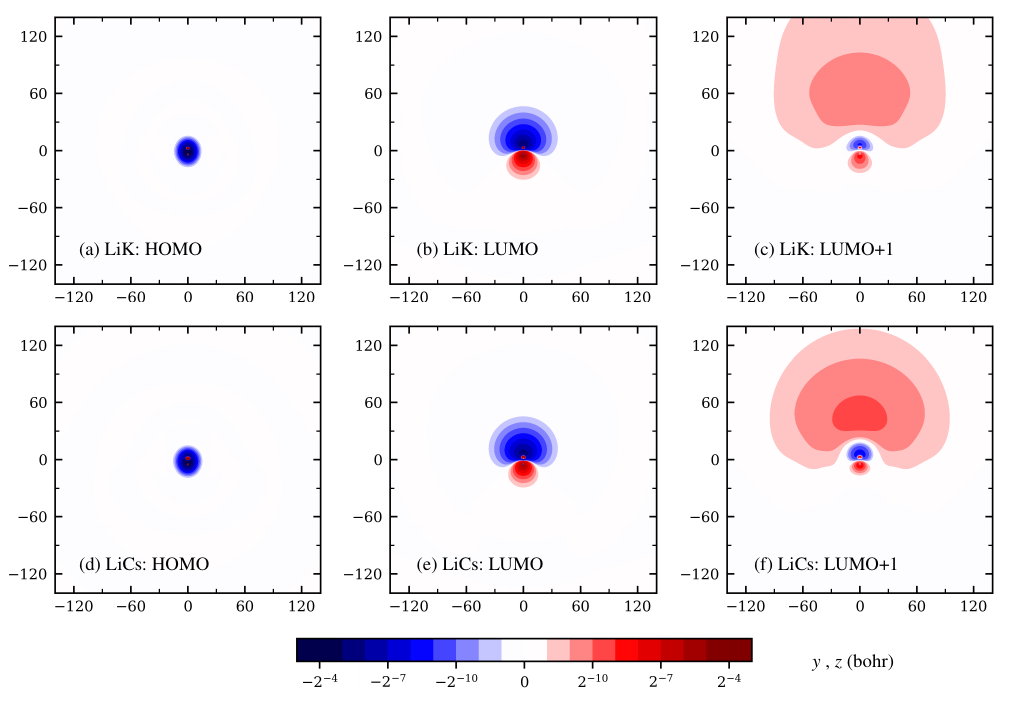}
\caption{Cross sections of selected Hartree--Fock molecular orbitals $\phi_i(x,y,z)$ of LiK (upper row) and LiCs (lower row) in the $yz$ plane of the $C_{2v}$ point group. The origin of the coordinate system is located at the molecular center of mass. Both spatial coordinates are given in bohr. \label{fig:orbitals}}
\end{figure*}
%************************************

%************FigII************************
\begin{figure} [!htb]
\centering
\includegraphics [width=\columnwidth] {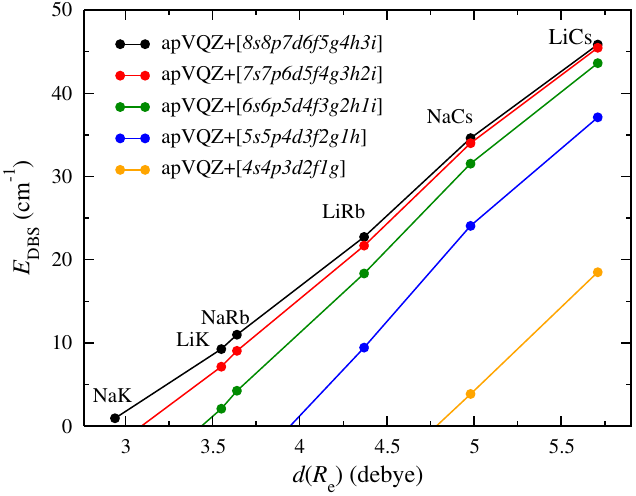}
\caption{Electron binding energies of the dipole-bound states of the NaK$^{-}$, LiK$^{-}$, NaRb$^{-}$, LiRb$^{-}$, NaCs$^{-}$ and LiCs$^{-}$ molecular anions, calculated using the EOM-EA-CCSD method with the aug-cc-pVQZ basis set (apVQZ) and systematically enlarged sets of additional diffuse functions. The energies are plotted as functions of the permanent electric dipole moments of the corresponding neutral molecules at their equilibrium geometries. \label{fig:DM-dE}}
\end{figure}
 %************************************

\subsection{Dipole-bound states}
\label{sec:DBS}

The long-range charge–dipole interaction allows a sufficiently polar molecule to bind an excess electron in a dipole-bound state (DBS). In contrast to a conventional valence-bound state, the excess-electron density in a DBS is localized predominantly outside the molecular valence region and extends over distances much larger than the bond length.

To illustrate this highly diffuse character, Fig.~\ref{fig:orbitals} compares selected canonical restricted Hartree--Fock molecular orbitals of neutral LiK and LiCs at their respective equilibrium geometries, calculated using the aug-cc-pVQZ basis sets augmented by the ($5s5p4d3f2g1h$) set of diffuse functions described in Sec.~\ref{sec:theory}. These species have closely related valence electronic structures but substantially different permanent electric dipole moments of 3.36~D and 5.28~D~\cite{LadjimiPRA24}, respectively. The highest occupied molecular orbital (HOMO) and the valence-like lowest unoccupied molecular orbital (LUMO) are localized mainly in the vicinity of the nuclei and have similar spatial structures in the two molecules. By contrast, the LUMO+1 is exceptionally diffuse and extends over tens of bohr beyond the molecular core. Within the dominant one-electron attachment picture, attachment to the LUMO produces the valence-bound ground state of the anion, whereas attachment to the LUMO+1 is associated with the DBS. The latter orbital is noticeably more diffuse in LiK than in LiCs, consistently with the weaker charge–dipole interaction and smaller electron binding energy in the less polar molecule.

We calculate the electron binding energies of the DBSs, $E_\text{DBS}$, supported by six polar alkali-metal molecules: NaK, LiK, NaRb, LiRb, NaCs, and LiCs. The calculations are performed at the equilibrium geometries of the corresponding neutral molecules. The dipole moments of the selected molecules range from 2.7~D to 5.3~D~\cite{LadjimiPRA24} and  exceed the critical value required for the formation of dipole-bound anion for an electron interacting with an ideal, nonrotating point dipole~\cite{FoxJCP66,LevyLeblondPR67,BrownJCP67}. Molecular rotation increases the effective critical dipole moment, whose precise value depends on the rotational constant, finite size of the molecular charge distribution, polarization, higher multipole moments, and short-range electron correlation. Nevertheless, dipole moments of approximately 2--2.5~D are generally sufficient to support molecular DBSs~\cite{ArdCPL09,CrawfordJCP77,GarrettCPL70,GarrettPRA71}. Among the remaining alkali-metal molecules, only some Fr-containing species are predicted to have sufficiently large dipole moments~\cite{LadjimiPRA24}.

The small binding energies and large spatial extent of DBSs impose stringent requirements on the one-electron basis set. In particular, the basis must accurately represent the electronic wave function far from both nuclei. We therefore augment the atomic orbital basis by even-tempered diffuse Gaussian functions placed on a ghost center at the bond midpoint. The hierarchy of seven increasingly extensive diffuse basis sets is specified in Sec.~\ref{sec:theory}. 

Figure~\ref{fig:DM-dE} presents the calculated electron binding energies as functions of the permanent dipole moments of the neutral molecules. For the least polar species, compact diffuse augmentations do not recover a bound state, whereas the DBSs of the more strongly polar molecules are already obtained with moderately extended basis sets. This behavior reflects the increasing spatial extent of the excess-electron wave function as the dipole moment approaches the binding threshold. The electron binding energies obtained with the two largest diffuse augmentations are similar, indicating satisfactory basis-set convergence. For the six chemically related molecules considered here, the calculated electron binding energies show an approximately monotonic, near-linear correlation with the permanent dipole moment, similar to the empirical trend reported for selected organic anions~\cite{QianJPCL19}. The generality of this trend remains to be established.

The combination of relatively large electron binding energies and small rotational constants makes the most polar alkali-metal molecules particularly promising systems for investigating rotationally excited dipole-bound states. For NaCs and LiCs, the ratio $E_\text{DBS}/B_e$ reaches several hundred, suggesting that a substantial number of rotational levels may remain below the electron-detachment threshold, providing a unique opportunity to study DBSs in high rotational states.

\subsection{Ground-state potential energy curves}

We compute the ground-state potential energy curves for all 57 diatomic molecular anions considered in this work: 21 alkali-metal anions composed of Li, Na, K, Rb, Cs, and Fr atoms, and 36 alkali-metal--alkaline-earth-metal anions additionally containing Be, Mg, Ca, Sr, Ba, or Ra. The PECs of the alkali-metal diatomic anions in their X$^{2}\Sigma^{+}$ ground electronic states are presented in Fig.~\ref{fig:Alk-Alk}, whereas those of the alkali-metal--alkaline-earth-metal anions in their X$^{1}\Sigma^{+}$ ground electronic states are shown in Fig.~\ref{fig:Alk-AlkE}.

All calculated PECs exhibit a smooth behavior with well-defined minima. Tables~\ref{table:Alk-Alk} and~\ref{table:Alk-AlkE} collect the resulting equilibrium molecular properties and spectroscopic constants: the equilibrium internuclear distance $R_e$, well depth $D_e$, harmonic vibrational constant $\omega_e$, first-order anharmonicity constant $\omega_e x_e$, equilibrium rotational constant $B_e$, permanent electric dipole moment $d_e=d(R_e)$, and the perpendicular and parallel components of the static electric dipole polarizability, $\alpha_e^{\perp}=\alpha_{\perp}(R_e)$ and $\alpha_e^{\parallel}=\alpha_{\parallel}(R_e)$, respectively. Electronic adiabatic electron affinities are also reported and are calculated as $E_{\mathrm{EA}}(AB)=D_e(AB^-)-D_e(AB)+\max\!\left[
E_{\mathrm{EA}}(A),
E_{\mathrm{EA}}(B)
\right]$, where \(E_{\mathrm{EA}}(A)\) and \(E_{\mathrm{EA}}(B)\) are the experimental electron affinities of the constituent atoms, $D_e(AB^-)$ is obtained in the present work, and $D_e(AB)$ is taken from Ref.~\cite{LadjimiPRA24}. Available experimental and theoretical literature values are included for comparison. The vibrational and rotational constants are evaluated using the atomic masses of the most abundant stable isotopes, with $^{223}$Fr and $^{228}$Ra used for Fr and Ra.

For most of the studied molecular anions, the present calculations constitute the first accurate predictions of the ground-state potential energy curves and spectroscopic constants. Full potential energy curves in a numerical form are collected in
the Supplemental Material~\cite{SM}.

\subsubsection{Alkali-metal diatomic anions}
 %******************FigI**************************
 \begin{figure*} [!htb]
%\centering
\includegraphics[width=1\textwidth]{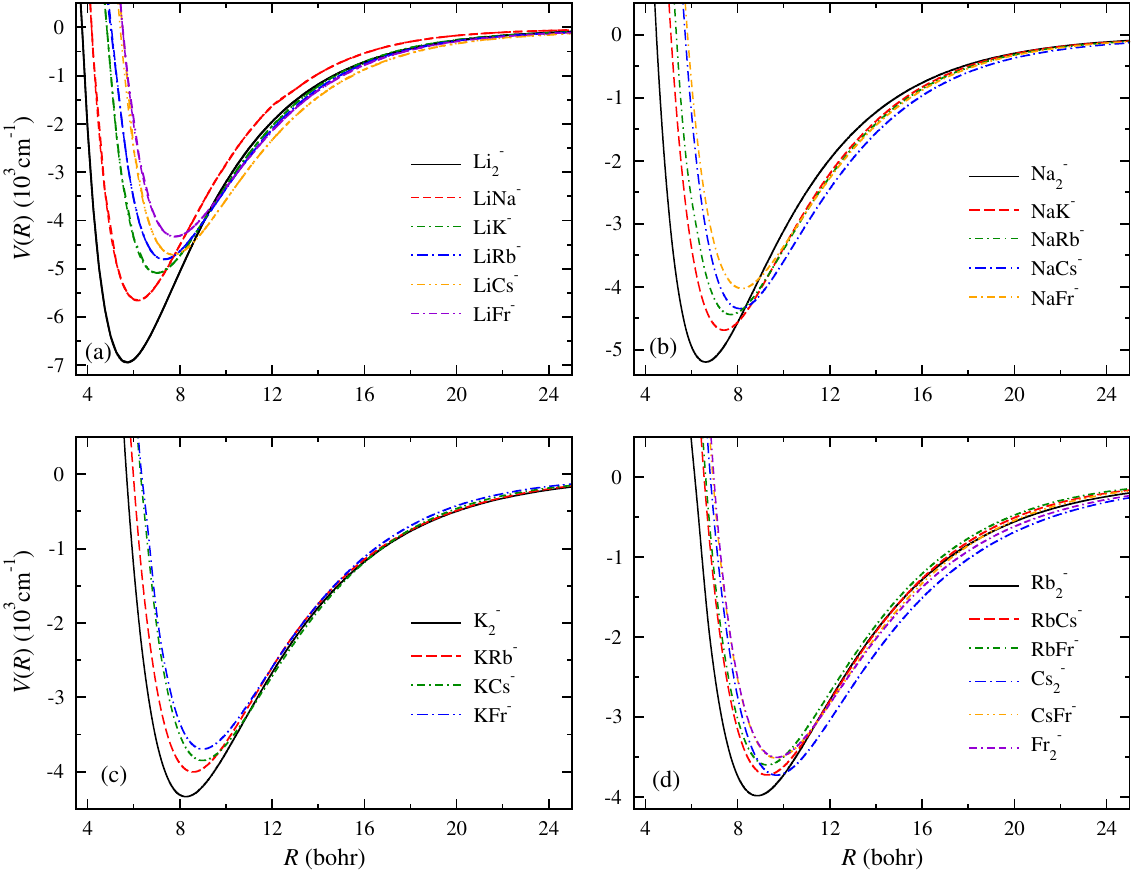}
\centering
\caption{Potential energy curves of alkali-metal diatomic molecular anions in the ground X$^2\Sigma^+$ electronic state.}
\label{fig:Alk-Alk}
\end{figure*}
%*****************************************

%******************************************TABII******************************************
\begin{table*}[!htb]
\begin{ruledtabular}
\caption{Characteristics of alkali-metal diatomic molecular anions in the X$^{2}\Sigma^{+}$ electronic ground state: equilibrium internuclear distance $R_e$, well depth $D_e$, harmonic vibrational constant $\omega_e$, first-order anharmonicity constant $\omega_ex_e$, equilibrium rotational constant $B_e$, equilibrium permanent electric dipole moment $d_e$, perpendicular and parallel components of the static electric dipole polarizability at equilibrium, $\alpha_e^{\perp}$ and $\alpha_e^{\parallel}$, respectively, electronic adiabatic electron affinity $E_\text{EA}$, and the lowest dissociation limit. Available experimental and theoretical results are included for comparison.}
\label{table:ener}
\begin{tabular*}{\textwidth}{@{\extracolsep{\fill}}llllllllllll}
 Anion & $R_e$ (bohr)& $D_e$ (cm${}^{-1}$)& $\omega_e$ (cm${}^{-1}$) & $\omega_ex_e$ (cm${}^{-1}$) & $B_e$ (cm${}^{-1}$) & $d_e$ (D) & $\alpha_e^\perp$ ($a_0^3$) & $\alpha_e^\parallel$ ($a_0^3$) & $E_\text{EA}$ (cm${}^{-1}$) & Diss. & Ref. \\
 \hline
  Li$_2^-$  
  & 5.713   & 6940      & 234.7   & 2.537 & 0.5258   & 0  & 629  & 1909 & 3408 &Li$^{-}$+ Li &  This Work\\
  & 5.813   & 5714      & 231.3   & 2.36  & 0.508    & -  & -    & - &-&&Theo.~\cite{MichelsCPL85}\\   
  & 6.0017  & -         & 212     & -     & -        & -  & -    & - & - & & Theo.~\cite{BoldyrevJCP93}\\
  & 5.787 & 6912  & 236.8 & 2.42 & - & -  & - & - & 3355 & & Theo.~\cite{HogreveEPJD00} \\
    & 5.858 & 6237 &  235.3  & 3.166  & 0.4652   & -  & -    & - & - & & Theo.~\cite{NasiriCTC17} \\
  & 5.779 & 6680 &      & -     & -        & -  & -    & - & 3253 & & Theo.~\cite{DunningJCC24} \\
  & 5.85(3) & 7152(200) & 232(35) & -     & 0.502(5) & -  & -    & - & 3465(73) &&Expt.~\cite{SarkasZPD94}\\
  LiNa$^{-}$ 
  &6.172    &5657       & 171.3   & 2.378 & 0.2939   & -3.234  & 669 & 1952 &3549 & Li$^{-}$+Na & This Work \\
  &6.446    &  -        & 153     & -     &-             &-   &- &-&- & &Theo.~\cite{BoldyrevJCP93}\\   
  LiK$^{-}$ 
  &7.010 &5088  &141.9  &0.932 & 0.2060 &-2.608 &1022 &1868&3868 &Li$^{-}$+ K&This Work\\  
  LiRb$^{-}$ 
  &7.319 &4808  &130.4   &0.874  &0.1734 &-4.656 &1079 &1985&3883 &Li$^{-}$+ Rb&This Work  \\    
  LiCs$^{-}$ 
  &7.707 &4708   &124.6   &0.984  &0.1521 &-5.299&1256 &2153 & 3868 &Li$^{-}$+ Cs&This Work  \\  
  LiFr$^{-}$ 
  &7.792 &4333   &119.7    &0.922 &0.1458 &-7.185 &1209 &2323&3850 & Li$^{-}$+ Fr&This Work  \\    
  Na$_2$$^{-}$ 
  &6.621 &5192 &104.4 & 0.670 & 0.1195 &0   &737 &2004 &3609 &Na$^{-}$+ Na&This Work \\    
  & - & 5004(160) & -  & -  & - & - & - & - & 3468(120) &  & Expt.~\cite{EatonCPL92}\\
  NaK$^{-}$ 
   &7.413 &4687 &83.6  &0.464  &0.0758 &-0.112  &909&2086 &3855 &Na$^{-}$+ K&This Work \\     
  & - & 4575(240) & -  & -  & - & - & - & - & 3751(240) &  & Expt.~\cite{EatonCPL92}\\
  NaRb$^{-}$  
  &7.709 &4440 &71.9 &0.398 & 0.0559&-2.918 & 1065 &2180 &3863 &Na$^{-}$+ Rb&This Work \\
  NaCs$^{-}$ 
  &8.097 &4347 &66.9   &0.355 & 0.0468 &-4.542 &1367 &2352 &3866 &Na$^{-}$+ Cs&This Work\\
  NaFr$^{-}$ 
  &8.195 &4026 &64.3   &0.426  &0.0429 &-6.916  &1246 &2383 &3843&Na$^{-}$+ Fr &This Work  \\ 
  K$_2$$^{-}$ 
  &8.266 &4336  &62.6 & 0.252 & 0.0452 &0 &1155 &2741&3957 &K$^{-}$+ K&This Work\\
  & - & 4145(160) & -  & -  & - & - & - & - & 4009(100) &  & Expt.~\cite{EatonCPL92}\\
  KRb$^{-}$  
  &8.579 &4006 &51.2 &   0.197 & 0.0306&-3.401&1241 &2867 &3871 &Rb$^{-}$+ K&This Work\\ 
  & - & 3897(160) & -  & -  & - & - & - & - & 3920(160) &  & Expt.~\cite{EatonCPL92}\\
  KCs$^{-}$
  &8.985  &3848  &46.3 &   0.116 & 0.0247 &-4.888 &1432 &3068  &3888 &K$^{-}$+ Cs&This Work  \\ 
  & - & 3540(160) & -  & -  & - & - & - & - & 3791(160) &  & Expt.~\cite{EatonCPL92}\\
  KFr$^{-}$ 
  &9.001 &3694 &43.8 &   0.139 & 0.0224 &-7.783  &1354 &3055&3892 &K$^{-}$+ Fr&This Work   \\  
  Rb$_2^-$ 
  &8.856  &3987  &38.2  & 0.119   & 0.0181 &0 &1321 &3040&3964 &Rb$^{-}$+ Rb&This Work \\  
  & 8.97  & 4137  & 28.3  & -  & - & - & - & - & - & & Theo.~\cite{KraussJCP90}\\ 
  &$$-$$  &4048  &$$-$$  &$$-$$  & $$-$$ &$$-$$&$$-$$ &-&$$-$$& &Expt.~\cite{MchughJCP89}\\
  & - & 4036(160) & -  & -  & - & - & - & - & 4017(120) &  & Expt.~\cite{EatonCPL92}\\  
  RbCs$^{-}$ 
  & 9.282 & 3725 & 33.9 &   0.085 & 0.0135 & -1.733 & 1489 & 3273 & 3885 & Rb$^{-}$+Cs & This Work\\ 
  & - & 3759(160) & -  & -  & - & - & - & - & 3855(160) &  & Expt.~\cite{EatonCPL92}\\   
  RbFr$^{-}$ 
  &9.288 &3601&30.9 &   0.041& 0.0113&-5.350 & 1449  &3296 &3892 &Rb$^{-}$+ Fr&This Work    \\ 
  Cs$_2$$^{-}$ 
  &9.691 &3728  &28.6   &0.067  & 0.0096 &0 &1627 &3620&3981 &Cs$^{-}$+ Cs&This Work\\ 
  & 9.72 &3621  &28.4   &0.042  & $$-$$ &$$-$$&$$-$$ &$$-$$&-& &Theo.~\cite{KraussJCP90}\\ 
  & - & 3629 & -  & -  & - & - & - & - & - &  & Expt.~\cite{MchughJCP89}\\
  & - & 3620(160) & -  & -  & - & - & - & - & 3783(120) &  & Expt.~\cite{EatonCPL92}\\   
  CsFr$^{-}$ 
  &9.688 &3515  &  25.6&   0.055 & 0.0077 &-3.525&1523 &3334 &3911 &Fr$^{-}$+ Cs&This Work\\ 
  Fr$_2$$^{-}$
  & 9.686 & 3506 & 22.0 & 0.042 & 0.0057 & 0 & 1505 & 3375 & 4009 & Fr$^-$+Fr & This Work\\
\end{tabular*}
\label{table:Alk-Alk}
\end{ruledtabular}
\end{table*}

%****************************************************************************

The alkali-metal diatomic molecular anions in their X$^{2}\Sigma^{+}$ electronic ground states are relatively strongly bound. Averaged over all 21 alkali-metal diatomic anions, the calculated well depth is $4389~\mathrm{cm}^{-1}$, approximately $500~\mathrm{cm}^{-1}$ larger than the average value of $3892~\mathrm{cm}^{-1}$ for the corresponding neutral molecules~\cite{LadjimiPRA24}. The well depths range from $D_e=6940~\mathrm{cm}^{-1}$ for the lightest and most strongly bound anion, Li$_2^{-}$, to $D_e=3506~\mathrm{cm}^{-1}$ for the least strongly bound anion, Fr$_2^{-}$. The equilibrium distances span $5.713$--$9.691~\mathrm{bohr}$, with an average value of $8.146~\mathrm{bohr}$; the limiting values correspond to Li$_2^{-}$ and Cs$_2^{-}$, respectively. 

As shown in Fig.~\ref{fig:Alk-Alk}, within each series in which one alkali-metal atom is fixed, the calculated well depth decreases systematically as the atomic number of the other alkali-metal atom increases. Overall, shallower potential wells are associated with longer equilibrium distances. This trend primarily reflects the increase in atomic size, which shifts the equilibrium distance outward; the resulting weakening of the leading charge-induced-dipole attraction, which scales asymptotically as $-\alpha/(2R^4)$, outweighs the general increase in the polarizability of the neutral fragment and leads to progressively shallower wells.

Table~S.1 of the Supplemental Material~\cite{SM} compares the spectroscopic constants obtained at the CCSD(T) and CCSD(T)$+\Delta$T levels, where $\Delta$T denotes the iterative-triples correction defined in Eq.~\eqref{eq:VSDT}. Inclusion of this correction increases the well depths by an average of $107~\mathrm{cm}^{-1}$ ($2.4\%$) and decreases the equilibrium distances by an average of $0.01~\mathrm{bohr}$ ($0.1\%$). Thus, post-CCSD(T) correlation effects modify the well depths at the few-percent level and should be included when quantitative accuracy at this level is required.

The calculated spectroscopic constants are compared with the available experimental and theoretical results in Table~\ref{table:Alk-Alk}. Li$_2^{-}$ was investigated experimentally by photoelectron spectroscopy, and its spectroscopic constants were extracted through a
Franck--Condon analysis~\cite{SarkasZPD94}. Our equilibrium distance is approximately $0.14~\mathrm{bohr}$ shorter than the reported experimental value, while the harmonic vibrational constant differs by approximately $3~\mathrm{cm}^{-1}$ and lies well within the relatively large experimental uncertainty. The calculated well depth is lower than the experimental result by $212~\mathrm{cm}^{-1}$, a
difference comparable to the quoted experimental uncertainty.
Photoelectron spectra have also been reported for Na$_2^{-}$, NaK$^{-}$, K$_2^{-}$, KRb$^{-}$, KCs$^{-}$, Rb$_2^{-}$, RbCs$^{-}$, and Cs$_2^{-}$~\cite{EatonCPL92}. Across all nine molecular anions for which experimental well depths are 
available, including Li$_2^{-}$, the mean absolute deviation of the present results is 146~cm$^{-1}$, or 3.4\%. The agreement is within 5\% for all systems except KCs$^{-}$, whose calculated well depth is larger than the experimental estimate by 308~cm$^{-1}$ (8.7\%), still within approximately twice the reported uncertainty. The corresponding electronic adiabatic electron affinities have mean absolute and relative deviations of 87~cm$^{-1}$ and 2.3\%, respectively, well within experimental uncertainty. Overall, the comparison indicates that the present interaction potentials and electron affinities reproduce the available measurements at the level of a few percent. 

%******************FigII**************************
\begin{figure*}[!tb]
\includegraphics[width=1\textwidth]{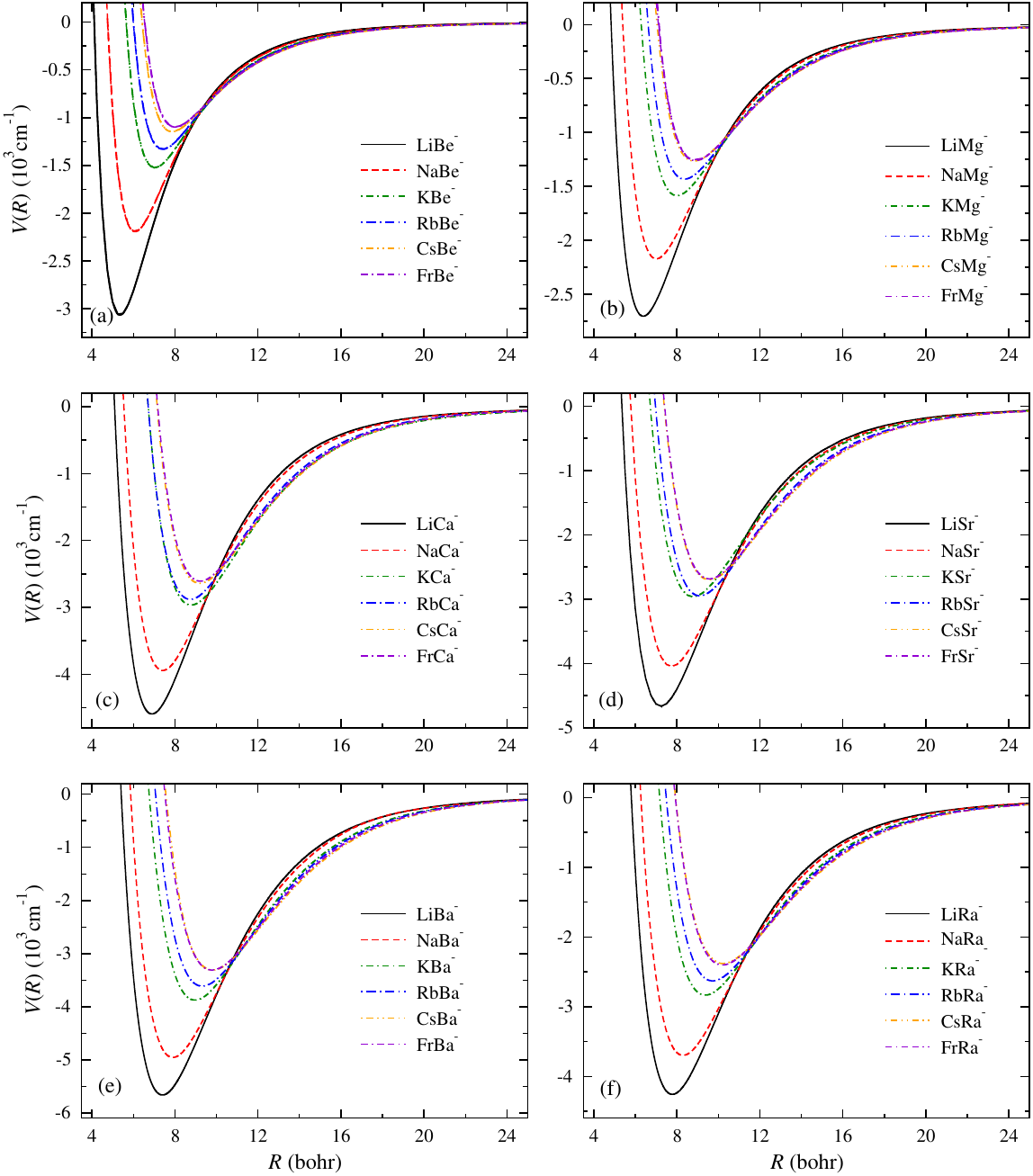}
\centering
\caption{Potential energy curves of all the alkali-metal--alkaline-earth-metal diatomic molecular anions in the ground X$^1\Sigma^+$ electronic state.}
\label{fig:Alk-AlkE}
\end{figure*}
 %*****************************************

%******************************************TABIII******************************************
\begin{table*} [!tb]
\begin{ruledtabular}
\caption{Characteristics of alkali-metal--alkaline-earth-metal diatomic molecular anions in the X$^{1}\Sigma^{+}$ electronic ground state: equilibrium internuclear distance $R_e$, well depth $D_e$, harmonic vibrational constant $\omega_e$, first-order anharmonicity constant $\omega_ex_e$, equilibrium rotational constant $B_e$, equilibrium permanent electric dipole moment $d_e$, perpendicular and parallel components of the static electric dipole polarizability, $\alpha_e^{\perp}$ and $\alpha_e^{\parallel}$, respectively, electronic adiabatic electron affinity $E_\text{EA}$, and the lowest dissociation limit. Available theoretical results are included for comparison.} 
\label{table:ener}
\begin{tabular*}{\textwidth}{@{\extracolsep{\fill}}llllllllllll}
 Anion & $R_e$ (bohr)& $D_e$ (cm${}^{-1}$)& $\omega_e$ (cm${}^{-1}$) & $\omega_ex_e$ (cm${}^{-1}$) & $B_e$ (cm${}^{-1}$) & $d_e$ (D) & $\alpha_e^\perp$ ($a_0^3$) & $\alpha_e^\parallel$ ($a_0^3$) & $E_\text{EA}$ (cm${}^{-1}$) & Diss. & Ref. \\
 \hline
  LiBe$^{-}$
  &5.353           &3066      &232.7       &5.665     &0.5326      &-3.833   &539 &881 &5498 & Li$^{-}$+ Be &This Work\\
  & 5.490 & 2904 & 269 & - & - & - & - & - & - & & Theo.~\cite{BauschlicherJCP92}\\
  &5.577           & -      &196       &-      &-              &-    &-  &- &-&&Theo.~\cite{BoldyrevJCP93}\\  
  NaBe$^{-}$ 
  &6.080         &2191     &144.5     &3.232    &0.2515   &-1.426   &638  & 1042  &5314 & Na$^{-}$+ Be&This Work\\  
    & 6.213 & 2014 & 143 & - & - & - & - & - & - & & Theo.~\cite{BauschlicherJCP92}\\ 
  KBe$^{-}$ 
  &7.036 &1524  &106.6 &   2.375 & 0.1662 &-0.399   &1163 &1438 &4662 & K$^{-}$+ Be &This Work\\   
  RbBe$^{-}$ 
  & 7.417 &1330  &92.4 &   2.199 & 0.1343 &0.361 &1345 &1624 &4462 & Rb$^{-}$+ Be&This Work\\  
  CsBe$^{-}$ 
  & 7.866 &1147  &82.2 &   2.079 & 0.1153 &0.770 &1720 &1941 &4218 & Cs$^{-}$+ Be&This Work\\ 
  FrBe$^{-}$ 
  &8.008 &1098 & 78.2 &   1.716 & 0.1084   &1.439 & 1512  &1815& 4314 & Fr$^{-}$+ Be &This Work\\   
  LiMg$^{-}$ 
  & 6.387 &2702   &147.8 &   2.577 & 0.2719  &-6.054 &483 &1025 & 6118&Li$^{-}$+ Mg&This Work\\
  & 6.412 & 2823 & 147 & - & - & - & - & - & - & & Theo.~\cite{BauschlicherJCP92}\\
  & 6.650 &-  & 124  & - & - & - & - & - & - & & Theo.~\cite{BoldyrevJCP93}\\        
  NaMg$^{-}$ 
   &7.016 &2171 &86.1  &0.953  &0.1042   &-3.838  &603  &1234 &5677 & Na$^{-}$+ Mg&This Work\\ 
   & 7.107 & 2178 & 89  & - & - & - & - & - & - & & Theo.~\cite{BauschlicherJCP92}\\
  KMg$^{-}$  
  &8.004 & 1585  &61.3 &   0.645 & 0.0633 &-3.619 &1050   &1681 & 4899 &K$^{-}$+ Mg&This Work\\  
  RbMg$^{-}$ 
  &8.367 &1433  &51.3&   0.599 & 0.0459   &-0.882 &1239  &1876 & 4685 &Rb$^{-}$+ Mg & This Work\\ 
 CsMg$^{-}$ 
  &8.834 &1265 & 45.1 &   0.523 & 0.0379 &0.074 &1547 &2184 &4425 & Cs$^{-}$+ Mg & This Work\\ 
 FrMg$^{-}$ 
 &8.891 &1260 &44.0 &   0.433 & 0.0352   &1.482   &1375  &2220 &4527 & Fr$^{-}$+ Mg &This Work\\ 
 LiCa$^{-}$  
 &6.891 &4592  &161.6 &   1.679 & 0.2124  &-4.399  &488  &1137  &7016 &Li$^{-}$+ Ca& This Work\\ 
 NaCa$^{-}$
 & 7.413 & 3943 &92.6 &   0.643 & 0.0751 &-2.107 &516     & 1314 & 6723& Na$^{-}$+ Ca&This Work\\ 
 KCa$^{-}$ 
 &8.399  &3092  &66.6 &   0.422  & 0.0432  &-1.744&895     &1843 & 5878 &K$^{-}$+ Ca&This Work\\
 &8.732&3047 &65.51 & -  & 0.0399  & -2.5$^a$ & -     &- &-&&Theo.~\cite{MoussaNJP21}\\ 
 RbCa$^{-}$ 
 &8.740 &2880 &53.7 &   0.219 & 0.0290  &1.677  &1031 & 2052 & 5659 & Rb$^{-}$+ Ca  & This Work\\ 
 CsCa$^{-}$ 
 &9.195 &2640  & 47.6 &   0.244 & 0.0232 &3.225  &1300 &2373 & 5346 &Cs$^{-}$+ Ca &This Work\\  
 FrCa$^{-}$ 
 &9.220 &2607 &45.5&   0.218 & 0.0209 &5.180  &1137  & 2285 & 5478 &Fr$^{-}$+ Ca&This Work\\ 
 LiSr$^{-}$ 
 &7.233 &4666 &148.7   &   1.165 & 0.1771 &-6.001 &589  &1194 & 7170 &Li$^{-}$+ Sr&This Work\\
 NaSr$^{-}$ 
 & 7.742 &4045  &80.0 &   0.361  & 0.0551 &-4.842 &640  &1333 & 6849  &Na$^{-}$+ Sr&This Work\\
 KSr$^{-}$
  &8.777 &2965 &54.2 &   0.248 & 0.0289 &-5.277 &917 &1951 & 5779 &K$^{-}$+ Sr &This Work\\  
 RbSr$^{-}$ 
 & 9.102 & 2947  &41.3 &   0.175 & 0.0168 &-1.744 &1035  &2157 & 5749 &Rb$^{-}$+ Sr &This Work\\  
 CsSr$^{-}$ 
 &9.572 &2697 &35.0 &   0.130 & 0.0124 &0.002  &1282  &2521 & 5428 &Cs$^{-}$+ Sr&This Work\\ 
 FrSr$^{-}$ 
 &9.573 &2685  &32.2 &   0.037 & 0.0104  &2.811  & 1156  &2376 & 5569  &Fr$^{-}$+ Sr &This Work\\
 LiBa$^{-}$ 
 &7.402 &5664  &154.7&   1.244 & 0.1646  &-5.360  &663 &1224 & 7306 &Li$^{-}$+ Ba &This Work\\ 
 NaBa$^{-}$ 
 &7.900 &4956  &81.6 &   0.406  & 0.0489  &-4.788 & 714  &1503 & 7174 &Na$^{-}$+ Ba  &This Work\\  
 KBa$^{-}$ 
 & 8.964 &3876 &54.6 &   0.221 & 0.0247  &-5.724  & 949 &1960 & 6373 &K$^{-}$+ Ba&This Work\\   
 RbBa$^{-}$ 
 & 9.316 & 3615  &39.5&   0.122 & 0.0132 &-2.781  &1052  &2163 & 6138 &Rb$^{-}$+ Ba  &This Work\\
 CsBa$^{-}$ 
 &9.808 &3309 &32.6 &   0.089 & 0.0092 & -1.004  &1258  &2519 & 5808 &Cs$^{-}$+ Ba  &This Work\\
 FrBa$^{-}$ 
 &9.782 &3312 &29.4 &   0.071 & 0.0074  &2.188  &1137  &2352 & 5983 &Fr$^{-}$+ Ba &This Work\\  
 LiRa$^{-}$ 
 & 7.789 &4259  &133.0 &   1.144  & 0.1459   &-8.423  &657  &1223 & 7084  &Li$^{-}$+ Ra&This Work\\ 
 NaRa$^{-}$ 
 &8.316 &3523  &67.9 &   0.379 & 0.0418 &-8.881  &729  &1477 & 6531 &Na$^{-}$+ Ra&This Work\\  
 KRa$^{-}$ 
 &9.375 &2835  &44.7 &   0.212 & 0.0206 &-10.317 &1009 &2104 & 5818 &K$^{-}$+ Ra &This Work\\
 RbRa$^{-}$ 
 &9.729 & 2630  &31.1 &   0.078 & 0.0103 &-8.154  & 1127 &2330 & 5583 &Rb$^{-}$+ Ra &This Work\\  
 CsRa$^{-}$ 
 &10.229 &2381  & 24.9 &   0.090 & 0.0069 & -6.807   &1371  &2723 & 5262 &Cs$^{-}$+ Ra &This Work\\ 
 FrRa$^{-}$ 
 &10.195 &2402  &21.9 &   0.063 & 0.0052 &-3.523  &1239  &2537 & 5407 &Fr$^{-}$+ Ra &This Work\\
\end{tabular*}
\label{table:Alk-AlkE}
\end{ruledtabular}
\begin{flushleft}
\footnotesize{$^a$ Value transformed to a coordinate system with its origin at the center of mass. }
\end{flushleft}
\end{table*}
%****************************************************************************

Earlier theoretical results are available only for a few of the alkali-metal diatomic anions and show varying levels of agreement with the present calculations. For Li$_2^{-}$, the earlier optimized-configuration-interaction calculation of Ref.~\cite{MichelsCPL85} underestimates the present well depth by $1226~\mathrm{cm}^{-1}$ and predicts an equilibrium distance longer by $0.100~\mathrm{bohr}$. Subsequent calculations~\cite{HogreveEPJD00,NasiriCTC17,DunningJCC24} yield equilibrium distances $0.066$--$0.145~\mathrm{bohr}$ longer than the present value and well depths lower by $28$--$703~\mathrm{cm}^{-1}$. The closest agreement is obtained in Ref.~\cite{HogreveEPJD00}, with differences of only $0.074~\mathrm{bohr}$ and $28~\mathrm{cm}^{-1}$ in $R_e$ and $D_e$, respectively. The harmonic vibrational constants reported in Refs.~\cite{HogreveEPJD00,NasiriCTC17} also agree closely with the present result, differing by $2.1$ and $0.6~\mathrm{cm}^{-1}$, respectively. For LiNa$^{-}$, our equilibrium distance is approximately $0.27~\mathrm{bohr}$ shorter than that reported in Ref.~\cite{BoldyrevJCP93}. Previous calculations for Rb$_2^{-}$ and Cs$_2^{-}$ employed compact effective potentials and core-polarization potentials, reducing each alkali-metal atom to a single explicitly treated valence electron~\cite{KraussJCP90}. Relative to those results, the present equilibrium distances differ by $0.114$ and $0.029~\mathrm{bohr}$, while the well depths differ by approximately $150$ and $107~\mathrm{cm}^{-1}$ for Rb$_2^{-}$ and Cs$_2^{-}$, respectively.

\subsubsection{Alkali-metal–alkaline-earth-metal diatomic anions}

The alkali-metal--alkaline-earth-metal molecular anions in their X$^{1}\Sigma^{+}$ electronic ground states are, on average, less strongly bound than the alkali-metal diatomic anions. Averaged over all 36 molecular anions, the calculated well depth is 2869~cm$^{-1}$, which is more than twice the average value of 1311~cm$^{-1}$ for the corresponding neutral molecules~\cite{LadjimiPRA24}, but approximately 1520~cm$^{-1}$ smaller than the average value for the alkali-metal diatomic anions. The well depths range from $D_e=1098$~cm$^{-1}$ for FrBe$^{-}$ to $D_e=5664$~cm$^{-1}$ for LiBa$^{-}$. The equilibrium distances span 5.353--10.229~bohr, with an average value of 8.331~bohr; the limiting values correspond to LiBe$^{-}$ and CsRa$^{-}$, respectively. Compared with the alkali-metal diatomic anions in their X$^{2}\Sigma^{+}$ electronic ground states, the alkali-metal--alkaline-earth-metal anions have slightly larger equilibrium distances, on average, by about 0.19~bohr.

As shown in Fig.~\ref{fig:Alk-AlkE}, for a fixed alkaline-earth-metal atom, the calculated well depth generally decreases as the atomic number of the alkali-metal atom increases, consistent with the size- and distance-dependent trend discussed above for alkali-metal diatomic anions. Accordingly, within each series, deeper potential wells are associated with shorter equilibrium distances. The Li-containing species are the most strongly bound and have the shortest equilibrium distances for every alkaline-earth-metal atom. At the heavy-alkali end of each series, the Cs- and Fr-containing anions have very similar potential energy curves. The Fr-containing species are slightly less strongly bound for Be, Mg, Ca, and Sr, whereas small reversals occur for Ba and Ra, with FrBa$^{-}$ and FrRa$^{-}$ being slightly more strongly bound than their Cs-containing counterparts. The variation with the alkaline-earth-metal atom is less regular. For a fixed alkali-metal atom, the equilibrium distance increases systematically from Be to Ra and would, by itself, favor progressively weaker binding. Instead, the well depth generally increases from Be or Mg to Ba and then decreases for Ra, broadly following the variation in the polarizability of the neutral alkaline-earth-metal atom. The Ba-to-Ra reduction reflects both the lower polarizability of Ra, associated with relativistic contraction and stabilization of its $7s$ shell, and its longer equilibrium distance. Consequently, the Ba-containing anions have the deepest potential wells in every series despite their relatively long equilibrium distances.

Table~S.2 of the Supplemental Material~\cite{SM} compares the spectroscopic constants obtained at the CCSD(T) and CCSD(T)$+\Delta$T levels. Inclusion of the iterative-triples correction increases the well depths by an average of 97~cm$^{-1}$ (3.4\%), and decreases the equilibrium distances by an average of 0.01~bohr (0.1\%). Iterative triple excitations therefore have a modest but systematic effect on the interaction energies and should be retained for quantitatively accurate predictions.

The calculated spectroscopic constants are compared in Table~\ref{table:Alk-AlkE} with previous theoretical results, which are available for only a few species; to our knowledge, no experimental spectroscopic data are currently available for these molecular anions. For LiBe$^{-}$, NaBe$^{-}$, LiMg$^{-}$, and NaMg$^{-}$, the equilibrium distances reported in Ref.~\cite{BauschlicherJCP92} differ from the present values by 0.025--0.137~bohr, while the well depths differ by 7--177~cm$^{-1}$ and the harmonic constants generally agree within a few cm$^{-1}$, except for LiBe$^{-}$. For LiBe$^{-}$ and LiMg$^{-}$, our equilibrium distances are shorter than those reported in Ref.~\cite{BoldyrevJCP93} by 0.22 and 0.26~bohr, respectively, corresponding to differences of approximately 4\%. For KCa$^{-}$, relative to the results of Ref.~\cite{MoussaNJP21}, the present well depth is larger by 45~cm$^{-1}$ (1.5\%), the equilibrium distance is shorter by 0.33~bohr (3.8\%), and the harmonic constants differ by only 1.1~cm$^{-1}$. Overall, the available theoretical results are in reasonable agreement with the present calculations, with the remaining differences likely reflecting the less complete treatment of electron correlation and the less extensive basis sets employed in the previous calculations.

\subsection{Convergence and accuracy}

We assess the convergence of the calculated interaction energies with respect to the one-electron basis set and the treatment of electron correlation for two representative systems: the open-shell KRb$^{-}$ anion in its X$^{2}\Sigma^{+}$ ground state and the closed-shell RbSr$^{-}$ anion in its X$^{1}\Sigma^{+}$ ground state.

%************FigIII************************
 \begin{figure} [!tb]
\centering
\includegraphics [width=\columnwidth] {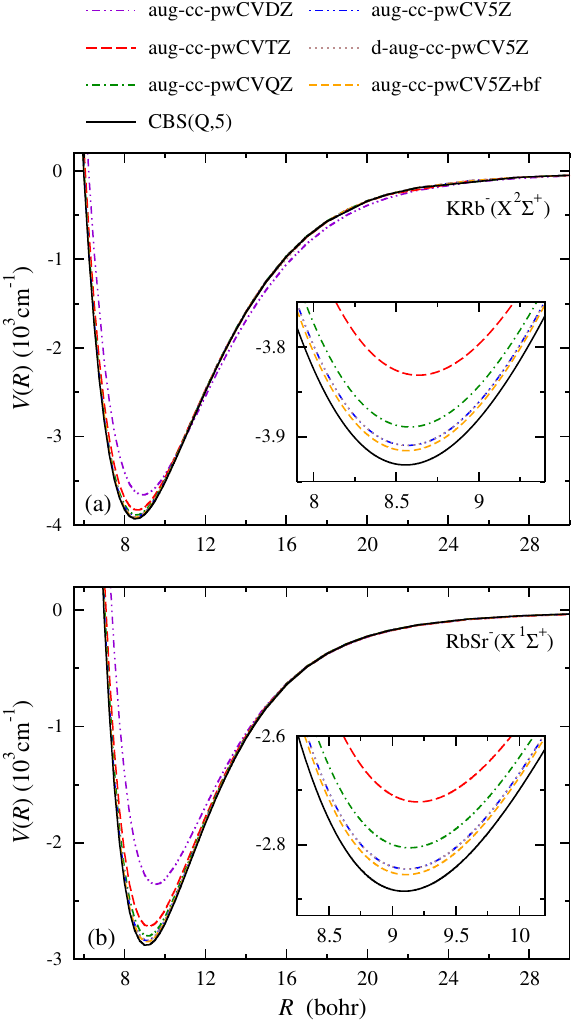}
\centering
\caption{Potential energy curves of the (a) KRb$^{-}$ and (b) RbSr$^{-}$ molecular anions in their X$^{2}\Sigma^{+}$ and X$^{1}\Sigma^{+}$ electronic ground states, respectively, computed at the CCSD(T) level using different basis sets.}
\label{Fig.3}
\end{figure}
 %************************************

 %************FigVI***********************
\begin{figure}[!tb]
\includegraphics[width=\columnwidth]{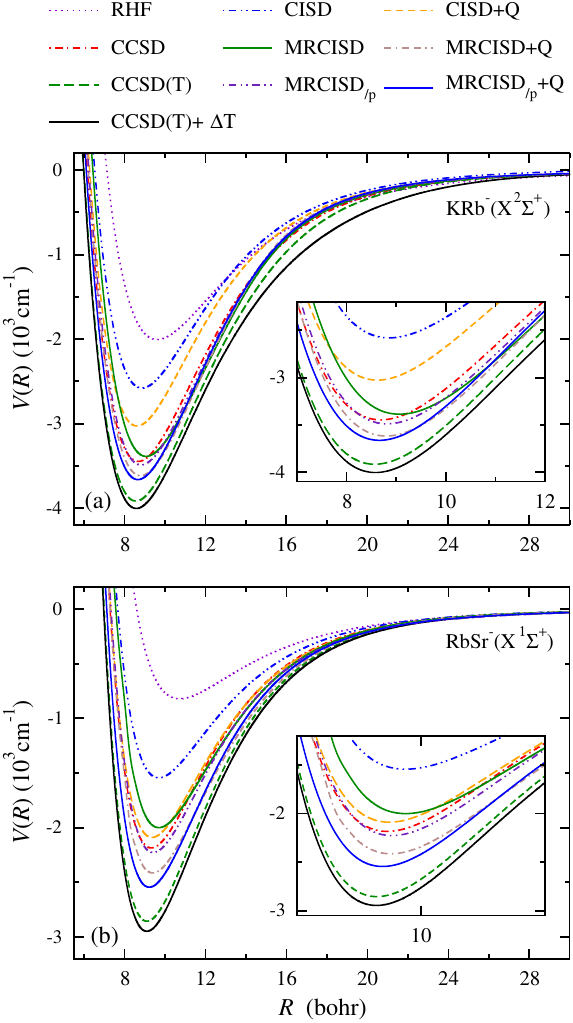}
\centering
\caption{Potential energy curves of the (a) KRb$^{-}$ and (b) RbSr$^{-}$ molecular anions in their X$^{2}\Sigma^{+}$ and X$^{1}\Sigma^{+}$ electronic ground states, respectively, computed using different electronic-structure methods. All calculations employ the aug-cc-pwCV5Z+bf basis set, except for the iterative-triples correction entering the composite CCSD(T)+$\Delta$T results, which is evaluated using the aug-cc-pwCVTZ basis set.}
\label{Fig.4}
\end{figure}
%*****************************************

Figure~\ref{Fig.3} presents counterpoise-corrected potential energy curves calculated at the CCSD(T) level using the aug-cc-pwCV$n$Z basis-set family, with $n=$ D, T, Q, and 5 \cite{HillJCP17,PrascherTCA11}. The interaction energies obtained with the quadruple- and quintuple-$\zeta$ basis sets are additionally extrapolated to the complete-basis-set (CBS) limit using the two-point $1/n^{3}$ formula \cite{HelgakerJCP97}. We also consider the aug-cc-pwCV5Z basis set supplemented by the bond functions, denoted aug-cc-pwCV5Z+bf. Because anions may require a more extensive radial description than neutral molecules or cations, we additionally test a doubly augmented basis set. The second set of diffuse exponents is generated by an even-tempered extension of the aug-cc-pwCV5Z basis \cite{UlusoyMPTC19}, resulting in the d-aug-cc-pwCV5Z basis set. This test determines whether the standard singly augmented basis is sufficient to describe the valence-bound ground states considered here.

For both molecular anions, the potential energy curves converge regularly with increasing cardinal number. Around the potential minima, the well depths obtained with the aug-cc-pwCVTZ basis set differ from the aug-cc-pwCVQZ results by 58~cm$^{-1}$ (1.5\%) and 84~cm$^{-1}$(3\%) for KRb$^-$ and RbSr$^-$, while the differences between the aug-cc-pwCVQZ and aug-cc-pwCV5Z results are 21~cm$^{-1}$(0.5\%) and 38~cm$^{-1}$(1.4\%), respectively. The aug-cc-pwCV5Z well depths are within 0.6\% and 1.4\% of the CBS-extrapolated values. Supplementing the quintuple-$\zeta$ basis with bond functions brings the calculated curves still closer to the CBS results: the corresponding well depths differ by 0.4\% and 1\%. Thus, the bond functions efficiently recover most of the remaining basis-set incompleteness in the interaction energy, in agreement with their established performance for weakly bound van der Waals complexes~\cite{TaoIRPC01} and metal-containing molecules~\cite{SmialkowskiPRA20,GronowskiPRA20,LadjimiPRA24}.

The potential energy curves obtained with the aug-cc-pwCV5Z and d-aug-cc-pwCV5Z basis sets are nearly indistinguishable. Additional diffuse augmentation therefore has a negligible effect on the interaction energies of the valence-bound ground states of KRb$^{-}$ and RbSr$^{-}$. The singly augmented basis sets are consequently sufficient for the ground-state calculations considered here, although substantially more diffuse functions remain necessary for the dipole-bound states discussed in Sec.~\ref{sec:DBS}.

Figure~\ref{Fig.4} examines convergence with respect to the electron-correlation treatment. Potential energy curves are compared at the RHF, CISD, CISD+Q, MRCISD, MRCISD+Q, MRCISD/$p$, MRCISD/$p$+Q, CCSD, CCSD(T), and composite CCSD(T)+$\Delta$T levels. The MRCISD calculations employ an active space comprising only the valence $s$ orbitals, whereas the MRCISD/$p$ calculations additionally include the lowest excited $p$ orbitals. Except for the iterative-triples correction, all calculations employ the aug-cc-pwCV5Z+bf basis. The CCSD(T)+$\Delta$T curves are obtained from Eqs.~\eqref{eq:Vint} and \eqref{eq:VSDT}, with the CCSDT--CCSD(T) difference evaluated in the aug-cc-pwCVTZ basis.

Electron correlation contributes substantially to the binding in both systems. Relative to RHF, CISD increases the well depths by approximately 578~cm$^{-1}$ for KRb$^{-}$ and 719~cm$^{-1}$ for RbSr$^{-}$, but still yields significantly shallower potentials than the coupled-cluster calculations. The Davidson correction increases the CISD and MRCISD well depths and generally brings them closer to the coupled-cluster results. Nevertheless, the noticeable spread among the different CI variants, including their dependence on the selected active space, makes them less suitable as a uniform reference for the present predominantly single-reference ground states.

Within the coupled-cluster hierarchy, the perturbative triple excitations included in CCSD(T) provide a substantial contribution relative to CCSD. Augmenting perturbative triples by the iterative CCSDT correction further deepens the potentials. For KRb$^{-}$, the $\Delta$T contribution increases the well depth by 111~cm$^{-1}$ (2.8\%), compared with 92~cm$^{-1}$ (3.2\%) for RbSr$^{-}$. Iterative triple excitations therefore produce a modest but systematic correction and should be included when accuracy at the few-percent level is sought.

These tests support the composite scheme adopted in Eq.~\eqref{eq:Vint}: the aug-cc-pwCV5Z basis supplemented by bondfunctions provides interaction energies close to the CBS limit, while the additive CCSDT--CCSD(T) correction accounts for the leading post-CCSD(T) correlation contribution. On the basis of the residual basis-set dependence and the magnitude of the higher-order correlation corrections, we expect the calculated well depths to be accurate to within a few percent.

\subsection{Permanent electric dipole moments and static electric dipole polarizabilities}

%************FigV************************
\begin{figure}[!tb]
\includegraphics [width=\columnwidth]{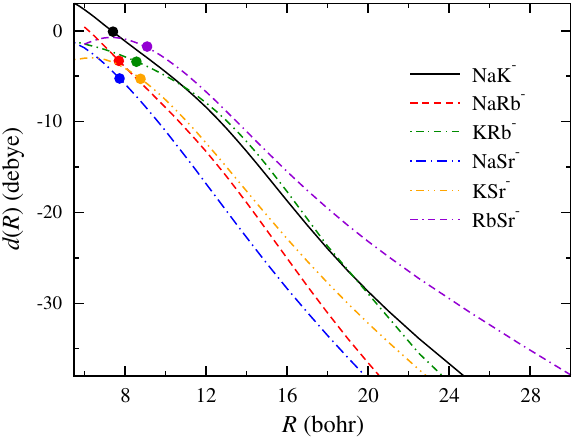}
\centering
\caption{Permanent electric dipole moments for selected diatomic molecular anions in their ground electronic states. The points indicate
values for equilibrium distances.}
\label{Fig_dm}
\end{figure}
 %************************************

Figure~\ref{Fig_dm} presents the permanent electric dipole moments $d(R)$ of selected heteronuclear molecular anions in their electronic
ground states. Because a molecular anion carries a nonzero total charge, its dipole moment depends on the choice of coordinate origin.
Throughout this work, the origin is placed at the molecular center of mass and the molecular axis is oriented according to the convention
defined in Sec.~\ref{sec:theory}. This choice is natural for describing the internal rovibrational dynamics and rotational Stark interactions of molecular ions~\cite{QuemenerCR12, LemeshkoMP13}. Consequently, both the signs and magnitudes of the reported dipole moments should be interpreted within this convention. For the homonuclear alkali-metal dimer anions, the dipole moment vanishes by inversion symmetry.

The dipole-moment curves in Fig.~\ref{Fig_dm} are smooth and become approximately linear at large internuclear distances. This behavior is
characteristic of charged heteronuclear molecules and differs qualitatively from that of neutral molecules, whose dipole moments vanish upon dissociation into neutral atoms~\cite{LadjimiPRA24}. At large $R$, the excess electron becomes localized on the fragment corresponding to the lowest dissociation limit. If the asymptote is $A^-+B$, the dipole moment relative to the center of mass behaves as
\begin{equation}
 |d(R)| \simeq e\frac{m_B}{m_A+m_B}R\,,
\end{equation} 
up to the sign determined by the orientation of the molecular axis. An analogous expression applies when the negative charge is localized on atom $B$. Thus, the asymptotic slopes of the curves are determined by both charge localization and the atomic masses.

The equilibrium dipole moments, $d_e\equiv d(R_e)$, are collected in Tables~\ref{table:Alk-Alk} and~\ref{table:Alk-AlkE}, and the
corresponding values for the selected species are marked by points in Fig.~\ref{Fig_dm}. For the heteronuclear alkali-metal diatomic anions in
their X$^2\Sigma^+$ electronic ground states, $|d_e|$ ranges from 0.112~D for NaK$^-$ to 7.783~D for KFr$^-$, with a mean absolute value
of 4.27~D. For the alkali-metal--alkaline-earth-metal anions in their X$^1\Sigma^+$ electronic ground states, $|d_e|$ ranges from an essentially vanishing 0.002~D for CsSr$^-$ to 10.317~D for KRa$^-$, with a mean absolute value of 3.64~D. The Ra-containing species are generally among the most strongly polar anions considered here, with KRa$^-$ having the largest equilibrium dipole moment in the entire set. We emphasize, however, that a small dipole moment of a charged molecule does not necessarily imply weak charge separation: it may instead result from the center of charge lying close to the center of mass.

For a linear molecule in a $\Sigma$ electronic state, the static electric dipole polarizability tensor has two independent components, $\alpha_e^\parallel\equiv\alpha_{zz}(R_e)$ and $\alpha_e^\perp\equiv\alpha_{xx}(R_e)=\alpha_{yy}(R_e)$, where the $z$ axis coincides with the molecular axis. The calculated values for the alkali-metal and alkali-metal--alkaline-earth-metal molecular anions are reported in Tables~\ref{table:Alk-Alk} and~\ref{table:Alk-AlkE}, respectively. Both components tend to increase as heavier and more easily polarizable atoms are introduced, and the results reveal substantial anisotropy of the electronic response. These polarizabilities determine quadratic Stark shifts and field-induced alignment and contribute to long-range induction interactions relevant to cold collisions and trapping in external fields~\cite{QuemenerCR12,LemeshkoMP13}. To our knowledge, the present results constitute the first predictions of these quantities for most of the molecular anions considered here.

\section{Summary and conclusions}
\label{sec:summary}

Despite their potential relevance to cold and ultracold physics and
chemistry, diatomic molecular anions composed of alkali-metal and
alkaline-earth-metal atoms remain considerably less explored than their
neutral and cationic counterparts. In this work, we have provided a
systematic, high-level theoretical characterization of 57 such species:
21 alkali-metal diatomic anions in their X$^{2}\Sigma^{+}$ electronic
ground states and 36 alkali-metal--alkaline-earth-metal anions in their
X$^{1}\Sigma^{+}$ electronic ground states. The calculations employed a
composite CCSD(T)+$\Delta$T approach, large core--valence Gaussian
basis sets augmented by bond functions, and small-core relativistic
energy-consistent pseudopotentials for the heavier elements.

For all considered molecular anions, we have calculated ground-state
potential energy curves, spectroscopic constants, permanent electric
dipole moments, and the parallel and perpendicular components of the
static electric dipole polarizability. All potential energy curves
exhibit well-defined minima. The alkali-metal diatomic anions are
relatively strongly bound, with an average well depth of
4389~cm$^{-1}$, compared with 3892~cm$^{-1}$ for the corresponding
neutral molecules. The alkali-metal--alkaline-earth-metal anions have
an average well depth of 2869~cm$^{-1}$, more than twice the
corresponding neutral-molecule average of 1311~cm$^{-1}$. For most of
the heteronuclear species, the present results constitute the first
accurate predictions of their ground-state potential energy curves,
spectroscopic constants, dipole moments, and static polarizabilities.

Our convergence analysis demonstrates that iterative triple
excitations produce modest but systematic corrections. On average, the
$\Delta$T contribution increases the well depths by 107~cm$^{-1}$
(2.4\%) for the alkali-metal diatomic anions and by 97~cm$^{-1}$ (3.4\%)
for the alkali-metal--alkaline-earth-metal anions. Post-CCSD(T)
correlation effects should therefore be included when quantitative
accuracy at the few-percent level is required. 

The calculated
equilibrium dipole moments span a broad range, with mean absolute
values of 4.27~D for the heteronuclear alkali-metal diatomic anions and
3.64~D for the alkali-metal--alkaline-earth-metal anions. KRa$^{-}$
has the largest equilibrium dipole moment among all investigated
species, 10.32~D. The static polarizabilities exhibit substantial
anisotropy and generally increase with the size and polarizability of
the constituent atoms.

Using the MRCISD method, we have additionally investigated the
lowest valence-excited A$^{2}\Sigma^{+}$ states of selected alkali-metal
molecular anions. These states cross the ground-state potentials of the
corresponding neutral molecules and are embedded in the
electron-detachment continuum over part of the internuclear-distance
range. The resulting temporary anionic states may provide resonant
pathways for low-energy electron capture and subsequent anion 
dissociation. Using the EOM-EA-CCSD method, we have also characterized dipole-bound states supported
by six polar alkali-metal molecules. Their binding energies increase
with the permanent dipole moment of the neutral molecular core, and
the most strongly polar species, particularly NaCs and LiCs, may
support numerous rotational levels below the electron-detachment
threshold.

These near-threshold anionic states may be accessed in collisions
between polar ground-state molecules and atoms prepared in highly
excited Rydberg states. In particular, transfer of the weakly bound
Rydberg electron to the molecule may populate valence-bound or
dipole-bound anionic states when the neutral and anionic electronic
states become nearly degenerate. The crossings between the ground
neutral and excited anionic potential energy curves predicted here
therefore identify internuclear distances at which resonant electron
attachment may be enhanced. A quantitative description of this process,
however, will require an explicit treatment of the Rydberg electron,
the ionic core, nuclear motion, and nonadiabatic couplings.

The present data provide a uniform reference for future
electronic-structure calculations and experimental searches for these
molecular anions. The calculated potential energy curves can be used
to determine rovibrational levels, assist the assignment of
photoelectron and molecular-anion spectra, and provide starting
potentials for refinement against future spectroscopic measurements.
More broadly, the predicted temporary, valence-bound, and dipole-bound
states identify promising molecular systems for studying low-energy
electron attachment and the interaction of ultracold polar molecules
with Rydberg atoms.

\begin{acknowledgments}
We gratefully acknowledge the National Science Centre Poland (grant no.~2020/38/E/ST2/00564) for the financial support and Poland’s high-performance computing infrastructure PLGrid (HPC Center: ACK Cyfronet AGH) for providing computer facilities and support (computational grant no.~PLG/2024/017844). S. A.~acknowledges the financial support (Bourses d'Alternance) from the Ministry of Higher Education and Scientific Research of Tunisia and the University of Monastir.
\end{acknowledgments}

\bibliography{sample.bib}

%merlin.mbs apsrev4-1.bst 2010-07-25 4.21a (PWD, AO, DPC) hacked
%Control: key (0)
%Control: author (8) initials jnrlst
%Control: editor formatted (1) identically to author
%Control: production of article title (-1) disabled
%Control: page (0) single
%Control: year (1) truncated
%Control: production of eprint (0) enabled
\begin{thebibliography}{146}%
\makeatletter
\providecommand \@ifxundefined [1]{%
 \@ifx{#1\undefined}
}%
\providecommand \@ifnum [1]{%
 \ifnum #1\expandafter \@firstoftwo
 \else \expandafter \@secondoftwo
 \fi
}%
\providecommand \@ifx [1]{%
 \ifx #1\expandafter \@firstoftwo
 \else \expandafter \@secondoftwo
 \fi
}%
\providecommand \natexlab [1]{#1}%
\providecommand \enquote  [1]{``#1''}%
\providecommand \bibnamefont  [1]{#1}%
\providecommand \bibfnamefont [1]{#1}%
\providecommand \citenamefont [1]{#1}%
\providecommand \href@noop [0]{\@secondoftwo}%
\providecommand \href [0]{\begingroup \@sanitize@url \@href}%
\providecommand \@href[1]{\@@startlink{#1}\@@href}%
\providecommand \@@href[1]{\endgroup#1\@@endlink}%
\providecommand \@sanitize@url [0]{\catcode `\\12\catcode `\$12\catcode
  `\&12\catcode `\#12\catcode `\^12\catcode `\_12\catcode `\%12\relax}%
\providecommand \@@startlink[1]{}%
\providecommand \@@endlink[0]{}%
\providecommand \url  [0]{\begingroup\@sanitize@url \@url }%
\providecommand \@url [1]{\endgroup\@href {#1}{\urlprefix }}%
\providecommand \urlprefix  [0]{URL }%
\providecommand \Eprint [0]{\href }%
\providecommand \doibase [0]{http://dx.doi.org/}%
\providecommand \selectlanguage [0]{\@gobble}%
\providecommand \bibinfo  [0]{\@secondoftwo}%
\providecommand \bibfield  [0]{\@secondoftwo}%
\providecommand \translation [1]{[#1]}%
\providecommand \BibitemOpen [0]{}%
\providecommand \bibitemStop [0]{}%
\providecommand \bibitemNoStop [0]{.\EOS\space}%
\providecommand \EOS [0]{\spacefactor3000\relax}%
\providecommand \BibitemShut  [1]{\csname bibitem#1\endcsname}%
\let\auto@bib@innerbib\@empty
%</preamble>
\bibitem [{\citenamefont {Bohn}\ \emph {et~al.}(2017)\citenamefont {Bohn},
  \citenamefont {Rey},\ and\ \citenamefont {Ye}}]{BohnScience17}%
  \BibitemOpen
  \bibfield  {author} {\bibinfo {author} {\bibfnamefont {J.~L.}\ \bibnamefont
  {Bohn}}, \bibinfo {author} {\bibfnamefont {A.~M.}\ \bibnamefont {Rey}}, \
  and\ \bibinfo {author} {\bibfnamefont {J.}~\bibnamefont {Ye}},\ }\href
  {\doibase 10.1126/science.aam6299} {\bibfield  {journal} {\bibinfo  {journal}
  {Science}\ }\textbf {\bibinfo {volume} {357}},\ \bibinfo {pages} {1002}
  (\bibinfo {year} {2017})}\BibitemShut {NoStop}%
\bibitem [{\citenamefont {Karman}\ \emph {et~al.}(2024)\citenamefont {Karman},
  \citenamefont {Tomza},\ and\ \citenamefont
  {P{\'e}rez-R{\'i}os}}]{KarmanNP24}%
  \BibitemOpen
  \bibfield  {author} {\bibinfo {author} {\bibfnamefont {T.}~\bibnamefont
  {Karman}}, \bibinfo {author} {\bibfnamefont {M.}~\bibnamefont {Tomza}}, \
  and\ \bibinfo {author} {\bibfnamefont {J.}~\bibnamefont
  {P{\'e}rez-R{\'i}os}},\ }\href {\doibase 10.1038/s41567-024-02467-3}
  {\bibfield  {journal} {\bibinfo  {journal} {Nat. Phys.}\ }\textbf {\bibinfo
  {volume} {20}},\ \bibinfo {pages} {722} (\bibinfo {year} {2024})}\BibitemShut
  {NoStop}%
\bibitem [{\citenamefont {DeMille}\ \emph {et~al.}(2017)\citenamefont
  {DeMille}, \citenamefont {Doyle},\ and\ \citenamefont
  {Sushkov}}]{DemilleScience17}%
  \BibitemOpen
  \bibfield  {author} {\bibinfo {author} {\bibfnamefont {D.}~\bibnamefont
  {DeMille}}, \bibinfo {author} {\bibfnamefont {J.~M.}\ \bibnamefont {Doyle}},
  \ and\ \bibinfo {author} {\bibfnamefont {A.~O.}\ \bibnamefont {Sushkov}},\
  }\href {\doibase 10.1126/science.aal3003} {\bibfield  {journal} {\bibinfo
  {journal} {Science}\ }\textbf {\bibinfo {volume} {357}},\ \bibinfo {pages}
  {990} (\bibinfo {year} {2017})}\BibitemShut {NoStop}%
\bibitem [{\citenamefont {DeMille}\ \emph {et~al.}(2024)\citenamefont
  {DeMille}, \citenamefont {Hutzler}, \citenamefont {Rey},\ and\ \citenamefont
  {Zelevinsky}}]{DeMilleNP24}%
  \BibitemOpen
  \bibfield  {author} {\bibinfo {author} {\bibfnamefont {D.}~\bibnamefont
  {DeMille}}, \bibinfo {author} {\bibfnamefont {N.~R.}\ \bibnamefont
  {Hutzler}}, \bibinfo {author} {\bibfnamefont {A.~M.}\ \bibnamefont {Rey}}, \
  and\ \bibinfo {author} {\bibfnamefont {T.}~\bibnamefont {Zelevinsky}},\
  }\href {\doibase 10.1038/s41567-024-02499-9} {\bibfield  {journal} {\bibinfo
  {journal} {Nat. Phys.}\ }\textbf {\bibinfo {volume} {20}},\ \bibinfo {pages}
  {741} (\bibinfo {year} {2024})}\BibitemShut {NoStop}%
\bibitem [{\citenamefont {Gross}\ and\ \citenamefont
  {Bloch}(2017)}]{GrossScience17}%
  \BibitemOpen
  \bibfield  {author} {\bibinfo {author} {\bibfnamefont {C.}~\bibnamefont
  {Gross}}\ and\ \bibinfo {author} {\bibfnamefont {I.}~\bibnamefont {Bloch}},\
  }\href {\doibase 10.1126/science.aal3837} {\bibfield  {journal} {\bibinfo
  {journal} {Science}\ }\textbf {\bibinfo {volume} {357}},\ \bibinfo {pages}
  {995} (\bibinfo {year} {2017})}\BibitemShut {NoStop}%
\bibitem [{\citenamefont {Cornish}\ \emph {et~al.}(2024)\citenamefont
  {Cornish}, \citenamefont {Tarbutt},\ and\ \citenamefont
  {Hazzard}}]{CornishNP24}%
  \BibitemOpen
  \bibfield  {author} {\bibinfo {author} {\bibfnamefont {S.~L.}\ \bibnamefont
  {Cornish}}, \bibinfo {author} {\bibfnamefont {M.~R.}\ \bibnamefont
  {Tarbutt}}, \ and\ \bibinfo {author} {\bibfnamefont {K.~R.~A.}\ \bibnamefont
  {Hazzard}},\ }\href {\doibase 10.1038/s41567-024-02453-9} {\bibfield
  {journal} {\bibinfo  {journal} {Nat. Phys.}\ }\textbf {\bibinfo {volume}
  {20}},\ \bibinfo {pages} {730} (\bibinfo {year} {2024})}\BibitemShut
  {NoStop}%
\bibitem [{\citenamefont {Tomza}\ \emph {et~al.}(2019)\citenamefont {Tomza},
  \citenamefont {Jachymski}, \citenamefont {Gerritsma}, \citenamefont
  {Negretti}, \citenamefont {Calarco}, \citenamefont {Idziaszek},\ and\
  \citenamefont {Julienne}}]{TomzaRMP19}%
  \BibitemOpen
  \bibfield  {author} {\bibinfo {author} {\bibfnamefont {M.}~\bibnamefont
  {Tomza}}, \bibinfo {author} {\bibfnamefont {K.}~\bibnamefont {Jachymski}},
  \bibinfo {author} {\bibfnamefont {R.}~\bibnamefont {Gerritsma}}, \bibinfo
  {author} {\bibfnamefont {A.}~\bibnamefont {Negretti}}, \bibinfo {author}
  {\bibfnamefont {T.}~\bibnamefont {Calarco}}, \bibinfo {author} {\bibfnamefont
  {Z.}~\bibnamefont {Idziaszek}}, \ and\ \bibinfo {author} {\bibfnamefont
  {P.~S.}\ \bibnamefont {Julienne}},\ }\href {\doibase
  10.1103/RevModPhys.91.035001} {\bibfield  {journal} {\bibinfo  {journal}
  {Rev. Mod. Phys.}\ }\textbf {\bibinfo {volume} {91}},\ \bibinfo {pages}
  {035001} (\bibinfo {year} {2019})}\BibitemShut {NoStop}%
\bibitem [{\citenamefont {Dei{\ss}}\ \emph {et~al.}(2024)\citenamefont
  {Dei{\ss}}, \citenamefont {Willitsch},\ and\ \citenamefont
  {Hecker~Denschlag}}]{DeissNP24}%
  \BibitemOpen
  \bibfield  {author} {\bibinfo {author} {\bibfnamefont {M.}~\bibnamefont
  {Dei{\ss}}}, \bibinfo {author} {\bibfnamefont {S.}~\bibnamefont {Willitsch}},
  \ and\ \bibinfo {author} {\bibfnamefont {J.}~\bibnamefont
  {Hecker~Denschlag}},\ }\href {\doibase 10.1038/s41567-024-02440-0} {\bibfield
   {journal} {\bibinfo  {journal} {Nat. Phys.}\ }\textbf {\bibinfo {volume}
  {20}},\ \bibinfo {pages} {713} (\bibinfo {year} {2024})}\BibitemShut
  {NoStop}%
\bibitem [{\citenamefont {Eschner}\ \emph {et~al.}(2003)\citenamefont
  {Eschner}, \citenamefont {Morigi}, \citenamefont {Schmidt-Kaler},\ and\
  \citenamefont {Blatt}}]{EschnerJOSAB03}%
  \BibitemOpen
  \bibfield  {author} {\bibinfo {author} {\bibfnamefont {J.}~\bibnamefont
  {Eschner}}, \bibinfo {author} {\bibfnamefont {G.}~\bibnamefont {Morigi}},
  \bibinfo {author} {\bibfnamefont {F.}~\bibnamefont {Schmidt-Kaler}}, \ and\
  \bibinfo {author} {\bibfnamefont {R.}~\bibnamefont {Blatt}},\ }\href
  {\doibase 10.1364/JOSAB.20.001003} {\bibfield  {journal} {\bibinfo  {journal}
  {J. Opt. Soc. Am. B}\ }\textbf {\bibinfo {volume} {20}},\ \bibinfo {pages}
  {1003} (\bibinfo {year} {2003})}\BibitemShut {NoStop}%
\bibitem [{\citenamefont {Grier}\ \emph {et~al.}(2009)\citenamefont {Grier},
  \citenamefont {Cetina}, \citenamefont {Oru\ifmmode \check{c}\else
  \v{c}\fi{}evi\ifmmode~\acute{c}\else \'{c}\fi{}},\ and\ \citenamefont
  {Vuleti\ifmmode~\acute{c}\else \'{c}\fi{}}}]{GrierPRL09}%
  \BibitemOpen
  \bibfield  {author} {\bibinfo {author} {\bibfnamefont {A.~T.}\ \bibnamefont
  {Grier}}, \bibinfo {author} {\bibfnamefont {M.}~\bibnamefont {Cetina}},
  \bibinfo {author} {\bibfnamefont {F.}~\bibnamefont {Oru\ifmmode
  \check{c}\else \v{c}\fi{}evi\ifmmode~\acute{c}\else \'{c}\fi{}}}, \ and\
  \bibinfo {author} {\bibfnamefont {V.}~\bibnamefont
  {Vuleti\ifmmode~\acute{c}\else \'{c}\fi{}}},\ }\href {\doibase
  10.1103/PhysRevLett.102.223201} {\bibfield  {journal} {\bibinfo  {journal}
  {Phys. Rev. Lett.}\ }\textbf {\bibinfo {volume} {102}},\ \bibinfo {pages}
  {223201} (\bibinfo {year} {2009})}\BibitemShut {NoStop}%
\bibitem [{\citenamefont {Zipkes}\ \emph {et~al.}(2010)\citenamefont {Zipkes},
  \citenamefont {Palzer}, \citenamefont {Sias},\ and\ \citenamefont
  {K{\"o}hl}}]{ZipkesNature10}%
  \BibitemOpen
  \bibfield  {author} {\bibinfo {author} {\bibfnamefont {C.}~\bibnamefont
  {Zipkes}}, \bibinfo {author} {\bibfnamefont {S.}~\bibnamefont {Palzer}},
  \bibinfo {author} {\bibfnamefont {C.}~\bibnamefont {Sias}}, \ and\ \bibinfo
  {author} {\bibfnamefont {M.}~\bibnamefont {K{\"o}hl}},\ }\href {\doibase
  10.1038/nature08865} {\bibfield  {journal} {\bibinfo  {journal} {Nature}\
  }\textbf {\bibinfo {volume} {464}},\ \bibinfo {pages} {388} (\bibinfo {year}
  {2010})}\BibitemShut {NoStop}%
\bibitem [{\citenamefont {Schmid}\ \emph {et~al.}(2010)\citenamefont {Schmid},
  \citenamefont {H\"arter},\ and\ \citenamefont
  {Hecker~Denschlag}}]{SchmidPRL10}%
  \BibitemOpen
  \bibfield  {author} {\bibinfo {author} {\bibfnamefont {S.}~\bibnamefont
  {Schmid}}, \bibinfo {author} {\bibfnamefont {A.}~\bibnamefont {H\"arter}}, \
  and\ \bibinfo {author} {\bibfnamefont {J.}~\bibnamefont {Hecker~Denschlag}},\
  }\href {\doibase 10.1103/PhysRevLett.105.133202} {\bibfield  {journal}
  {\bibinfo  {journal} {Phys. Rev. Lett.}\ }\textbf {\bibinfo {volume} {105}},\
  \bibinfo {pages} {133202} (\bibinfo {year} {2010})}\BibitemShut {NoStop}%
\bibitem [{\citenamefont {Schmidt}\ \emph {et~al.}(2020)\citenamefont
  {Schmidt}, \citenamefont {Weckesser}, \citenamefont {Thielemann},
  \citenamefont {Schaetz},\ and\ \citenamefont {Karpa}}]{SchmidtPRL20}%
  \BibitemOpen
  \bibfield  {author} {\bibinfo {author} {\bibfnamefont {J.}~\bibnamefont
  {Schmidt}}, \bibinfo {author} {\bibfnamefont {P.}~\bibnamefont {Weckesser}},
  \bibinfo {author} {\bibfnamefont {F.}~\bibnamefont {Thielemann}}, \bibinfo
  {author} {\bibfnamefont {T.}~\bibnamefont {Schaetz}}, \ and\ \bibinfo
  {author} {\bibfnamefont {L.}~\bibnamefont {Karpa}},\ }\href {\doibase
  10.1103/PhysRevLett.124.053402} {\bibfield  {journal} {\bibinfo  {journal}
  {Phys. Rev. Lett.}\ }\textbf {\bibinfo {volume} {124}},\ \bibinfo {pages}
  {053402} (\bibinfo {year} {2020})}\BibitemShut {NoStop}%
\bibitem [{\citenamefont {Hall}\ \emph {et~al.}(2011)\citenamefont {Hall},
  \citenamefont {Aymar}, \citenamefont {Bouloufa-Maafa}, \citenamefont
  {Dulieu},\ and\ \citenamefont {Willitsch}}]{HallPRL11}%
  \BibitemOpen
  \bibfield  {author} {\bibinfo {author} {\bibfnamefont {F.~H.~J.}\
  \bibnamefont {Hall}}, \bibinfo {author} {\bibfnamefont {M.}~\bibnamefont
  {Aymar}}, \bibinfo {author} {\bibfnamefont {N.}~\bibnamefont
  {Bouloufa-Maafa}}, \bibinfo {author} {\bibfnamefont {O.}~\bibnamefont
  {Dulieu}}, \ and\ \bibinfo {author} {\bibfnamefont {S.}~\bibnamefont
  {Willitsch}},\ }\href {\doibase 10.1103/PhysRevLett.107.243202} {\bibfield
  {journal} {\bibinfo  {journal} {Phys. Rev. Lett.}\ }\textbf {\bibinfo
  {volume} {107}},\ \bibinfo {pages} {243202} (\bibinfo {year}
  {2011})}\BibitemShut {NoStop}%
\bibitem [{\citenamefont {Haze}\ \emph {et~al.}(2013)\citenamefont {Haze},
  \citenamefont {Hata}, \citenamefont {Fujinaga},\ and\ \citenamefont
  {Mukaiyama}}]{HazePRA13}%
  \BibitemOpen
  \bibfield  {author} {\bibinfo {author} {\bibfnamefont {S.}~\bibnamefont
  {Haze}}, \bibinfo {author} {\bibfnamefont {S.}~\bibnamefont {Hata}}, \bibinfo
  {author} {\bibfnamefont {M.}~\bibnamefont {Fujinaga}}, \ and\ \bibinfo
  {author} {\bibfnamefont {T.}~\bibnamefont {Mukaiyama}},\ }\href {\doibase
  10.1103/PhysRevA.87.052715} {\bibfield  {journal} {\bibinfo  {journal} {Phys.
  Rev. A}\ }\textbf {\bibinfo {volume} {87}},\ \bibinfo {pages} {052715}
  (\bibinfo {year} {2013})}\BibitemShut {NoStop}%
\bibitem [{\citenamefont {Smith}\ \emph {et~al.}(2014)\citenamefont {Smith},
  \citenamefont {Goodman}, \citenamefont {Sivarajah}, \citenamefont {Wells},
  \citenamefont {Banerjee}, \citenamefont {C{\^o}t{\'e}}, \citenamefont
  {Michels}, \citenamefont {Mongtomery},\ and\ \citenamefont
  {Narducci}}]{SmithAPB14}%
  \BibitemOpen
  \bibfield  {author} {\bibinfo {author} {\bibfnamefont {W.}~\bibnamefont
  {Smith}}, \bibinfo {author} {\bibfnamefont {D.}~\bibnamefont {Goodman}},
  \bibinfo {author} {\bibfnamefont {I.}~\bibnamefont {Sivarajah}}, \bibinfo
  {author} {\bibfnamefont {J.}~\bibnamefont {Wells}}, \bibinfo {author}
  {\bibfnamefont {S.}~\bibnamefont {Banerjee}}, \bibinfo {author}
  {\bibfnamefont {R.}~\bibnamefont {C{\^o}t{\'e}}}, \bibinfo {author}
  {\bibfnamefont {H.}~\bibnamefont {Michels}}, \bibinfo {author} {\bibfnamefont
  {J.}~\bibnamefont {Mongtomery}}, \ and\ \bibinfo {author} {\bibfnamefont
  {F.}~\bibnamefont {Narducci}},\ }\href {\doibase 10.1007/s00340-013-5672-2}
  {\bibfield  {journal} {\bibinfo  {journal} {Appl. Phys. B}\ }\textbf
  {\bibinfo {volume} {114}},\ \bibinfo {pages} {75} (\bibinfo {year}
  {2014})}\BibitemShut {NoStop}%
\bibitem [{\citenamefont {Meir}\ \emph {et~al.}(2016)\citenamefont {Meir},
  \citenamefont {Sikorsky}, \citenamefont {Ben-Shlomi}, \citenamefont
  {Akerman}, \citenamefont {Dallal},\ and\ \citenamefont {Ozeri}}]{MeirPRL16}%
  \BibitemOpen
  \bibfield  {author} {\bibinfo {author} {\bibfnamefont {Z.}~\bibnamefont
  {Meir}}, \bibinfo {author} {\bibfnamefont {T.}~\bibnamefont {Sikorsky}},
  \bibinfo {author} {\bibfnamefont {R.}~\bibnamefont {Ben-Shlomi}}, \bibinfo
  {author} {\bibfnamefont {N.}~\bibnamefont {Akerman}}, \bibinfo {author}
  {\bibfnamefont {Y.}~\bibnamefont {Dallal}}, \ and\ \bibinfo {author}
  {\bibfnamefont {R.}~\bibnamefont {Ozeri}},\ }\href {\doibase
  10.1103/PhysRevLett.117.243401} {\bibfield  {journal} {\bibinfo  {journal}
  {Phys. Rev. Lett.}\ }\textbf {\bibinfo {volume} {117}},\ \bibinfo {pages}
  {243401} (\bibinfo {year} {2016})}\BibitemShut {NoStop}%
\bibitem [{\citenamefont {Dutta}\ \emph {et~al.}(2017)\citenamefont {Dutta},
  \citenamefont {Sawant},\ and\ \citenamefont {Rangwala}}]{DuttaPRL17}%
  \BibitemOpen
  \bibfield  {author} {\bibinfo {author} {\bibfnamefont {S.}~\bibnamefont
  {Dutta}}, \bibinfo {author} {\bibfnamefont {R.}~\bibnamefont {Sawant}}, \
  and\ \bibinfo {author} {\bibfnamefont {S.~A.}\ \bibnamefont {Rangwala}},\
  }\href {\doibase 10.1103/PhysRevLett.118.113401} {\bibfield  {journal}
  {\bibinfo  {journal} {Phys. Rev. Lett.}\ }\textbf {\bibinfo {volume} {118}},\
  \bibinfo {pages} {113401} (\bibinfo {year} {2017})}\BibitemShut {NoStop}%
\bibitem [{\citenamefont {Joger}\ \emph {et~al.}(2017)\citenamefont {Joger},
  \citenamefont {F{\"u}rst}, \citenamefont {Ewald}, \citenamefont {Feldker},
  \citenamefont {Tomza},\ and\ \citenamefont {Gerritsma}}]{JogerPRA17}%
  \BibitemOpen
  \bibfield  {author} {\bibinfo {author} {\bibfnamefont {J.}~\bibnamefont
  {Joger}}, \bibinfo {author} {\bibfnamefont {H.}~\bibnamefont {F{\"u}rst}},
  \bibinfo {author} {\bibfnamefont {N.}~\bibnamefont {Ewald}}, \bibinfo
  {author} {\bibfnamefont {T.}~\bibnamefont {Feldker}}, \bibinfo {author}
  {\bibfnamefont {M.}~\bibnamefont {Tomza}}, \ and\ \bibinfo {author}
  {\bibfnamefont {R.}~\bibnamefont {Gerritsma}},\ }\href {\doibase
  10.1103/PhysRevA.96.030703} {\bibfield  {journal} {\bibinfo  {journal} {Phys.
  Rev. A}\ }\textbf {\bibinfo {volume} {96}},\ \bibinfo {pages} {030703}
  (\bibinfo {year} {2017})}\BibitemShut {NoStop}%
\bibitem [{\citenamefont {Weckesser}\ \emph {et~al.}(2021)\citenamefont
  {Weckesser}, \citenamefont {Thielemann}, \citenamefont {Wiater},
  \citenamefont {Wojciechowska}, \citenamefont {Karpa}, \citenamefont
  {Jachymski}, \citenamefont {Tomza}, \citenamefont {Walker},\ and\
  \citenamefont {Schaetz}}]{WeckesserNature21}%
  \BibitemOpen
  \bibfield  {author} {\bibinfo {author} {\bibfnamefont {P.}~\bibnamefont
  {Weckesser}}, \bibinfo {author} {\bibfnamefont {F.}~\bibnamefont
  {Thielemann}}, \bibinfo {author} {\bibfnamefont {D.}~\bibnamefont {Wiater}},
  \bibinfo {author} {\bibfnamefont {A.}~\bibnamefont {Wojciechowska}}, \bibinfo
  {author} {\bibfnamefont {L.}~\bibnamefont {Karpa}}, \bibinfo {author}
  {\bibfnamefont {K.}~\bibnamefont {Jachymski}}, \bibinfo {author}
  {\bibfnamefont {M.}~\bibnamefont {Tomza}}, \bibinfo {author} {\bibfnamefont
  {T.}~\bibnamefont {Walker}}, \ and\ \bibinfo {author} {\bibfnamefont
  {T.}~\bibnamefont {Schaetz}},\ }\href {\doibase 10.1038/s41586-021-04112-y}
  {\bibfield  {journal} {\bibinfo  {journal} {Nature}\ }\textbf {\bibinfo
  {volume} {600}},\ \bibinfo {pages} {429} (\bibinfo {year}
  {2021})}\BibitemShut {NoStop}%
\bibitem [{\citenamefont {Hall}\ \emph {et~al.}(2013)\citenamefont {Hall},
  \citenamefont {Aymar}, \citenamefont {Raoult}, \citenamefont {Dulieu},\ and\
  \citenamefont {Willitsch}}]{HallMP13}%
  \BibitemOpen
  \bibfield  {author} {\bibinfo {author} {\bibfnamefont {F.~H.}\ \bibnamefont
  {Hall}}, \bibinfo {author} {\bibfnamefont {M.}~\bibnamefont {Aymar}},
  \bibinfo {author} {\bibfnamefont {M.}~\bibnamefont {Raoult}}, \bibinfo
  {author} {\bibfnamefont {O.}~\bibnamefont {Dulieu}}, \ and\ \bibinfo {author}
  {\bibfnamefont {S.}~\bibnamefont {Willitsch}},\ }\href {\doibase
  10.1080/00268976.2013.770930} {\bibfield  {journal} {\bibinfo  {journal}
  {Mol. Phys.}\ }\textbf {\bibinfo {volume} {111}},\ \bibinfo {pages} {1683}
  (\bibinfo {year} {2013})}\BibitemShut {NoStop}%
\bibitem [{\citenamefont {Sikorsky}\ \emph {et~al.}(2018)\citenamefont
  {Sikorsky}, \citenamefont {Meir}, \citenamefont {Ben-Shlomi}, \citenamefont
  {Akerman},\ and\ \citenamefont {Ozeri}}]{SikorskyNC18}%
  \BibitemOpen
  \bibfield  {author} {\bibinfo {author} {\bibfnamefont {T.}~\bibnamefont
  {Sikorsky}}, \bibinfo {author} {\bibfnamefont {Z.}~\bibnamefont {Meir}},
  \bibinfo {author} {\bibfnamefont {R.}~\bibnamefont {Ben-Shlomi}}, \bibinfo
  {author} {\bibfnamefont {N.}~\bibnamefont {Akerman}}, \ and\ \bibinfo
  {author} {\bibfnamefont {R.}~\bibnamefont {Ozeri}},\ }\href {\doibase
  10.1038/s41467-018-03373-y} {\bibfield  {journal} {\bibinfo  {journal} {Nat.
  Commun.}\ }\textbf {\bibinfo {volume} {9}},\ \bibinfo {pages} {920} (\bibinfo
  {year} {2018})}\BibitemShut {NoStop}%
\bibitem [{\citenamefont {Feldker}\ \emph {et~al.}(2020)\citenamefont
  {Feldker}, \citenamefont {F{\"u}rst}, \citenamefont {Hirzler}, \citenamefont
  {Ewald}, \citenamefont {Mazzanti}, \citenamefont {Wiater}, \citenamefont
  {Tomza},\ and\ \citenamefont {Gerritsma}}]{FeldkerNP20}%
  \BibitemOpen
  \bibfield  {author} {\bibinfo {author} {\bibfnamefont {T.}~\bibnamefont
  {Feldker}}, \bibinfo {author} {\bibfnamefont {H.}~\bibnamefont {F{\"u}rst}},
  \bibinfo {author} {\bibfnamefont {H.}~\bibnamefont {Hirzler}}, \bibinfo
  {author} {\bibfnamefont {N.}~\bibnamefont {Ewald}}, \bibinfo {author}
  {\bibfnamefont {M.}~\bibnamefont {Mazzanti}}, \bibinfo {author}
  {\bibfnamefont {D.}~\bibnamefont {Wiater}}, \bibinfo {author} {\bibfnamefont
  {M.}~\bibnamefont {Tomza}}, \ and\ \bibinfo {author} {\bibfnamefont
  {R.}~\bibnamefont {Gerritsma}},\ }\href {\doibase 10.1038/s41567-019-0772-5}
  {\bibfield  {journal} {\bibinfo  {journal} {Nat. Phys.}\ }\textbf {\bibinfo
  {volume} {16}},\ \bibinfo {pages} {413} (\bibinfo {year} {2020})}\BibitemShut
  {NoStop}%
\bibitem [{\citenamefont {{\'S}mia{\l}kowski}\ and\ \citenamefont
  {Tomza}(2020)}]{SmialkowskiPRA20}%
  \BibitemOpen
  \bibfield  {author} {\bibinfo {author} {\bibfnamefont {M.}~\bibnamefont
  {{\'S}mia{\l}kowski}}\ and\ \bibinfo {author} {\bibfnamefont
  {M.}~\bibnamefont {Tomza}},\ }\href {\doibase 10.1103/PhysRevA.101.012501}
  {\bibfield  {journal} {\bibinfo  {journal} {Phys. Rev. A}\ }\textbf {\bibinfo
  {volume} {101}},\ \bibinfo {pages} {012501} (\bibinfo {year}
  {2020})}\BibitemShut {NoStop}%
\bibitem [{\citenamefont {da~Silva~Jr}\ \emph {et~al.}(2015)\citenamefont
  {da~Silva~Jr}, \citenamefont {Raoult}, \citenamefont {Aymar},\ and\
  \citenamefont {Dulieu}}]{daSilvaNJP2015}%
  \BibitemOpen
  \bibfield  {author} {\bibinfo {author} {\bibfnamefont {H.}~\bibnamefont
  {da~Silva~Jr}}, \bibinfo {author} {\bibfnamefont {M.}~\bibnamefont {Raoult}},
  \bibinfo {author} {\bibfnamefont {M.}~\bibnamefont {Aymar}}, \ and\ \bibinfo
  {author} {\bibfnamefont {O.}~\bibnamefont {Dulieu}},\ }\href {\doibase
  10.1088/1367-2630/17/4/045015} {\bibfield  {journal} {\bibinfo  {journal}
  {New J. Phys.}\ }\textbf {\bibinfo {volume} {17}},\ \bibinfo {pages} {045015}
  (\bibinfo {year} {2015})}\BibitemShut {NoStop}%
\bibitem [{\citenamefont {Zrafi}\ \emph {et~al.}(2020)\citenamefont {Zrafi},
  \citenamefont {Ladjimi}, \citenamefont {Said}, \citenamefont {Berriche},\
  and\ \citenamefont {Tomza}}]{ZrafiNJP20}%
  \BibitemOpen
  \bibfield  {author} {\bibinfo {author} {\bibfnamefont {W.}~\bibnamefont
  {Zrafi}}, \bibinfo {author} {\bibfnamefont {H.}~\bibnamefont {Ladjimi}},
  \bibinfo {author} {\bibfnamefont {H.}~\bibnamefont {Said}}, \bibinfo {author}
  {\bibfnamefont {H.}~\bibnamefont {Berriche}}, \ and\ \bibinfo {author}
  {\bibfnamefont {M.}~\bibnamefont {Tomza}},\ }\href {\doibase
  10.1088/1367-2630/ab9429} {\bibfield  {journal} {\bibinfo  {journal} {New J.
  Phys.}\ }\textbf {\bibinfo {volume} {22}},\ \bibinfo {pages} {073015}
  (\bibinfo {year} {2020})}\BibitemShut {NoStop}%
\bibitem [{\citenamefont {Idziaszek}\ \emph {et~al.}(2009)\citenamefont
  {Idziaszek}, \citenamefont {Calarco}, \citenamefont {Julienne},\ and\
  \citenamefont {Simoni}}]{IdziaszekPRA09}%
  \BibitemOpen
  \bibfield  {author} {\bibinfo {author} {\bibfnamefont {Z.}~\bibnamefont
  {Idziaszek}}, \bibinfo {author} {\bibfnamefont {T.}~\bibnamefont {Calarco}},
  \bibinfo {author} {\bibfnamefont {P.~S.}\ \bibnamefont {Julienne}}, \ and\
  \bibinfo {author} {\bibfnamefont {A.}~\bibnamefont {Simoni}},\ }\href
  {\doibase 10.1103/PhysRevA.79.010702} {\bibfield  {journal} {\bibinfo
  {journal} {Phys. Rev. A}\ }\textbf {\bibinfo {volume} {79}},\ \bibinfo
  {pages} {010702} (\bibinfo {year} {2009})}\BibitemShut {NoStop}%
\bibitem [{\citenamefont {Tomza}\ \emph {et~al.}(2015)\citenamefont {Tomza},
  \citenamefont {Koch},\ and\ \citenamefont {Moszynski}}]{TomzaPRA15}%
  \BibitemOpen
  \bibfield  {author} {\bibinfo {author} {\bibfnamefont {M.}~\bibnamefont
  {Tomza}}, \bibinfo {author} {\bibfnamefont {C.~P.}\ \bibnamefont {Koch}}, \
  and\ \bibinfo {author} {\bibfnamefont {R.}~\bibnamefont {Moszynski}},\ }\href
  {\doibase 10.1103/PhysRevA.91.042706} {\bibfield  {journal} {\bibinfo
  {journal} {Phys. Rev. A}\ }\textbf {\bibinfo {volume} {91}},\ \bibinfo
  {pages} {042706} (\bibinfo {year} {2015})}\BibitemShut {NoStop}%
\bibitem [{\citenamefont {M\o{}lhave}\ and\ \citenamefont
  {Drewsen}(2000)}]{MolhavePRA00}%
  \BibitemOpen
  \bibfield  {author} {\bibinfo {author} {\bibfnamefont {K.}~\bibnamefont
  {M\o{}lhave}}\ and\ \bibinfo {author} {\bibfnamefont {M.}~\bibnamefont
  {Drewsen}},\ }\href {\doibase 10.1103/PhysRevA.62.011401} {\bibfield
  {journal} {\bibinfo  {journal} {Phys. Rev. A}\ }\textbf {\bibinfo {volume}
  {62}},\ \bibinfo {pages} {011401} (\bibinfo {year} {2000})}\BibitemShut
  {NoStop}%
\bibitem [{\citenamefont {Hansen}\ \emph {et~al.}(2012)\citenamefont {Hansen},
  \citenamefont {Sørensen}, \citenamefont {Staanum},\ and\ \citenamefont
  {Drewsen}}]{HansenAC12}%
  \BibitemOpen
  \bibfield  {author} {\bibinfo {author} {\bibfnamefont {A.~K.}\ \bibnamefont
  {Hansen}}, \bibinfo {author} {\bibfnamefont {M.~A.}\ \bibnamefont
  {Sørensen}}, \bibinfo {author} {\bibfnamefont {P.~F.}\ \bibnamefont
  {Staanum}}, \ and\ \bibinfo {author} {\bibfnamefont {M.}~\bibnamefont
  {Drewsen}},\ }\href {\doibase 10.1002/anie.201203550} {\bibfield  {journal}
  {\bibinfo  {journal} {Angew. Chem. Int. Ed.}\ }\textbf {\bibinfo {volume}
  {51}},\ \bibinfo {pages} {7960} (\bibinfo {year} {2012})}\BibitemShut
  {NoStop}%
\bibitem [{\citenamefont {Patel}\ \emph {et~al.}(2026)\citenamefont {Patel},
  \citenamefont {Sardar}, \citenamefont {Saraladevi}, \citenamefont {Tomza},\
  and\ \citenamefont {Brown}}]{PatelJPCL26}%
  \BibitemOpen
  \bibfield  {author} {\bibinfo {author} {\bibfnamefont {S.}~\bibnamefont
  {Patel}}, \bibinfo {author} {\bibfnamefont {D.}~\bibnamefont {Sardar}},
  \bibinfo {author} {\bibfnamefont {J.}~\bibnamefont {Saraladevi}}, \bibinfo
  {author} {\bibfnamefont {M.}~\bibnamefont {Tomza}}, \ and\ \bibinfo {author}
  {\bibfnamefont {K.~R.}\ \bibnamefont {Brown}},\ }\href {\doibase
  10.1021/acs.jpclett.6c00828} {\bibfield  {journal} {\bibinfo  {journal} {J.
  Phys. Chem. Lett.}\ }\textbf {\bibinfo {volume} {17}},\ \bibinfo {pages}
  {6574} (\bibinfo {year} {2026})}\BibitemShut {NoStop}%
\bibitem [{\citenamefont {Sullivan}\ \emph {et~al.}(2011)\citenamefont
  {Sullivan}, \citenamefont {Rellergert}, \citenamefont {Kotochigova},
  \citenamefont {Chen}, \citenamefont {Schowalter},\ and\ \citenamefont
  {Hudson}}]{SullivanPCCP11}%
  \BibitemOpen
  \bibfield  {author} {\bibinfo {author} {\bibfnamefont {S.~T.}\ \bibnamefont
  {Sullivan}}, \bibinfo {author} {\bibfnamefont {W.~G.}\ \bibnamefont
  {Rellergert}}, \bibinfo {author} {\bibfnamefont {S.}~\bibnamefont
  {Kotochigova}}, \bibinfo {author} {\bibfnamefont {K.}~\bibnamefont {Chen}},
  \bibinfo {author} {\bibfnamefont {S.~J.}\ \bibnamefont {Schowalter}}, \ and\
  \bibinfo {author} {\bibfnamefont {E.~R.}\ \bibnamefont {Hudson}},\ }\href
  {\doibase 10.1039/C1CP21205B} {\bibfield  {journal} {\bibinfo  {journal}
  {Phys. Chem. Chem. Phys.}\ }\textbf {\bibinfo {volume} {13}},\ \bibinfo
  {pages} {18859} (\bibinfo {year} {2011})}\BibitemShut {NoStop}%
\bibitem [{\citenamefont {Jyothi}\ \emph {et~al.}(2016)\citenamefont {Jyothi},
  \citenamefont {Ray}, \citenamefont {Dutta}, \citenamefont {Allouche},
  \citenamefont {Vexiau}, \citenamefont {Dulieu},\ and\ \citenamefont
  {Rangwala}}]{JyothiPRL16}%
  \BibitemOpen
  \bibfield  {author} {\bibinfo {author} {\bibfnamefont {S.}~\bibnamefont
  {Jyothi}}, \bibinfo {author} {\bibfnamefont {T.}~\bibnamefont {Ray}},
  \bibinfo {author} {\bibfnamefont {S.}~\bibnamefont {Dutta}}, \bibinfo
  {author} {\bibfnamefont {A.~R.}\ \bibnamefont {Allouche}}, \bibinfo {author}
  {\bibfnamefont {R.}~\bibnamefont {Vexiau}}, \bibinfo {author} {\bibfnamefont
  {O.}~\bibnamefont {Dulieu}}, \ and\ \bibinfo {author} {\bibfnamefont {S.~A.}\
  \bibnamefont {Rangwala}},\ }\href {\doibase 10.1103/PhysRevLett.117.213002}
  {\bibfield  {journal} {\bibinfo  {journal} {Phys. Rev. Lett.}\ }\textbf
  {\bibinfo {volume} {117}},\ \bibinfo {pages} {213002} (\bibinfo {year}
  {2016})}\BibitemShut {NoStop}%
\bibitem [{\citenamefont {Rellergert}\ \emph {et~al.}(2013)\citenamefont
  {Rellergert}, \citenamefont {Sullivan}, \citenamefont {Schowalter},
  \citenamefont {Kotochigova}, \citenamefont {Chen},\ and\ \citenamefont
  {Hudson}}]{RellergertN13}%
  \BibitemOpen
  \bibfield  {author} {\bibinfo {author} {\bibfnamefont {W.~G.}\ \bibnamefont
  {Rellergert}}, \bibinfo {author} {\bibfnamefont {S.~T.}\ \bibnamefont
  {Sullivan}}, \bibinfo {author} {\bibfnamefont {S.~J.}\ \bibnamefont
  {Schowalter}}, \bibinfo {author} {\bibfnamefont {S.}~\bibnamefont
  {Kotochigova}}, \bibinfo {author} {\bibfnamefont {K.}~\bibnamefont {Chen}}, \
  and\ \bibinfo {author} {\bibfnamefont {E.~R.}\ \bibnamefont {Hudson}},\
  }\href {\doibase 10.1038/nature11937} {\bibfield  {journal} {\bibinfo
  {journal} {Nature}\ }\textbf {\bibinfo {volume} {495}},\ \bibinfo {pages}
  {490} (\bibinfo {year} {2013})}\BibitemShut {NoStop}%
\bibitem [{\citenamefont {Hansen}\ \emph {et~al.}(2014)\citenamefont {Hansen},
  \citenamefont {Versolato}, \citenamefont {K{\l}osowski}, \citenamefont
  {Kristensen}, \citenamefont {Gingell}, \citenamefont {Schwarz}, \citenamefont
  {Windberger}, \citenamefont {Ullrich}, \citenamefont {L{\'o}pez-Urrutia},\
  and\ \citenamefont {Drewsen}}]{Hansen14}%
  \BibitemOpen
  \bibfield  {author} {\bibinfo {author} {\bibfnamefont {A.~K.}\ \bibnamefont
  {Hansen}}, \bibinfo {author} {\bibfnamefont {O.}~\bibnamefont {Versolato}},
  \bibinfo {author} {\bibfnamefont {{\L}.}~\bibnamefont {K{\l}osowski}},
  \bibinfo {author} {\bibfnamefont {S.}~\bibnamefont {Kristensen}}, \bibinfo
  {author} {\bibfnamefont {A.}~\bibnamefont {Gingell}}, \bibinfo {author}
  {\bibfnamefont {M.}~\bibnamefont {Schwarz}}, \bibinfo {author} {\bibfnamefont
  {A.}~\bibnamefont {Windberger}}, \bibinfo {author} {\bibfnamefont
  {J.}~\bibnamefont {Ullrich}}, \bibinfo {author} {\bibfnamefont {J.~C.}\
  \bibnamefont {L{\'o}pez-Urrutia}}, \ and\ \bibinfo {author} {\bibfnamefont
  {M.}~\bibnamefont {Drewsen}},\ }\href {\doibase 10.1038/nature12996}
  {\bibfield  {journal} {\bibinfo  {journal} {Nature}\ }\textbf {\bibinfo
  {volume} {508}},\ \bibinfo {pages} {76} (\bibinfo {year} {2014})}\BibitemShut
  {NoStop}%
\bibitem [{\citenamefont {Simons}(2008)}]{SimonsJPCA08}%
  \BibitemOpen
  \bibfield  {author} {\bibinfo {author} {\bibfnamefont {J.}~\bibnamefont
  {Simons}},\ }\href {\doibase 10.1021/jp711490b} {\bibfield  {journal}
  {\bibinfo  {journal} {J. Phys. Chem. A}\ }\textbf {\bibinfo {volume} {112}},\
  \bibinfo {pages} {6401} (\bibinfo {year} {2008})}\BibitemShut {NoStop}%
\bibitem [{\citenamefont {Simons}(2023)}]{SimonsJPCA23}%
  \BibitemOpen
  \bibfield  {author} {\bibinfo {author} {\bibfnamefont {J.}~\bibnamefont
  {Simons}},\ }\href {\doibase 10.1021/acs.jpca.3c01564} {\bibfield  {journal}
  {\bibinfo  {journal} {J. Phys. Chem. A}\ }\textbf {\bibinfo {volume} {127}},\
  \bibinfo {pages} {3940} (\bibinfo {year} {2023})}\BibitemShut {NoStop}%
\bibitem [{\citenamefont {Simons}(2011)}]{SimonsARPC11}%
  \BibitemOpen
  \bibfield  {author} {\bibinfo {author} {\bibfnamefont {J.}~\bibnamefont
  {Simons}},\ }\href {\doibase 10.1146/annurev-physchem-032210-103547}
  {\bibfield  {journal} {\bibinfo  {journal} {Annu. Rev. Phys. Chem.}\ }\textbf
  {\bibinfo {volume} {62}},\ \bibinfo {pages} {107} (\bibinfo {year}
  {2011})}\BibitemShut {NoStop}%
\bibitem [{\citenamefont {Kellerbauer}\ and\ \citenamefont
  {Walz}(2006)}]{KellerbauerNJP06}%
  \BibitemOpen
  \bibfield  {author} {\bibinfo {author} {\bibfnamefont {A.}~\bibnamefont
  {Kellerbauer}}\ and\ \bibinfo {author} {\bibfnamefont {J.}~\bibnamefont
  {Walz}},\ }\href {\doibase 10.1088/1367-2630/8/3/045} {\bibfield  {journal}
  {\bibinfo  {journal} {New J. Phys.}\ }\textbf {\bibinfo {volume} {8}},\
  \bibinfo {pages} {45} (\bibinfo {year} {2006})}\BibitemShut {NoStop}%
\bibitem [{\citenamefont {Cerchiari}\ \emph {et~al.}(2018)\citenamefont
  {Cerchiari}, \citenamefont {Kellerbauer}, \citenamefont {Safronova},
  \citenamefont {Safronova},\ and\ \citenamefont {Yzombard}}]{CerchiariPRL18}%
  \BibitemOpen
  \bibfield  {author} {\bibinfo {author} {\bibfnamefont {G.}~\bibnamefont
  {Cerchiari}}, \bibinfo {author} {\bibfnamefont {A.}~\bibnamefont
  {Kellerbauer}}, \bibinfo {author} {\bibfnamefont {M.~S.}\ \bibnamefont
  {Safronova}}, \bibinfo {author} {\bibfnamefont {U.~I.}\ \bibnamefont
  {Safronova}}, \ and\ \bibinfo {author} {\bibfnamefont {P.}~\bibnamefont
  {Yzombard}},\ }\href {\doibase 10.1103/PhysRevLett.120.133205} {\bibfield
  {journal} {\bibinfo  {journal} {Phys. Rev. Lett.}\ }\textbf {\bibinfo
  {volume} {120}},\ \bibinfo {pages} {133205} (\bibinfo {year}
  {2018})}\BibitemShut {NoStop}%
\bibitem [{\citenamefont {Pegg}(2004)}]{PeggRPP04}%
  \BibitemOpen
  \bibfield  {author} {\bibinfo {author} {\bibfnamefont {D.~J.}\ \bibnamefont
  {Pegg}},\ }\href {\doibase 10.1088/0034-4885/67/6/R02} {\bibfield  {journal}
  {\bibinfo  {journal} {Rep. Prog. Phys.}\ }\textbf {\bibinfo {volume} {67}},\
  \bibinfo {pages} {857} (\bibinfo {year} {2004})}\BibitemShut {NoStop}%
\bibitem [{\citenamefont {Bilodeau}\ and\ \citenamefont
  {Haugen}(2000)}]{BilodeauPRL00}%
  \BibitemOpen
  \bibfield  {author} {\bibinfo {author} {\bibfnamefont {R.~C.}\ \bibnamefont
  {Bilodeau}}\ and\ \bibinfo {author} {\bibfnamefont {H.~K.}\ \bibnamefont
  {Haugen}},\ }\href {\doibase 10.1103/PhysRevLett.85.534} {\bibfield
  {journal} {\bibinfo  {journal} {Phys. Rev. Lett.}\ }\textbf {\bibinfo
  {volume} {85}},\ \bibinfo {pages} {534} (\bibinfo {year} {2000})}\BibitemShut
  {NoStop}%
\bibitem [{\citenamefont {Walter}\ \emph {et~al.}(2007)\citenamefont {Walter},
  \citenamefont {Gibson}, \citenamefont {Janczak}, \citenamefont {Starr},
  \citenamefont {Snedden}, \citenamefont {Field~III},\ and\ \citenamefont
  {Andersson}}]{WalterPRL07}%
  \BibitemOpen
  \bibfield  {author} {\bibinfo {author} {\bibfnamefont {C.~W.}\ \bibnamefont
  {Walter}}, \bibinfo {author} {\bibfnamefont {N.~D.}\ \bibnamefont {Gibson}},
  \bibinfo {author} {\bibfnamefont {C.~M.}\ \bibnamefont {Janczak}}, \bibinfo
  {author} {\bibfnamefont {K.~A.}\ \bibnamefont {Starr}}, \bibinfo {author}
  {\bibfnamefont {A.~P.}\ \bibnamefont {Snedden}}, \bibinfo {author}
  {\bibfnamefont {R.~L.}\ \bibnamefont {Field~III}}, \ and\ \bibinfo {author}
  {\bibfnamefont {P.}~\bibnamefont {Andersson}},\ }\href {\doibase
  10.1103/PhysRevA.76.052702} {\bibfield  {journal} {\bibinfo  {journal} {Phys.
  Rev. A}\ }\textbf {\bibinfo {volume} {76}},\ \bibinfo {pages} {052702}
  (\bibinfo {year} {2007})}\BibitemShut {NoStop}%
\bibitem [{\citenamefont {O'Malley}\ and\ \citenamefont
  {Beck}(2010)}]{OMalleyPRA10}%
  \BibitemOpen
  \bibfield  {author} {\bibinfo {author} {\bibfnamefont {S.~M.}\ \bibnamefont
  {O'Malley}}\ and\ \bibinfo {author} {\bibfnamefont {D.~R.}\ \bibnamefont
  {Beck}},\ }\href {\doibase 10.1103/PhysRevA.81.032503} {\bibfield  {journal}
  {\bibinfo  {journal} {Phys. Rev. A}\ }\textbf {\bibinfo {volume} {81}},\
  \bibinfo {pages} {032503} (\bibinfo {year} {2010})}\BibitemShut {NoStop}%
\bibitem [{\citenamefont {Walter}\ \emph {et~al.}(2014)\citenamefont {Walter},
  \citenamefont {Gibson}, \citenamefont {Matyas}, \citenamefont {Crocker},
  \citenamefont {Dungan}, \citenamefont {Matola},\ and\ \citenamefont
  {Rohl\'en}}]{WalterPRL14}%
  \BibitemOpen
  \bibfield  {author} {\bibinfo {author} {\bibfnamefont {C.~W.}\ \bibnamefont
  {Walter}}, \bibinfo {author} {\bibfnamefont {N.~D.}\ \bibnamefont {Gibson}},
  \bibinfo {author} {\bibfnamefont {D.~J.}\ \bibnamefont {Matyas}}, \bibinfo
  {author} {\bibfnamefont {C.}~\bibnamefont {Crocker}}, \bibinfo {author}
  {\bibfnamefont {K.~A.}\ \bibnamefont {Dungan}}, \bibinfo {author}
  {\bibfnamefont {B.~R.}\ \bibnamefont {Matola}}, \ and\ \bibinfo {author}
  {\bibfnamefont {J.}~\bibnamefont {Rohl\'en}},\ }\href {\doibase
  10.1103/PhysRevLett.113.063001} {\bibfield  {journal} {\bibinfo  {journal}
  {Phys. Rev. Lett.}\ }\textbf {\bibinfo {volume} {113}},\ \bibinfo {pages}
  {063001} (\bibinfo {year} {2014})}\BibitemShut {NoStop}%
\bibitem [{\citenamefont {Tang}\ \emph {et~al.}(2019)\citenamefont {Tang},
  \citenamefont {Si}, \citenamefont {Fei}, \citenamefont {Fu}, \citenamefont
  {Lu}, \citenamefont {Brage}, \citenamefont {Liu}, \citenamefont {Chen},\ and\
  \citenamefont {Ning}}]{TangPRL19}%
  \BibitemOpen
  \bibfield  {author} {\bibinfo {author} {\bibfnamefont {R.}~\bibnamefont
  {Tang}}, \bibinfo {author} {\bibfnamefont {R.}~\bibnamefont {Si}}, \bibinfo
  {author} {\bibfnamefont {Z.}~\bibnamefont {Fei}}, \bibinfo {author}
  {\bibfnamefont {X.}~\bibnamefont {Fu}}, \bibinfo {author} {\bibfnamefont
  {Y.}~\bibnamefont {Lu}}, \bibinfo {author} {\bibfnamefont {T.}~\bibnamefont
  {Brage}}, \bibinfo {author} {\bibfnamefont {H.}~\bibnamefont {Liu}}, \bibinfo
  {author} {\bibfnamefont {C.}~\bibnamefont {Chen}}, \ and\ \bibinfo {author}
  {\bibfnamefont {C.}~\bibnamefont {Ning}},\ }\href {\doibase
  10.1103/PhysRevLett.123.203002} {\bibfield  {journal} {\bibinfo  {journal}
  {Phys. Rev. Lett.}\ }\textbf {\bibinfo {volume} {123}},\ \bibinfo {pages}
  {203002} (\bibinfo {year} {2019})}\BibitemShut {NoStop}%
\bibitem [{\citenamefont {Yzombard}\ \emph {et~al.}(2015)\citenamefont
  {Yzombard}, \citenamefont {Hamamda}, \citenamefont {Gerber}, \citenamefont
  {Doser},\ and\ \citenamefont {Comparat}}]{YzombardPRL15}%
  \BibitemOpen
  \bibfield  {author} {\bibinfo {author} {\bibfnamefont {P.}~\bibnamefont
  {Yzombard}}, \bibinfo {author} {\bibfnamefont {M.}~\bibnamefont {Hamamda}},
  \bibinfo {author} {\bibfnamefont {S.}~\bibnamefont {Gerber}}, \bibinfo
  {author} {\bibfnamefont {M.}~\bibnamefont {Doser}}, \ and\ \bibinfo {author}
  {\bibfnamefont {D.}~\bibnamefont {Comparat}},\ }\href {\doibase
  10.1103/PhysRevLett.114.213001} {\bibfield  {journal} {\bibinfo  {journal}
  {Phys. Rev. Lett.}\ }\textbf {\bibinfo {volume} {114}},\ \bibinfo {pages}
  {213001} (\bibinfo {year} {2015})}\BibitemShut {NoStop}%
\bibitem [{\citenamefont {N\"otzold}\ \emph {et~al.}(2022)\citenamefont
  {N\"otzold}, \citenamefont {Wild}, \citenamefont {Lochmann},\ and\
  \citenamefont {Wester}}]{NotzoldPRA22}%
  \BibitemOpen
  \bibfield  {author} {\bibinfo {author} {\bibfnamefont {M.}~\bibnamefont
  {N\"otzold}}, \bibinfo {author} {\bibfnamefont {R.}~\bibnamefont {Wild}},
  \bibinfo {author} {\bibfnamefont {C.}~\bibnamefont {Lochmann}}, \ and\
  \bibinfo {author} {\bibfnamefont {R.}~\bibnamefont {Wester}},\ }\href
  {\doibase 10.1103/PhysRevA.106.023111} {\bibfield  {journal} {\bibinfo
  {journal} {Phys. Rev. A}\ }\textbf {\bibinfo {volume} {106}},\ \bibinfo
  {pages} {023111} (\bibinfo {year} {2022})}\BibitemShut {NoStop}%
\bibitem [{\citenamefont {Rienstra-Kiracofe}\ \emph {et~al.}(2002)\citenamefont
  {Rienstra-Kiracofe}, \citenamefont {Tschumper}, \citenamefont {Schaefer},
  \citenamefont {Nandi},\ and\ \citenamefont {Ellison}}]{RienstraCR02}%
  \BibitemOpen
  \bibfield  {author} {\bibinfo {author} {\bibfnamefont {J.~C.}\ \bibnamefont
  {Rienstra-Kiracofe}}, \bibinfo {author} {\bibfnamefont {G.~S.}\ \bibnamefont
  {Tschumper}}, \bibinfo {author} {\bibfnamefont {H.~F.}\ \bibnamefont
  {Schaefer}}, \bibinfo {author} {\bibfnamefont {S.}~\bibnamefont {Nandi}}, \
  and\ \bibinfo {author} {\bibfnamefont {G.~B.}\ \bibnamefont {Ellison}},\
  }\href {\doibase 10.1021/cr990044u} {\bibfield  {journal} {\bibinfo
  {journal} {Chem. Rev.}\ }\textbf {\bibinfo {volume} {102}},\ \bibinfo {pages}
  {231} (\bibinfo {year} {2002})}\BibitemShut {NoStop}%
\bibitem [{\citenamefont {Desfran{\c{c}}ois}\ \emph {et~al.}(1996)\citenamefont
  {Desfran{\c{c}}ois}, \citenamefont {Abdoul-Carime},\ and\ \citenamefont
  {Schermann}}]{FranccoisIJMPB96}%
  \BibitemOpen
  \bibfield  {author} {\bibinfo {author} {\bibfnamefont {C.}~\bibnamefont
  {Desfran{\c{c}}ois}}, \bibinfo {author} {\bibfnamefont {H.}~\bibnamefont
  {Abdoul-Carime}}, \ and\ \bibinfo {author} {\bibfnamefont {J.-P.}\
  \bibnamefont {Schermann}},\ }\href {\doibase 10.1142/S0217979296000520}
  {\bibfield  {journal} {\bibinfo  {journal} {Int. J. Mod. Phys. B}\ }\textbf
  {\bibinfo {volume} {10}},\ \bibinfo {pages} {1339} (\bibinfo {year}
  {1996})}\BibitemShut {NoStop}%
\bibitem [{\citenamefont {Jordan}\ and\ \citenamefont
  {Wang}(2003)}]{JordanARPC03}%
  \BibitemOpen
  \bibfield  {author} {\bibinfo {author} {\bibfnamefont {K.~D.}\ \bibnamefont
  {Jordan}}\ and\ \bibinfo {author} {\bibfnamefont {F.}~\bibnamefont {Wang}},\
  }\href {\doibase 10.1146/annurev.physchem.54.011002.103851} {\bibfield
  {journal} {\bibinfo  {journal} {Annu. Rev. Phys. Chem.}\ }\textbf {\bibinfo
  {volume} {54}},\ \bibinfo {pages} {367} (\bibinfo {year} {2003})}\BibitemShut
  {NoStop}%
\bibitem [{\citenamefont {Lu}\ \emph {et~al.}(2021)\citenamefont {Lu},
  \citenamefont {Tang},\ and\ \citenamefont {Ning}}]{LuJPCL21}%
  \BibitemOpen
  \bibfield  {author} {\bibinfo {author} {\bibfnamefont {Y.}~\bibnamefont
  {Lu}}, \bibinfo {author} {\bibfnamefont {R.}~\bibnamefont {Tang}}, \ and\
  \bibinfo {author} {\bibfnamefont {C.}~\bibnamefont {Ning}},\ }\href {\doibase
  10.1021/acs.jpclett.1c01726} {\bibfield  {journal} {\bibinfo  {journal} {J.
  Phys. Chem. Lett.}\ }\textbf {\bibinfo {volume} {12}},\ \bibinfo {pages}
  {5897} (\bibinfo {year} {2021})}\BibitemShut {NoStop}%
\bibitem [{\citenamefont {Tauch}\ \emph {et~al.}(2023)\citenamefont {Tauch},
  \citenamefont {Hassan}, \citenamefont {N{\"o}tzold}, \citenamefont {Endres},
  \citenamefont {Wester},\ and\ \citenamefont {Weidem{\"u}ller}}]{TauchNP23}%
  \BibitemOpen
  \bibfield  {author} {\bibinfo {author} {\bibfnamefont {J.}~\bibnamefont
  {Tauch}}, \bibinfo {author} {\bibfnamefont {S.~Z.}\ \bibnamefont {Hassan}},
  \bibinfo {author} {\bibfnamefont {M.}~\bibnamefont {N{\"o}tzold}}, \bibinfo
  {author} {\bibfnamefont {E.~S.}\ \bibnamefont {Endres}}, \bibinfo {author}
  {\bibfnamefont {R.}~\bibnamefont {Wester}}, \ and\ \bibinfo {author}
  {\bibfnamefont {M.}~\bibnamefont {Weidem{\"u}ller}},\ }\href {\doibase
  10.1038/s41567-023-02084-6} {\bibfield  {journal} {\bibinfo  {journal} {Nat.
  Phys.}\ }\textbf {\bibinfo {volume} {19}},\ \bibinfo {pages} {1270} (\bibinfo
  {year} {2023})}\BibitemShut {NoStop}%
\bibitem [{\citenamefont {Hauser}\ \emph {et~al.}(2015)\citenamefont {Hauser},
  \citenamefont {Lee}, \citenamefont {Carelli}, \citenamefont {Spieler},
  \citenamefont {Lakhmanskaya}, \citenamefont {Endres}, \citenamefont {Kumar},
  \citenamefont {Gianturco},\ and\ \citenamefont {Wester}}]{HauserNP15}%
  \BibitemOpen
  \bibfield  {author} {\bibinfo {author} {\bibfnamefont {D.}~\bibnamefont
  {Hauser}}, \bibinfo {author} {\bibfnamefont {S.}~\bibnamefont {Lee}},
  \bibinfo {author} {\bibfnamefont {F.}~\bibnamefont {Carelli}}, \bibinfo
  {author} {\bibfnamefont {S.}~\bibnamefont {Spieler}}, \bibinfo {author}
  {\bibfnamefont {O.}~\bibnamefont {Lakhmanskaya}}, \bibinfo {author}
  {\bibfnamefont {E.~S.}\ \bibnamefont {Endres}}, \bibinfo {author}
  {\bibfnamefont {S.~S.}\ \bibnamefont {Kumar}}, \bibinfo {author}
  {\bibfnamefont {F.}~\bibnamefont {Gianturco}}, \ and\ \bibinfo {author}
  {\bibfnamefont {R.}~\bibnamefont {Wester}},\ }\href {\doibase
  10.1038/nphys3326} {\bibfield  {journal} {\bibinfo  {journal} {Nat. Phys.}\
  }\textbf {\bibinfo {volume} {11}},\ \bibinfo {pages} {467} (\bibinfo {year}
  {2015})}\BibitemShut {NoStop}%
\bibitem [{\citenamefont {Deiglmayr}\ \emph {et~al.}(2012)\citenamefont
  {Deiglmayr}, \citenamefont {G{\"o}ritz}, \citenamefont {Best}, \citenamefont
  {Weidem{\"u}ller},\ and\ \citenamefont {Wester}}]{DeiglmayrPRA12}%
  \BibitemOpen
  \bibfield  {author} {\bibinfo {author} {\bibfnamefont {J.}~\bibnamefont
  {Deiglmayr}}, \bibinfo {author} {\bibfnamefont {A.}~\bibnamefont
  {G{\"o}ritz}}, \bibinfo {author} {\bibfnamefont {T.}~\bibnamefont {Best}},
  \bibinfo {author} {\bibfnamefont {M.}~\bibnamefont {Weidem{\"u}ller}}, \ and\
  \bibinfo {author} {\bibfnamefont {R.}~\bibnamefont {Wester}},\ }\href
  {\doibase 10.1103/PhysRevA.86.043438} {\bibfield  {journal} {\bibinfo
  {journal} {Phys. Rev. A}\ }\textbf {\bibinfo {volume} {86}},\ \bibinfo
  {pages} {043438} (\bibinfo {year} {2012})}\BibitemShut {NoStop}%
\bibitem [{\citenamefont {Hassan}\ \emph
  {et~al.}(2022{\natexlab{a}})\citenamefont {Hassan}, \citenamefont {Tauch},
  \citenamefont {Kas}, \citenamefont {N{\"o}tzold}, \citenamefont {Carrera},
  \citenamefont {Endres}, \citenamefont {Wester},\ and\ \citenamefont
  {Weidem{\"u}ller}}]{HassanNC22}%
  \BibitemOpen
  \bibfield  {author} {\bibinfo {author} {\bibfnamefont {S.~Z.}\ \bibnamefont
  {Hassan}}, \bibinfo {author} {\bibfnamefont {J.}~\bibnamefont {Tauch}},
  \bibinfo {author} {\bibfnamefont {M.}~\bibnamefont {Kas}}, \bibinfo {author}
  {\bibfnamefont {M.}~\bibnamefont {N{\"o}tzold}}, \bibinfo {author}
  {\bibfnamefont {H.~L.}\ \bibnamefont {Carrera}}, \bibinfo {author}
  {\bibfnamefont {E.~S.}\ \bibnamefont {Endres}}, \bibinfo {author}
  {\bibfnamefont {R.}~\bibnamefont {Wester}}, \ and\ \bibinfo {author}
  {\bibfnamefont {M.}~\bibnamefont {Weidem{\"u}ller}},\ }\href {\doibase
  10.1038/s41467-022-28382-w} {\bibfield  {journal} {\bibinfo  {journal} {Nat.
  Commun.}\ }\textbf {\bibinfo {volume} {13}},\ \bibinfo {pages} {818}
  (\bibinfo {year} {2022}{\natexlab{a}})}\BibitemShut {NoStop}%
\bibitem [{\citenamefont {Hassan}\ \emph
  {et~al.}(2022{\natexlab{b}})\citenamefont {Hassan}, \citenamefont {Tauch},
  \citenamefont {Kas}, \citenamefont {Nötzold}, \citenamefont {Wester},\ and\
  \citenamefont {Weidemüller}}]{HassanJCP22}%
  \BibitemOpen
  \bibfield  {author} {\bibinfo {author} {\bibfnamefont {S.~Z.}\ \bibnamefont
  {Hassan}}, \bibinfo {author} {\bibfnamefont {J.}~\bibnamefont {Tauch}},
  \bibinfo {author} {\bibfnamefont {M.}~\bibnamefont {Kas}}, \bibinfo {author}
  {\bibfnamefont {M.}~\bibnamefont {Nötzold}}, \bibinfo {author}
  {\bibfnamefont {R.}~\bibnamefont {Wester}}, \ and\ \bibinfo {author}
  {\bibfnamefont {M.}~\bibnamefont {Weidemüller}},\ }\href {\doibase
  10.1063/5.0082734} {\bibfield  {journal} {\bibinfo  {journal} {J. Chem.
  Phys.}\ }\textbf {\bibinfo {volume} {156}},\ \bibinfo {pages} {094304}
  (\bibinfo {year} {2022}{\natexlab{b}})}\BibitemShut {NoStop}%
\bibitem [{\citenamefont {Gonz{\'a}lez-S{\'a}nchez}\ \emph
  {et~al.}(2015)\citenamefont {Gonz{\'a}lez-S{\'a}nchez}, \citenamefont
  {Carelli}, \citenamefont {Gianturco},\ and\ \citenamefont
  {Wester}}]{GonzalezCP15}%
  \BibitemOpen
  \bibfield  {author} {\bibinfo {author} {\bibfnamefont {L.}~\bibnamefont
  {Gonz{\'a}lez-S{\'a}nchez}}, \bibinfo {author} {\bibfnamefont
  {F.}~\bibnamefont {Carelli}}, \bibinfo {author} {\bibfnamefont
  {F.}~\bibnamefont {Gianturco}}, \ and\ \bibinfo {author} {\bibfnamefont
  {R.}~\bibnamefont {Wester}},\ }\href {\doibase
  10.1016/j.chemphys.2015.05.027} {\bibfield  {journal} {\bibinfo  {journal}
  {Chem. Phys.}\ }\textbf {\bibinfo {volume} {462}},\ \bibinfo {pages} {111}
  (\bibinfo {year} {2015})}\BibitemShut {NoStop}%
\bibitem [{\citenamefont {Kas}\ \emph {et~al.}(2017)\citenamefont {Kas},
  \citenamefont {Loreau}, \citenamefont {Liévin},\ and\ \citenamefont
  {Vaeck}}]{KasJCP17}%
  \BibitemOpen
  \bibfield  {author} {\bibinfo {author} {\bibfnamefont {M.}~\bibnamefont
  {Kas}}, \bibinfo {author} {\bibfnamefont {J.}~\bibnamefont {Loreau}},
  \bibinfo {author} {\bibfnamefont {J.}~\bibnamefont {Liévin}}, \ and\
  \bibinfo {author} {\bibfnamefont {N.}~\bibnamefont {Vaeck}},\ }\href
  {\doibase 10.1063/1.4983627} {\bibfield  {journal} {\bibinfo  {journal} {J.
  Chem. Phys.}\ }\textbf {\bibinfo {volume} {146}},\ \bibinfo {pages} {194309}
  (\bibinfo {year} {2017})}\BibitemShut {NoStop}%
\bibitem [{\citenamefont {Tomza}(2017)}]{TomzaPCCP17}%
  \BibitemOpen
  \bibfield  {author} {\bibinfo {author} {\bibfnamefont {M.}~\bibnamefont
  {Tomza}},\ }\href {\doibase 10.1039/C7CP02127E} {\bibfield  {journal}
  {\bibinfo  {journal} {Phys. Chem. Chem. Phys.}\ }\textbf {\bibinfo {volume}
  {19}},\ \bibinfo {pages} {16512} (\bibinfo {year} {2017})}\BibitemShut
  {NoStop}%
\bibitem [{\citenamefont {Kas}\ \emph {et~al.}(2019)\citenamefont {Kas},
  \citenamefont {Loreau}, \citenamefont {Li\'evin},\ and\ \citenamefont
  {Vaeck}}]{KasPRA19}%
  \BibitemOpen
  \bibfield  {author} {\bibinfo {author} {\bibfnamefont {M.}~\bibnamefont
  {Kas}}, \bibinfo {author} {\bibfnamefont {J.}~\bibnamefont {Loreau}},
  \bibinfo {author} {\bibfnamefont {J.}~\bibnamefont {Li\'evin}}, \ and\
  \bibinfo {author} {\bibfnamefont {N.}~\bibnamefont {Vaeck}},\ }\href
  {\doibase 10.1103/PhysRevA.99.042702} {\bibfield  {journal} {\bibinfo
  {journal} {Phys. Rev. A}\ }\textbf {\bibinfo {volume} {99}},\ \bibinfo
  {pages} {042702} (\bibinfo {year} {2019})}\BibitemShut {NoStop}%
\bibitem [{\citenamefont {Byrd}\ \emph {et~al.}(2013)\citenamefont {Byrd},
  \citenamefont {Michels}, \citenamefont {Montgomery~Jr},\ and\ \citenamefont
  {C{\^o}t{\'e}}}]{ByrdPRA13}%
  \BibitemOpen
  \bibfield  {author} {\bibinfo {author} {\bibfnamefont {J.~N.}\ \bibnamefont
  {Byrd}}, \bibinfo {author} {\bibfnamefont {H.~H.}\ \bibnamefont {Michels}},
  \bibinfo {author} {\bibfnamefont {J.~A.}\ \bibnamefont {Montgomery~Jr}}, \
  and\ \bibinfo {author} {\bibfnamefont {R.}~\bibnamefont {C{\^o}t{\'e}}},\
  }\href {\doibase 10.1103/PhysRevA.88.032710} {\bibfield  {journal} {\bibinfo
  {journal} {Phys. Rev. A}\ }\textbf {\bibinfo {volume} {88}},\ \bibinfo
  {pages} {032710} (\bibinfo {year} {2013})}\BibitemShut {NoStop}%
\bibitem [{\citenamefont {Guttridge}\ \emph {et~al.}(2023)\citenamefont
  {Guttridge}, \citenamefont {Ruttley}, \citenamefont {Baldock}, \citenamefont
  {Gonz\'alez-F\'erez}, \citenamefont {Sadeghpour}, \citenamefont {Adams},\
  and\ \citenamefont {Cornish}}]{GuttridgePRL23}%
  \BibitemOpen
  \bibfield  {author} {\bibinfo {author} {\bibfnamefont {A.}~\bibnamefont
  {Guttridge}}, \bibinfo {author} {\bibfnamefont {D.~K.}\ \bibnamefont
  {Ruttley}}, \bibinfo {author} {\bibfnamefont {A.~C.}\ \bibnamefont
  {Baldock}}, \bibinfo {author} {\bibfnamefont {R.}~\bibnamefont
  {Gonz\'alez-F\'erez}}, \bibinfo {author} {\bibfnamefont {H.~R.}\ \bibnamefont
  {Sadeghpour}}, \bibinfo {author} {\bibfnamefont {C.~S.}\ \bibnamefont
  {Adams}}, \ and\ \bibinfo {author} {\bibfnamefont {S.~L.}\ \bibnamefont
  {Cornish}},\ }\href {\doibase 10.1103/PhysRevLett.131.013401} {\bibfield
  {journal} {\bibinfo  {journal} {Phys. Rev. Lett.}\ }\textbf {\bibinfo
  {volume} {131}},\ \bibinfo {pages} {013401} (\bibinfo {year}
  {2023})}\BibitemShut {NoStop}%
\bibitem [{\citenamefont {Zhu}\ \emph {et~al.}(2025)\citenamefont {Zhu},
  \citenamefont {Luke}, \citenamefont {Shaham}, \citenamefont {Liu},\ and\
  \citenamefont {Ni}}]{ZhuPRL25}%
  \BibitemOpen
  \bibfield  {author} {\bibinfo {author} {\bibfnamefont {L.}~\bibnamefont
  {Zhu}}, \bibinfo {author} {\bibfnamefont {J.}~\bibnamefont {Luke}}, \bibinfo
  {author} {\bibfnamefont {R.}~\bibnamefont {Shaham}}, \bibinfo {author}
  {\bibfnamefont {Y.-X.}\ \bibnamefont {Liu}}, \ and\ \bibinfo {author}
  {\bibfnamefont {K.-K.}\ \bibnamefont {Ni}},\ }\href {\doibase
  10.1103/48rk-sxfs} {\bibfield  {journal} {\bibinfo  {journal} {Phys. Rev.
  Lett.}\ }\textbf {\bibinfo {volume} {135}},\ \bibinfo {pages} {153001}
  (\bibinfo {year} {2025})}\BibitemShut {NoStop}%
\bibitem [{\citenamefont {Patsch}\ \emph {et~al.}(2022)\citenamefont {Patsch},
  \citenamefont {Zeppenfeld},\ and\ \citenamefont {Koch}}]{PatschJPCL22}%
  \BibitemOpen
  \bibfield  {author} {\bibinfo {author} {\bibfnamefont {S.}~\bibnamefont
  {Patsch}}, \bibinfo {author} {\bibfnamefont {M.}~\bibnamefont {Zeppenfeld}},
  \ and\ \bibinfo {author} {\bibfnamefont {C.~P.}\ \bibnamefont {Koch}},\
  }\href {\doibase 10.1021/acs.jpclett.2c02521} {\bibfield  {journal} {\bibinfo
   {journal} {J. Phys. Chem. Lett.}\ }\textbf {\bibinfo {volume} {13}},\
  \bibinfo {pages} {10728} (\bibinfo {year} {2022})}\BibitemShut {NoStop}%
\bibitem [{\citenamefont {Zou}\ \emph {et~al.}(2026)\citenamefont {Zou},
  \citenamefont {Wang}, \citenamefont {Gonz\'alez-F\'erez}, \citenamefont
  {Sadeghpour},\ and\ \citenamefont {Hogan}}]{ZouPRL26}%
  \BibitemOpen
  \bibfield  {author} {\bibinfo {author} {\bibfnamefont {J.}~\bibnamefont
  {Zou}}, \bibinfo {author} {\bibfnamefont {R.~R.~W.}\ \bibnamefont {Wang}},
  \bibinfo {author} {\bibfnamefont {R.}~\bibnamefont {Gonz\'alez-F\'erez}},
  \bibinfo {author} {\bibfnamefont {H.~R.}\ \bibnamefont {Sadeghpour}}, \ and\
  \bibinfo {author} {\bibfnamefont {S.~D.}\ \bibnamefont {Hogan}},\ }\href
  {\doibase 10.1103/k9d5-1jcc} {\bibfield  {journal} {\bibinfo  {journal}
  {Phys. Rev. Lett.}\ }\textbf {\bibinfo {volume} {136}},\ \bibinfo {pages}
  {113402} (\bibinfo {year} {2026})}\BibitemShut {NoStop}%
\bibitem [{\citenamefont {Huber}\ and\ \citenamefont
  {B\"uchler}(2012)}]{HuberPRL12}%
  \BibitemOpen
  \bibfield  {author} {\bibinfo {author} {\bibfnamefont {S.~D.}\ \bibnamefont
  {Huber}}\ and\ \bibinfo {author} {\bibfnamefont {H.~P.}\ \bibnamefont
  {B\"uchler}},\ }\href {\doibase 10.1103/PhysRevLett.108.193006} {\bibfield
  {journal} {\bibinfo  {journal} {Phys. Rev. Lett.}\ }\textbf {\bibinfo
  {volume} {108}},\ \bibinfo {pages} {193006} (\bibinfo {year}
  {2012})}\BibitemShut {NoStop}%
\bibitem [{\citenamefont {Zhao}\ \emph {et~al.}(2012)\citenamefont {Zhao},
  \citenamefont {Glaetzle}, \citenamefont {Pupillo},\ and\ \citenamefont
  {Zoller}}]{ZhaoPRL12}%
  \BibitemOpen
  \bibfield  {author} {\bibinfo {author} {\bibfnamefont {B.}~\bibnamefont
  {Zhao}}, \bibinfo {author} {\bibfnamefont {A.~W.}\ \bibnamefont {Glaetzle}},
  \bibinfo {author} {\bibfnamefont {G.}~\bibnamefont {Pupillo}}, \ and\
  \bibinfo {author} {\bibfnamefont {P.}~\bibnamefont {Zoller}},\ }\href
  {\doibase 10.1103/PhysRevLett.108.193007} {\bibfield  {journal} {\bibinfo
  {journal} {Phys. Rev. Lett.}\ }\textbf {\bibinfo {volume} {108}},\ \bibinfo
  {pages} {193007} (\bibinfo {year} {2012})}\BibitemShut {NoStop}%
\bibitem [{\citenamefont {Zhang}\ \emph {et~al.}(2024)\citenamefont {Zhang},
  \citenamefont {Rittenhouse}, \citenamefont {Tscherbul}, \citenamefont
  {Sadeghpour},\ and\ \citenamefont {Hutzler}}]{ZhangPRL24}%
  \BibitemOpen
  \bibfield  {author} {\bibinfo {author} {\bibfnamefont {C.}~\bibnamefont
  {Zhang}}, \bibinfo {author} {\bibfnamefont {S.~T.}\ \bibnamefont
  {Rittenhouse}}, \bibinfo {author} {\bibfnamefont {T.~V.}\ \bibnamefont
  {Tscherbul}}, \bibinfo {author} {\bibfnamefont {H.~R.}\ \bibnamefont
  {Sadeghpour}}, \ and\ \bibinfo {author} {\bibfnamefont {N.~R.}\ \bibnamefont
  {Hutzler}},\ }\href {\doibase 10.1103/PhysRevLett.132.033001} {\bibfield
  {journal} {\bibinfo  {journal} {Phys. Rev. Lett.}\ }\textbf {\bibinfo
  {volume} {132}},\ \bibinfo {pages} {033001} (\bibinfo {year}
  {2024})}\BibitemShut {NoStop}%
\bibitem [{\citenamefont {Zhang}\ and\ \citenamefont
  {Tarbutt}(2022)}]{ZhangPRXQ22}%
  \BibitemOpen
  \bibfield  {author} {\bibinfo {author} {\bibfnamefont {C.}~\bibnamefont
  {Zhang}}\ and\ \bibinfo {author} {\bibfnamefont {M.}~\bibnamefont
  {Tarbutt}},\ }\href {\doibase 10.1103/PRXQuantum.3.030340} {\bibfield
  {journal} {\bibinfo  {journal} {PRX Quantum}\ }\textbf {\bibinfo {volume}
  {3}},\ \bibinfo {pages} {030340} (\bibinfo {year} {2022})}\BibitemShut
  {NoStop}%
\bibitem [{\citenamefont {Wang}\ \emph {et~al.}(2022)\citenamefont {Wang},
  \citenamefont {Williams}, \citenamefont {Picard}, \citenamefont {Yao},\ and\
  \citenamefont {Ni}}]{WangPRXQ22}%
  \BibitemOpen
  \bibfield  {author} {\bibinfo {author} {\bibfnamefont {K.}~\bibnamefont
  {Wang}}, \bibinfo {author} {\bibfnamefont {C.~P.}\ \bibnamefont {Williams}},
  \bibinfo {author} {\bibfnamefont {L.~R.}\ \bibnamefont {Picard}}, \bibinfo
  {author} {\bibfnamefont {N.~Y.}\ \bibnamefont {Yao}}, \ and\ \bibinfo
  {author} {\bibfnamefont {K.-K.}\ \bibnamefont {Ni}},\ }\href {\doibase
  10.1103/PRXQuantum.3.030339} {\bibfield  {journal} {\bibinfo  {journal} {PRX
  Quantum}\ }\textbf {\bibinfo {volume} {3}},\ \bibinfo {pages} {030339}
  (\bibinfo {year} {2022})}\BibitemShut {NoStop}%
\bibitem [{\citenamefont {Bai}\ \emph {et~al.}(2026)\citenamefont {Bai},
  \citenamefont {Wei}, \citenamefont {Zhang}, \citenamefont {Li},\ and\
  \citenamefont {Shao}}]{Bai2026}%
  \BibitemOpen
  \bibfield  {author} {\bibinfo {author} {\bibfnamefont {Y.-H.}\ \bibnamefont
  {Bai}}, \bibinfo {author} {\bibfnamefont {Y.}~\bibnamefont {Wei}}, \bibinfo
  {author} {\bibfnamefont {C.}~\bibnamefont {Zhang}}, \bibinfo {author}
  {\bibfnamefont {W.}~\bibnamefont {Li}}, \ and\ \bibinfo {author}
  {\bibfnamefont {X.-Q.}\ \bibnamefont {Shao}},\ }\href@noop {} {\bibfield
  {journal} {\bibinfo  {journal} {arXiv preprint arXiv:2603.29349}\ } (\bibinfo
  {year} {2026})}\BibitemShut {NoStop}%
\bibitem [{\citenamefont {Zhang}\ \emph {et~al.}(2026)\citenamefont {Zhang},
  \citenamefont {Murciano}, \citenamefont {Tantivasadakarn},\ and\
  \citenamefont {Finkelstein}}]{Zhang2026}%
  \BibitemOpen
  \bibfield  {author} {\bibinfo {author} {\bibfnamefont {C.}~\bibnamefont
  {Zhang}}, \bibinfo {author} {\bibfnamefont {S.}~\bibnamefont {Murciano}},
  \bibinfo {author} {\bibfnamefont {N.}~\bibnamefont {Tantivasadakarn}}, \ and\
  \bibinfo {author} {\bibfnamefont {R.}~\bibnamefont {Finkelstein}},\
  }\href@noop {} {\bibfield  {journal} {\bibinfo  {journal} {arXiv preprint
  arXiv:2602.12909}\ } (\bibinfo {year} {2026})}\BibitemShut {NoStop}%
\bibitem [{\citenamefont {Dobrzyniecki}\ and\ \citenamefont
  {Tomza}(2023)}]{DobrzynieckiPRA23}%
  \BibitemOpen
  \bibfield  {author} {\bibinfo {author} {\bibfnamefont {J.}~\bibnamefont
  {Dobrzyniecki}}\ and\ \bibinfo {author} {\bibfnamefont {M.}~\bibnamefont
  {Tomza}},\ }\href {\doibase 10.1103/PhysRevA.108.052618} {\bibfield
  {journal} {\bibinfo  {journal} {Phys. Rev. A}\ }\textbf {\bibinfo {volume}
  {108}},\ \bibinfo {pages} {052618} (\bibinfo {year} {2023})}\BibitemShut
  {NoStop}%
\bibitem [{\citenamefont {Sarkas}\ \emph {et~al.}(1994)\citenamefont {Sarkas},
  \citenamefont {Arnold}, \citenamefont {Hendricks}, \citenamefont {Slager},\
  and\ \citenamefont {Bowen}}]{SarkasZPD94}%
  \BibitemOpen
  \bibfield  {author} {\bibinfo {author} {\bibfnamefont {H.}~\bibnamefont
  {Sarkas}}, \bibinfo {author} {\bibfnamefont {S.}~\bibnamefont {Arnold}},
  \bibinfo {author} {\bibfnamefont {J.}~\bibnamefont {Hendricks}}, \bibinfo
  {author} {\bibfnamefont {V.}~\bibnamefont {Slager}}, \ and\ \bibinfo {author}
  {\bibfnamefont {K.}~\bibnamefont {Bowen}},\ }\href {\doibase
  10.1007/BF01437139} {\bibfield  {journal} {\bibinfo  {journal} {Zeitschrift
  f{\"u}r Phys. D At. Mol. Clust.}\ }\textbf {\bibinfo {volume} {29}},\
  \bibinfo {pages} {209} (\bibinfo {year} {1994})}\BibitemShut {NoStop}%
\bibitem [{\citenamefont {McHugh}\ \emph {et~al.}(1989)\citenamefont {McHugh},
  \citenamefont {Eaton}, \citenamefont {Lee}, \citenamefont {Sarkas},
  \citenamefont {Kidder}, \citenamefont {Snodgrass}, \citenamefont {Manaa},\
  and\ \citenamefont {Bowen}}]{MchughJCP89}%
  \BibitemOpen
  \bibfield  {author} {\bibinfo {author} {\bibfnamefont {K.}~\bibnamefont
  {McHugh}}, \bibinfo {author} {\bibfnamefont {J.}~\bibnamefont {Eaton}},
  \bibinfo {author} {\bibfnamefont {G.}~\bibnamefont {Lee}}, \bibinfo {author}
  {\bibfnamefont {H.}~\bibnamefont {Sarkas}}, \bibinfo {author} {\bibfnamefont
  {L.}~\bibnamefont {Kidder}}, \bibinfo {author} {\bibfnamefont
  {J.}~\bibnamefont {Snodgrass}}, \bibinfo {author} {\bibfnamefont
  {M.}~\bibnamefont {Manaa}}, \ and\ \bibinfo {author} {\bibfnamefont
  {K.}~\bibnamefont {Bowen}},\ }\href {\doibase 10.1063/1.456861} {\bibfield
  {journal} {\bibinfo  {journal} {J. Chem. Phys.}\ }\textbf {\bibinfo {volume}
  {91}},\ \bibinfo {pages} {3792} (\bibinfo {year} {1989})}\BibitemShut
  {NoStop}%
\bibitem [{\citenamefont {Eaton}\ \emph {et~al.}(1992)\citenamefont {Eaton},
  \citenamefont {Sarkas}, \citenamefont {Arnold}, \citenamefont {McHugh},\ and\
  \citenamefont {Bowen}}]{EatonCPL92}%
  \BibitemOpen
  \bibfield  {author} {\bibinfo {author} {\bibfnamefont {J.}~\bibnamefont
  {Eaton}}, \bibinfo {author} {\bibfnamefont {H.}~\bibnamefont {Sarkas}},
  \bibinfo {author} {\bibfnamefont {S.}~\bibnamefont {Arnold}}, \bibinfo
  {author} {\bibfnamefont {K.}~\bibnamefont {McHugh}}, \ and\ \bibinfo {author}
  {\bibfnamefont {K.}~\bibnamefont {Bowen}},\ }\href {\doibase
  10.1016/0009-2614(92)85697-9} {\bibfield  {journal} {\bibinfo  {journal}
  {Chem. Phys. Lett.}\ }\textbf {\bibinfo {volume} {193}},\ \bibinfo {pages}
  {141} (\bibinfo {year} {1992})}\BibitemShut {NoStop}%
\bibitem [{\citenamefont {Andersen}\ and\ \citenamefont
  {Simons}(1976)}]{AndersenJCP76}%
  \BibitemOpen
  \bibfield  {author} {\bibinfo {author} {\bibfnamefont {E.}~\bibnamefont
  {Andersen}}\ and\ \bibinfo {author} {\bibfnamefont {J.}~\bibnamefont
  {Simons}},\ }\href {\doibase 10.1063/1.432086} {\bibfield  {journal}
  {\bibinfo  {journal} {J. Chem. Phys.}\ }\textbf {\bibinfo {volume} {64}},\
  \bibinfo {pages} {4548} (\bibinfo {year} {1976})}\BibitemShut {NoStop}%
\bibitem [{\citenamefont {Dixon}\ \emph {et~al.}(1977)\citenamefont {Dixon},
  \citenamefont {Gole},\ and\ \citenamefont {Jordan}}]{DixonJCP77}%
  \BibitemOpen
  \bibfield  {author} {\bibinfo {author} {\bibfnamefont {D.~A.}\ \bibnamefont
  {Dixon}}, \bibinfo {author} {\bibfnamefont {J.~L.}\ \bibnamefont {Gole}}, \
  and\ \bibinfo {author} {\bibfnamefont {K.~D.}\ \bibnamefont {Jordan}},\
  }\href {\doibase 10.1063/1.433978} {\bibfield  {journal} {\bibinfo  {journal}
  {J. Chem. Phys.}\ }\textbf {\bibinfo {volume} {66}},\ \bibinfo {pages} {567}
  (\bibinfo {year} {1977})}\BibitemShut {NoStop}%
\bibitem [{\citenamefont {Shepard}\ \emph {et~al.}(1978)\citenamefont
  {Shepard}, \citenamefont {Jordan},\ and\ \citenamefont
  {Simons}}]{ShepardJCP78}%
  \BibitemOpen
  \bibfield  {author} {\bibinfo {author} {\bibfnamefont {R.}~\bibnamefont
  {Shepard}}, \bibinfo {author} {\bibfnamefont {K.~D.}\ \bibnamefont {Jordan}},
  \ and\ \bibinfo {author} {\bibfnamefont {J.}~\bibnamefont {Simons}},\ }\href
  {\doibase 10.1063/1.436718} {\bibfield  {journal} {\bibinfo  {journal} {J.
  Chem. Phys.}\ }\textbf {\bibinfo {volume} {69}},\ \bibinfo {pages} {1788}
  (\bibinfo {year} {1978})}\BibitemShut {NoStop}%
\bibitem [{\citenamefont {Sunil}\ and\ \citenamefont
  {Jordan}(1984)}]{SunilCPL84}%
  \BibitemOpen
  \bibfield  {author} {\bibinfo {author} {\bibfnamefont {K.}~\bibnamefont
  {Sunil}}\ and\ \bibinfo {author} {\bibfnamefont {K.}~\bibnamefont {Jordan}},\
  }\href {\doibase 10.1016/0009-2614(84)80076-7} {\bibfield  {journal}
  {\bibinfo  {journal} {Chem. Phys. Lett.}\ }\textbf {\bibinfo {volume}
  {104}},\ \bibinfo {pages} {343} (\bibinfo {year} {1984})}\BibitemShut
  {NoStop}%
\bibitem [{\citenamefont {Konowalow}\ and\ \citenamefont
  {Fish}(1984)}]{KonowalowCPL84}%
  \BibitemOpen
  \bibfield  {author} {\bibinfo {author} {\bibfnamefont {D.~D.}\ \bibnamefont
  {Konowalow}}\ and\ \bibinfo {author} {\bibfnamefont {J.~L.}\ \bibnamefont
  {Fish}},\ }\href {\doibase 10.1016/0009-2614(84)80197-9} {\bibfield
  {journal} {\bibinfo  {journal} {Chem. Phys. Lett.}\ }\textbf {\bibinfo
  {volume} {104}},\ \bibinfo {pages} {210} (\bibinfo {year}
  {1984})}\BibitemShut {NoStop}%
\bibitem [{\citenamefont {Michels}\ \emph {et~al.}(1985)\citenamefont
  {Michels}, \citenamefont {Hobbs},\ and\ \citenamefont
  {Wright}}]{MichelsCPL85}%
  \BibitemOpen
  \bibfield  {author} {\bibinfo {author} {\bibfnamefont {H.}~\bibnamefont
  {Michels}}, \bibinfo {author} {\bibfnamefont {R.}~\bibnamefont {Hobbs}}, \
  and\ \bibinfo {author} {\bibfnamefont {L.}~\bibnamefont {Wright}},\ }\href
  {\doibase 10.1016/0009-2614(85)85268-4} {\bibfield  {journal} {\bibinfo
  {journal} {Chem. Phys. Lett.}\ }\textbf {\bibinfo {volume} {118}},\ \bibinfo
  {pages} {67} (\bibinfo {year} {1985})}\BibitemShut {NoStop}%
\bibitem [{\citenamefont {Boldyrev}\ \emph {et~al.}(1993)\citenamefont
  {Boldyrev}, \citenamefont {Simons},\ and\ \citenamefont
  {Schleyer}}]{BoldyrevJCP93}%
  \BibitemOpen
  \bibfield  {author} {\bibinfo {author} {\bibfnamefont {A.~I.}\ \bibnamefont
  {Boldyrev}}, \bibinfo {author} {\bibfnamefont {J.}~\bibnamefont {Simons}}, \
  and\ \bibinfo {author} {\bibfnamefont {P.~v.~R.}\ \bibnamefont {Schleyer}},\
  }\href {\doibase 10.1063/1.465600} {\bibfield  {journal} {\bibinfo  {journal}
  {J. Chem. Phys.}\ }\textbf {\bibinfo {volume} {99}},\ \bibinfo {pages} {8793}
  (\bibinfo {year} {1993})}\BibitemShut {NoStop}%
\bibitem [{\citenamefont {Petch}\ \emph {et~al.}(1995)\citenamefont {Petch},
  \citenamefont {Cooper}, \citenamefont {Gerratt}, \citenamefont {Karadakov},\
  and\ \citenamefont {Raimondi}}]{PetchJCS95}%
  \BibitemOpen
  \bibfield  {author} {\bibinfo {author} {\bibfnamefont {B.}~\bibnamefont
  {Petch}}, \bibinfo {author} {\bibfnamefont {D.~L.}\ \bibnamefont {Cooper}},
  \bibinfo {author} {\bibfnamefont {J.}~\bibnamefont {Gerratt}}, \bibinfo
  {author} {\bibfnamefont {P.~B.}\ \bibnamefont {Karadakov}}, \ and\ \bibinfo
  {author} {\bibfnamefont {M.}~\bibnamefont {Raimondi}},\ }\href {\doibase
  10.1039/FT9959103751} {\bibfield  {journal} {\bibinfo  {journal} {J. Chem.
  Soc.{,} Faraday Trans.}\ }\textbf {\bibinfo {volume} {91}},\ \bibinfo {pages}
  {3751} (\bibinfo {year} {1995})}\BibitemShut {NoStop}%
\bibitem [{\citenamefont {Hogreve}(2000)}]{HogreveEPJD00}%
  \BibitemOpen
  \bibfield  {author} {\bibinfo {author} {\bibfnamefont {H.}~\bibnamefont
  {Hogreve}},\ }\href {\doibase 10.1007/s100530050011} {\bibfield  {journal}
  {\bibinfo  {journal} {Eur. Phys. J. D}\ }\textbf {\bibinfo {volume} {8}},\
  \bibinfo {pages} {85} (\bibinfo {year} {2000})}\BibitemShut {NoStop}%
\bibitem [{\citenamefont {Nasiri}\ and\ \citenamefont
  {Zahedi}(2017)}]{NasiriCTC17}%
  \BibitemOpen
  \bibfield  {author} {\bibinfo {author} {\bibfnamefont {S.}~\bibnamefont
  {Nasiri}}\ and\ \bibinfo {author} {\bibfnamefont {M.}~\bibnamefont
  {Zahedi}},\ }\href {\doibase 10.1016/j.comptc.2017.05.015} {\bibfield
  {journal} {\bibinfo  {journal} {Comput. Theor. Chem.}\ }\textbf {\bibinfo
  {volume} {1114}},\ \bibinfo {pages} {106} (\bibinfo {year}
  {2017})}\BibitemShut {NoStop}%
\bibitem [{\citenamefont {Dunning~Jr}\ and\ \citenamefont
  {Xu}(2024)}]{DunningJCC24}%
  \BibitemOpen
  \bibfield  {author} {\bibinfo {author} {\bibfnamefont {T.~H.}\ \bibnamefont
  {Dunning~Jr}}\ and\ \bibinfo {author} {\bibfnamefont {L.~T.}\ \bibnamefont
  {Xu}},\ }\href {\doibase 10.1002/jcc.27246} {\bibfield  {journal} {\bibinfo
  {journal} {J. Comput. Chem.}\ }\textbf {\bibinfo {volume} {45}},\ \bibinfo
  {pages} {405} (\bibinfo {year} {2024})}\BibitemShut {NoStop}%
\bibitem [{\citenamefont {Partridge}\ \emph {et~al.}(1983)\citenamefont
  {Partridge}, \citenamefont {Dixon}, \citenamefont {Walch}, \citenamefont
  {Bauschlicher~Jr.},\ and\ \citenamefont {Gole}}]{PartridgeJCP83}%
  \BibitemOpen
  \bibfield  {author} {\bibinfo {author} {\bibfnamefont {H.}~\bibnamefont
  {Partridge}}, \bibinfo {author} {\bibfnamefont {D.~A.}\ \bibnamefont
  {Dixon}}, \bibinfo {author} {\bibfnamefont {S.~P.}\ \bibnamefont {Walch}},
  \bibinfo {author} {\bibfnamefont {C.~W.}\ \bibnamefont {Bauschlicher~Jr.}}, \
  and\ \bibinfo {author} {\bibfnamefont {J.~L.}\ \bibnamefont {Gole}},\ }\href
  {\doibase 10.1063/1.445962} {\bibfield  {journal} {\bibinfo  {journal} {J.
  Chem. Phys.}\ }\textbf {\bibinfo {volume} {79}},\ \bibinfo {pages} {1859}
  (\bibinfo {year} {1983})}\BibitemShut {NoStop}%
\bibitem [{\citenamefont {Krauss}\ and\ \citenamefont
  {Stevens}(1990)}]{KraussJCP90}%
  \BibitemOpen
  \bibfield  {author} {\bibinfo {author} {\bibfnamefont {M.}~\bibnamefont
  {Krauss}}\ and\ \bibinfo {author} {\bibfnamefont {W.}~\bibnamefont
  {Stevens}},\ }\href {\doibase 10.1063/1.458756} {\bibfield  {journal}
  {\bibinfo  {journal} {J. Chem. Phys.}\ }\textbf {\bibinfo {volume} {93}},\
  \bibinfo {pages} {4236} (\bibinfo {year} {1990})}\BibitemShut {NoStop}%
\bibitem [{\citenamefont {Bauschlicher~Jr.}\ \emph {et~al.}(1992)\citenamefont
  {Bauschlicher~Jr.}, \citenamefont {Langhoff},\ and\ \citenamefont
  {Partridge}}]{BauschlicherJCP92}%
  \BibitemOpen
  \bibfield  {author} {\bibinfo {author} {\bibfnamefont {C.~W.}\ \bibnamefont
  {Bauschlicher~Jr.}}, \bibinfo {author} {\bibfnamefont {S.~R.}\ \bibnamefont
  {Langhoff}}, \ and\ \bibinfo {author} {\bibfnamefont {H.}~\bibnamefont
  {Partridge}},\ }\href {\doibase 10.1063/1.462160} {\bibfield  {journal}
  {\bibinfo  {journal} {J. Chem. Phys.}\ }\textbf {\bibinfo {volume} {96}},\
  \bibinfo {pages} {1240} (\bibinfo {year} {1992})}\BibitemShut {NoStop}%
\bibitem [{\citenamefont {Moussa}\ \emph {et~al.}(2021)\citenamefont {Moussa},
  \citenamefont {El-Kork},\ and\ \citenamefont {Korek}}]{MoussaNJP21}%
  \BibitemOpen
  \bibfield  {author} {\bibinfo {author} {\bibfnamefont {A.}~\bibnamefont
  {Moussa}}, \bibinfo {author} {\bibfnamefont {N.}~\bibnamefont {El-Kork}}, \
  and\ \bibinfo {author} {\bibfnamefont {M.}~\bibnamefont {Korek}},\ }\href
  {\doibase 10.1088/1367-2630/abd50d} {\bibfield  {journal} {\bibinfo
  {journal} {New J. Phys.}\ }\textbf {\bibinfo {volume} {23}},\ \bibinfo
  {pages} {013017} (\bibinfo {year} {2021})}\BibitemShut {NoStop}%
\bibitem [{\citenamefont {Gronowski}\ \emph {et~al.}(2020)\citenamefont
  {Gronowski}, \citenamefont {Koza},\ and\ \citenamefont
  {Tomza}}]{GronowskiPRA20}%
  \BibitemOpen
  \bibfield  {author} {\bibinfo {author} {\bibfnamefont {M.}~\bibnamefont
  {Gronowski}}, \bibinfo {author} {\bibfnamefont {A.~M.}\ \bibnamefont {Koza}},
  \ and\ \bibinfo {author} {\bibfnamefont {M.}~\bibnamefont {Tomza}},\ }\href
  {\doibase 10.1103/PhysRevA.102.020801} {\bibfield  {journal} {\bibinfo
  {journal} {Phys. Rev. A}\ }\textbf {\bibinfo {volume} {102}},\ \bibinfo
  {pages} {020801} (\bibinfo {year} {2020})}\BibitemShut {NoStop}%
\bibitem [{\citenamefont {Karman}\ \emph {et~al.}(2023)\citenamefont {Karman},
  \citenamefont {Gronowski}, \citenamefont {Tomza}, \citenamefont {Park},
  \citenamefont {Son}, \citenamefont {Lu}, \citenamefont {Jamison},\ and\
  \citenamefont {Ketterle}}]{KarmanPRA23}%
  \BibitemOpen
  \bibfield  {author} {\bibinfo {author} {\bibfnamefont {T.}~\bibnamefont
  {Karman}}, \bibinfo {author} {\bibfnamefont {M.}~\bibnamefont {Gronowski}},
  \bibinfo {author} {\bibfnamefont {M.}~\bibnamefont {Tomza}}, \bibinfo
  {author} {\bibfnamefont {J.~J.}\ \bibnamefont {Park}}, \bibinfo {author}
  {\bibfnamefont {H.}~\bibnamefont {Son}}, \bibinfo {author} {\bibfnamefont
  {Y.-K.}\ \bibnamefont {Lu}}, \bibinfo {author} {\bibfnamefont {A.~O.}\
  \bibnamefont {Jamison}}, \ and\ \bibinfo {author} {\bibfnamefont
  {W.}~\bibnamefont {Ketterle}},\ }\href {\doibase 10.1103/PhysRevA.108.023309}
  {\bibfield  {journal} {\bibinfo  {journal} {Phys. Rev. A}\ }\textbf {\bibinfo
  {volume} {108}},\ \bibinfo {pages} {023309} (\bibinfo {year}
  {2023})}\BibitemShut {NoStop}%
\bibitem [{\citenamefont {Ladjimi}\ and\ \citenamefont
  {Tomza}(2023)}]{LadjimiPRA23}%
  \BibitemOpen
  \bibfield  {author} {\bibinfo {author} {\bibfnamefont {H.}~\bibnamefont
  {Ladjimi}}\ and\ \bibinfo {author} {\bibfnamefont {M.}~\bibnamefont
  {Tomza}},\ }\href {\doibase 10.1103/PhysRevA.108.L021302} {\bibfield
  {journal} {\bibinfo  {journal} {Phys. Rev. A}\ }\textbf {\bibinfo {volume}
  {108}},\ \bibinfo {pages} {L021302} (\bibinfo {year} {2023})}\BibitemShut
  {NoStop}%
\bibitem [{\citenamefont {Ladjimi}\ and\ \citenamefont
  {Tomza}(2024)}]{LadjimiPRA24}%
  \BibitemOpen
  \bibfield  {author} {\bibinfo {author} {\bibfnamefont {H.}~\bibnamefont
  {Ladjimi}}\ and\ \bibinfo {author} {\bibfnamefont {M.}~\bibnamefont
  {Tomza}},\ }\href {\doibase 10.1103/PhysRevA.109.052814} {\bibfield
  {journal} {\bibinfo  {journal} {Phys. Rev. A}\ }\textbf {\bibinfo {volume}
  {109}},\ \bibinfo {pages} {052814} (\bibinfo {year} {2024})}\BibitemShut
  {NoStop}%
\bibitem [{\citenamefont {Bartlett}\ and\ \citenamefont
  {Musia{\l}}(2007)}]{BartlettRMP07}%
  \BibitemOpen
  \bibfield  {author} {\bibinfo {author} {\bibfnamefont {R.~J.}\ \bibnamefont
  {Bartlett}}\ and\ \bibinfo {author} {\bibfnamefont {M.}~\bibnamefont
  {Musia{\l}}},\ }\href {\doibase 10.1103/RevModPhys.79.291} {\bibfield
  {journal} {\bibinfo  {journal} {Rev. Mod. Phys.}\ }\textbf {\bibinfo {volume}
  {79}},\ \bibinfo {pages} {291} (\bibinfo {year} {2007})}\BibitemShut
  {NoStop}%
\bibitem [{\citenamefont {Prascher}\ \emph {et~al.}(2011)\citenamefont
  {Prascher}, \citenamefont {Woon}, \citenamefont {Peterson}, \citenamefont
  {Dunning},\ and\ \citenamefont {Wilson}}]{PrascherTCA11}%
  \BibitemOpen
  \bibfield  {author} {\bibinfo {author} {\bibfnamefont {B.~P.}\ \bibnamefont
  {Prascher}}, \bibinfo {author} {\bibfnamefont {D.~E.}\ \bibnamefont {Woon}},
  \bibinfo {author} {\bibfnamefont {K.~A.}\ \bibnamefont {Peterson}}, \bibinfo
  {author} {\bibfnamefont {T.~H.}\ \bibnamefont {Dunning}}, \ and\ \bibinfo
  {author} {\bibfnamefont {A.~K.}\ \bibnamefont {Wilson}},\ }\href {\doibase
  10.1007/s00214-010-0764-0} {\bibfield  {journal} {\bibinfo  {journal} {Theor.
  Chem. Acc.}\ }\textbf {\bibinfo {volume} {128}},\ \bibinfo {pages} {69}
  (\bibinfo {year} {2011})}\BibitemShut {NoStop}%
\bibitem [{\citenamefont {Hill}\ and\ \citenamefont
  {Peterson}(2017)}]{HillJCP17}%
  \BibitemOpen
  \bibfield  {author} {\bibinfo {author} {\bibfnamefont {J.~G.}\ \bibnamefont
  {Hill}}\ and\ \bibinfo {author} {\bibfnamefont {K.~A.}\ \bibnamefont
  {Peterson}},\ }\href {\doibase 10.1063/1.5010587} {\bibfield  {journal}
  {\bibinfo  {journal} {J. Chem. Phys.}\ }\textbf {\bibinfo {volume} {147}},\
  \bibinfo {pages} {244106} (\bibinfo {year} {2017})}\BibitemShut {NoStop}%
\bibitem [{Bon(g 02)}]{Bond2024}%
  \BibitemOpen
  \href@noop {} {\  (\bibinfo {year} {$s$: 0.6, 0.2, 0.067, 0.02, $p$: 0.6,
  0.2, 0.067, 0.02, $d$: 0.4, 0.13, 0.04, $f$: 0.4, 0.13, 0.04, $g$:
  0.2})}\BibitemShut {NoStop}%
\bibitem [{\citenamefont {Tao}\ and\ \citenamefont {Pan}(1992)}]{TaoJCP92}%
  \BibitemOpen
  \bibfield  {author} {\bibinfo {author} {\bibfnamefont {F.-M.}\ \bibnamefont
  {Tao}}\ and\ \bibinfo {author} {\bibfnamefont {Y.-K.}\ \bibnamefont {Pan}},\
  }\href {\doibase 10.1063/1.463852} {\bibfield  {journal} {\bibinfo  {journal}
  {J. Chem. Phys.}\ }\textbf {\bibinfo {volume} {97}},\ \bibinfo {pages} {4989}
  (\bibinfo {year} {1992})}\BibitemShut {NoStop}%
\bibitem [{\citenamefont {Boys}\ and\ \citenamefont
  {Bernardi}(1970)}]{BoysMP70}%
  \BibitemOpen
  \bibfield  {author} {\bibinfo {author} {\bibfnamefont {S.~F.}\ \bibnamefont
  {Boys}}\ and\ \bibinfo {author} {\bibfnamefont {F.}~\bibnamefont
  {Bernardi}},\ }\href {\doibase 10.1080/00268977000101561} {\bibfield
  {journal} {\bibinfo  {journal} {Mol. Phys.}\ }\textbf {\bibinfo {volume}
  {19}},\ \bibinfo {pages} {553} (\bibinfo {year} {1970})}\BibitemShut
  {NoStop}%
\bibitem [{\citenamefont {Dolg}\ and\ \citenamefont {Cao}(2012)}]{DolgCR12}%
  \BibitemOpen
  \bibfield  {author} {\bibinfo {author} {\bibfnamefont {M.}~\bibnamefont
  {Dolg}}\ and\ \bibinfo {author} {\bibfnamefont {X.}~\bibnamefont {Cao}},\
  }\href {\doibase 10.1021/cr2001383} {\bibfield  {journal} {\bibinfo
  {journal} {Chem. Rev.}\ }\textbf {\bibinfo {volume} {112}},\ \bibinfo {pages}
  {403} (\bibinfo {year} {2012})}\BibitemShut {NoStop}%
\bibitem [{\citenamefont {Lim}\ \emph {et~al.}(2005)\citenamefont {Lim},
  \citenamefont {Schwerdtfeger}, \citenamefont {S{\"o}hnel},\ and\
  \citenamefont {Stoll}}]{LimJCP05}%
  \BibitemOpen
  \bibfield  {author} {\bibinfo {author} {\bibfnamefont {I.~S.}\ \bibnamefont
  {Lim}}, \bibinfo {author} {\bibfnamefont {P.}~\bibnamefont {Schwerdtfeger}},
  \bibinfo {author} {\bibfnamefont {T.}~\bibnamefont {S{\"o}hnel}}, \ and\
  \bibinfo {author} {\bibfnamefont {H.}~\bibnamefont {Stoll}},\ }\href
  {\doibase 10.1063/1.1869979} {\bibfield  {journal} {\bibinfo  {journal} {J.
  Chem. Phys.}\ }\textbf {\bibinfo {volume} {122}},\ \bibinfo {pages} {034107}
  (\bibinfo {year} {2005})}\BibitemShut {NoStop}%
\bibitem [{\citenamefont {Lim}\ \emph {et~al.}(2006)\citenamefont {Lim},
  \citenamefont {Stoll},\ and\ \citenamefont {Schwerdtfeger}}]{LimJCP06}%
  \BibitemOpen
  \bibfield  {author} {\bibinfo {author} {\bibfnamefont {I.~S.}\ \bibnamefont
  {Lim}}, \bibinfo {author} {\bibfnamefont {H.}~\bibnamefont {Stoll}}, \ and\
  \bibinfo {author} {\bibfnamefont {P.}~\bibnamefont {Schwerdtfeger}},\ }\href
  {\doibase 10.1063/1.2148945} {\bibfield  {journal} {\bibinfo  {journal} {J.
  Chem. Phys.}\ }\textbf {\bibinfo {volume} {124}},\ \bibinfo {pages} {034107}
  (\bibinfo {year} {2006})}\BibitemShut {NoStop}%
\bibitem [{\citenamefont {Helgaker}\ \emph {et~al.}(1997)\citenamefont
  {Helgaker}, \citenamefont {Klopper}, \citenamefont {Koch},\ and\
  \citenamefont {Noga}}]{HelgakerJCP97}%
  \BibitemOpen
  \bibfield  {author} {\bibinfo {author} {\bibfnamefont {T.}~\bibnamefont
  {Helgaker}}, \bibinfo {author} {\bibfnamefont {W.}~\bibnamefont {Klopper}},
  \bibinfo {author} {\bibfnamefont {H.}~\bibnamefont {Koch}}, \ and\ \bibinfo
  {author} {\bibfnamefont {J.}~\bibnamefont {Noga}},\ }\href {\doibase
  10.1063/1.473863} {\bibfield  {journal} {\bibinfo  {journal} {J. Chem.
  Phys.}\ }\textbf {\bibinfo {volume} {106}},\ \bibinfo {pages} {9639}
  (\bibinfo {year} {1997})}\BibitemShut {NoStop}%
\bibitem [{\citenamefont {Helgaker}\ \emph {et~al.}(2000)\citenamefont
  {Helgaker}, \citenamefont {Jorgensen},\ and\ \citenamefont
  {Olsen}}]{Helgaker00}%
  \BibitemOpen
  \bibinfo {editor} {\bibfnamefont {T.}~\bibnamefont {Helgaker}}, \bibinfo
  {editor} {\bibfnamefont {P.}~\bibnamefont {Jorgensen}}, \ and\ \bibinfo
  {editor} {\bibfnamefont {J.}~\bibnamefont {Olsen}},\ eds.,\ \href@noop {}
  {\emph {\bibinfo {title} {{Molecular Electronic-Structure Theory}}}}\
  (\bibinfo  {publisher} {Wiley},\ \bibinfo {year} {2000})\BibitemShut
  {NoStop}%
\bibitem [{\citenamefont {Nooijen}\ and\ \citenamefont
  {Bartlett}(1995)}]{NooijenJCP95}%
  \BibitemOpen
  \bibfield  {author} {\bibinfo {author} {\bibfnamefont {M.}~\bibnamefont
  {Nooijen}}\ and\ \bibinfo {author} {\bibfnamefont {R.~J.}\ \bibnamefont
  {Bartlett}},\ }\href {\doibase 10.1063/1.468592} {\bibfield  {journal}
  {\bibinfo  {journal} {J. Chem. Phys.}\ }\textbf {\bibinfo {volume} {102}},\
  \bibinfo {pages} {3629} (\bibinfo {year} {1995})}\BibitemShut {NoStop}%
\bibitem [{\citenamefont {Kaufmann}\ \emph {et~al.}(1989)\citenamefont
  {Kaufmann}, \citenamefont {Baumeister},\ and\ \citenamefont
  {Jungen}}]{KaufmannJPB89}%
  \BibitemOpen
  \bibfield  {author} {\bibinfo {author} {\bibfnamefont {K.}~\bibnamefont
  {Kaufmann}}, \bibinfo {author} {\bibfnamefont {W.}~\bibnamefont
  {Baumeister}}, \ and\ \bibinfo {author} {\bibfnamefont {M.}~\bibnamefont
  {Jungen}},\ }\href {\doibase 10.1088/0953-4075/22/14/007} {\bibfield
  {journal} {\bibinfo  {journal} {J. Phys. B: At. Mol. Opt. Phys.}\ }\textbf
  {\bibinfo {volume} {22}},\ \bibinfo {pages} {2223} (\bibinfo {year}
  {1989})}\BibitemShut {NoStop}%
\bibitem [{SM()}]{SM}%
  \BibitemOpen
  \href@noop {} {}\bibinfo {note} {See Supplemental Material at [URL will be
  inserted by publisher] for the calculated potential energy curves, developed
  basis sets, and electron binding energies of the dipole-bound states in a
  numerical form, as well as the spectroscopic parameters at the CCSD(T)
  level.}\BibitemShut {Stop}%
\bibitem [{\citenamefont {Matthews}\ \emph {et~al.}(2020)\citenamefont
  {Matthews}, \citenamefont {Cheng}, \citenamefont {Harding}, \citenamefont
  {Lipparini}, \citenamefont {Stopkowicz}, \citenamefont {Jagau}, \citenamefont
  {Szalay}, \citenamefont {Gauss},\ and\ \citenamefont
  {Stanton}}]{MatthewsJCP20}%
  \BibitemOpen
  \bibfield  {author} {\bibinfo {author} {\bibfnamefont {D.~A.}\ \bibnamefont
  {Matthews}}, \bibinfo {author} {\bibfnamefont {L.}~\bibnamefont {Cheng}},
  \bibinfo {author} {\bibfnamefont {M.~E.}\ \bibnamefont {Harding}}, \bibinfo
  {author} {\bibfnamefont {F.}~\bibnamefont {Lipparini}}, \bibinfo {author}
  {\bibfnamefont {S.}~\bibnamefont {Stopkowicz}}, \bibinfo {author}
  {\bibfnamefont {T.-C.}\ \bibnamefont {Jagau}}, \bibinfo {author}
  {\bibfnamefont {P.~G.}\ \bibnamefont {Szalay}}, \bibinfo {author}
  {\bibfnamefont {J.}~\bibnamefont {Gauss}}, \ and\ \bibinfo {author}
  {\bibfnamefont {J.~F.}\ \bibnamefont {Stanton}},\ }\href {\doibase
  10.1063/5.0004837} {\bibfield  {journal} {\bibinfo  {journal} {J. Chem.
  Phys.}\ }\textbf {\bibinfo {volume} {152}},\ \bibinfo {pages} {041601}
  (\bibinfo {year} {2020})}\BibitemShut {NoStop}%
\bibitem [{\citenamefont {K{\'a}llay}\ \emph {et~al.}(2020)\citenamefont
  {K{\'a}llay}, \citenamefont {Nagy}, \citenamefont {Mester}, \citenamefont
  {Rolik}, \citenamefont {Samu}, \citenamefont {Csontos}, \citenamefont
  {Cs{\'o}ka}, \citenamefont {Szab{\'o}}, \citenamefont {Gyevi-Nagy},
  \citenamefont {H{\'e}gely} \emph {et~al.}}]{KallayJCP20}%
  \BibitemOpen
  \bibfield  {author} {\bibinfo {author} {\bibfnamefont {M.}~\bibnamefont
  {K{\'a}llay}}, \bibinfo {author} {\bibfnamefont {P.~R.}\ \bibnamefont
  {Nagy}}, \bibinfo {author} {\bibfnamefont {D.}~\bibnamefont {Mester}},
  \bibinfo {author} {\bibfnamefont {Z.}~\bibnamefont {Rolik}}, \bibinfo
  {author} {\bibfnamefont {G.}~\bibnamefont {Samu}}, \bibinfo {author}
  {\bibfnamefont {J.}~\bibnamefont {Csontos}}, \bibinfo {author} {\bibfnamefont
  {J.}~\bibnamefont {Cs{\'o}ka}}, \bibinfo {author} {\bibfnamefont {P.~B.}\
  \bibnamefont {Szab{\'o}}}, \bibinfo {author} {\bibfnamefont {L.}~\bibnamefont
  {Gyevi-Nagy}}, \bibinfo {author} {\bibfnamefont {B.}~\bibnamefont
  {H{\'e}gely}},  \emph {et~al.},\ }\href {\doibase 10.1063/1.5142048}
  {\bibfield  {journal} {\bibinfo  {journal} {J. Chem. Phys.}\ }\textbf
  {\bibinfo {volume} {152}},\ \bibinfo {pages} {074107} (\bibinfo {year}
  {2020})}\BibitemShut {NoStop}%
\bibitem [{\citenamefont {Werner}\ \emph {et~al.}()\citenamefont {Werner},
  \citenamefont {Knowles} \emph {et~al.}}]{MOLPRO_brief}%
  \BibitemOpen
  \bibfield  {author} {\bibinfo {author} {\bibfnamefont {H.-J.}\ \bibnamefont
  {Werner}}, \bibinfo {author} {\bibfnamefont {P.~J.}\ \bibnamefont {Knowles}},
   \emph {et~al.},\ }\href@noop {} {\enquote {\bibinfo {title} {Molpro, version
  2019.2, a package of ab initio programs},}\ }\bibinfo {note} {See
  https://www.molpro.net}\BibitemShut {NoStop}%
\bibitem [{\citenamefont {Werner}\ \emph {et~al.}(2012)\citenamefont {Werner},
  \citenamefont {Knowles}, \citenamefont {Knizia}, \citenamefont {Manby},\ and\
  \citenamefont {Sch{\"u}tz}}]{WernerWIRCMS12}%
  \BibitemOpen
  \bibfield  {author} {\bibinfo {author} {\bibfnamefont {H.-J.}\ \bibnamefont
  {Werner}}, \bibinfo {author} {\bibfnamefont {P.~J.}\ \bibnamefont {Knowles}},
  \bibinfo {author} {\bibfnamefont {G.}~\bibnamefont {Knizia}}, \bibinfo
  {author} {\bibfnamefont {F.~R.}\ \bibnamefont {Manby}}, \ and\ \bibinfo
  {author} {\bibfnamefont {M.}~\bibnamefont {Sch{\"u}tz}},\ }\href {\doibase
  10.1002/wcms.82} {\bibfield  {journal} {\bibinfo  {journal} {WIREs Comput.
  Mol. Sci.}\ }\textbf {\bibinfo {volume} {2}},\ \bibinfo {pages} {242}
  (\bibinfo {year} {2012})}\BibitemShut {NoStop}%
\bibitem [{\citenamefont {Kramida}\ \emph {et~al.}(2022)\citenamefont
  {Kramida}, \citenamefont {{Yu.~Ralchenko}}, \citenamefont {Reader},\ and\
  \citenamefont {{NIST ASD Team}}}]{NIST_ASD}%
  \BibitemOpen
  \bibfield  {author} {\bibinfo {author} {\bibfnamefont {A.}~\bibnamefont
  {Kramida}}, \bibinfo {author} {\bibnamefont {{Yu.~Ralchenko}}}, \bibinfo
  {author} {\bibfnamefont {J.}~\bibnamefont {Reader}}, \ and\ \bibinfo {author}
  {\bibnamefont {{NIST ASD Team}}},\ }\href {\doibase
  www.nist.gov/pml/atomic-spectra-database} {}\bibinfo {howpublished} {{NIST
  Atomic Spectra Database (ver. 5.12), [Online]. Available:
  {\tt{physics.nist.gov/asd}} [2022, January 31]. National Institute of
  Standards and Technology, Gaithersburg, MD.}} (\bibinfo {year}
  {2022})\BibitemShut {NoStop}%
\bibitem [{\citenamefont {Haeffler}\ \emph {et~al.}(1996)\citenamefont
  {Haeffler}, \citenamefont {Hanstorp}, \citenamefont {Kiyan}, \citenamefont
  {Klinkm{\"u}ller}, \citenamefont {Ljungblad},\ and\ \citenamefont
  {Pegg}}]{HaefflerPRA96}%
  \BibitemOpen
  \bibfield  {author} {\bibinfo {author} {\bibfnamefont {G.}~\bibnamefont
  {Haeffler}}, \bibinfo {author} {\bibfnamefont {D.}~\bibnamefont {Hanstorp}},
  \bibinfo {author} {\bibfnamefont {I.}~\bibnamefont {Kiyan}}, \bibinfo
  {author} {\bibfnamefont {A.~E.}\ \bibnamefont {Klinkm{\"u}ller}}, \bibinfo
  {author} {\bibfnamefont {U.}~\bibnamefont {Ljungblad}}, \ and\ \bibinfo
  {author} {\bibfnamefont {D.~J.}\ \bibnamefont {Pegg}},\ }\href {\doibase
  10.1103/PhysRevA.53.4127} {\bibfield  {journal} {\bibinfo  {journal} {Phys.
  Rev. A}\ }\textbf {\bibinfo {volume} {53}},\ \bibinfo {pages} {4127}
  (\bibinfo {year} {1996})}\BibitemShut {NoStop}%
\bibitem [{\citenamefont {Miffre}\ \emph {et~al.}(2006)\citenamefont {Miffre},
  \citenamefont {Jacquey}, \citenamefont {B{\"u}chner}, \citenamefont
  {Tr{\'e}nec},\ and\ \citenamefont {Vigu{\'e}}}]{MiffreEPJD06}%
  \BibitemOpen
  \bibfield  {author} {\bibinfo {author} {\bibfnamefont {A.}~\bibnamefont
  {Miffre}}, \bibinfo {author} {\bibfnamefont {M.}~\bibnamefont {Jacquey}},
  \bibinfo {author} {\bibfnamefont {M.}~\bibnamefont {B{\"u}chner}}, \bibinfo
  {author} {\bibfnamefont {G.}~\bibnamefont {Tr{\'e}nec}}, \ and\ \bibinfo
  {author} {\bibfnamefont {J.}~\bibnamefont {Vigu{\'e}}},\ }\href {\doibase
  10.1140/epjd/e2006-00015-5} {\bibfield  {journal} {\bibinfo  {journal} {Eur.
  Phys. J. D}\ }\textbf {\bibinfo {volume} {38}},\ \bibinfo {pages} {353}
  (\bibinfo {year} {2006})}\BibitemShut {NoStop}%
\bibitem [{\citenamefont {McNeill}\ \emph {et~al.}(2020)\citenamefont
  {McNeill}, \citenamefont {Peterson},\ and\ \citenamefont
  {Dixon}}]{McneillJCP20}%
  \BibitemOpen
  \bibfield  {author} {\bibinfo {author} {\bibfnamefont {A.~S.}\ \bibnamefont
  {McNeill}}, \bibinfo {author} {\bibfnamefont {K.~A.}\ \bibnamefont
  {Peterson}}, \ and\ \bibinfo {author} {\bibfnamefont {D.~A.}\ \bibnamefont
  {Dixon}},\ }\href {\doibase 10.1063/5.0026876} {\bibfield  {journal}
  {\bibinfo  {journal} {J. Chem. Phys.}\ }\textbf {\bibinfo {volume} {153}},\
  \bibinfo {pages} {174304} (\bibinfo {year} {2020})}\BibitemShut {NoStop}%
\bibitem [{\citenamefont {Sahoo}(2020)}]{SahooPRA20}%
  \BibitemOpen
  \bibfield  {author} {\bibinfo {author} {\bibfnamefont {B.}~\bibnamefont
  {Sahoo}},\ }\href {\doibase 10.1103/PhysRevA.102.022820} {\bibfield
  {journal} {\bibinfo  {journal} {Phys. Rev. A}\ }\textbf {\bibinfo {volume}
  {102}},\ \bibinfo {pages} {022820} (\bibinfo {year} {2020})}\BibitemShut
  {NoStop}%
\bibitem [{\citenamefont {Andersen}\ \emph {et~al.}(1999)\citenamefont
  {Andersen}, \citenamefont {Haugen},\ and\ \citenamefont
  {Hotop}}]{AndersenJPC99}%
  \BibitemOpen
  \bibfield  {author} {\bibinfo {author} {\bibfnamefont {T.}~\bibnamefont
  {Andersen}}, \bibinfo {author} {\bibfnamefont {H.}~\bibnamefont {Haugen}}, \
  and\ \bibinfo {author} {\bibfnamefont {H.}~\bibnamefont {Hotop}},\ }\href
  {\doibase 10.1063/1.556047} {\bibfield  {journal} {\bibinfo  {journal} {J.
  Phys. Chem. Ref. Data}\ }\textbf {\bibinfo {volume} {28}},\ \bibinfo {pages}
  {1511} (\bibinfo {year} {1999})}\BibitemShut {NoStop}%
\bibitem [{\citenamefont {Ekstrom}\ \emph {et~al.}(1995)\citenamefont
  {Ekstrom}, \citenamefont {Schmiedmayer}, \citenamefont {Chapman},
  \citenamefont {Hammond},\ and\ \citenamefont {Pritchard}}]{EkstromPRA95}%
  \BibitemOpen
  \bibfield  {author} {\bibinfo {author} {\bibfnamefont {C.~R.}\ \bibnamefont
  {Ekstrom}}, \bibinfo {author} {\bibfnamefont {J.}~\bibnamefont
  {Schmiedmayer}}, \bibinfo {author} {\bibfnamefont {M.~S.}\ \bibnamefont
  {Chapman}}, \bibinfo {author} {\bibfnamefont {T.~D.}\ \bibnamefont
  {Hammond}}, \ and\ \bibinfo {author} {\bibfnamefont {D.~E.}\ \bibnamefont
  {Pritchard}},\ }\href {\doibase 10.1103/PhysRevA.51.3883} {\bibfield
  {journal} {\bibinfo  {journal} {Phys. Rev. A}\ }\textbf {\bibinfo {volume}
  {51}},\ \bibinfo {pages} {3883} (\bibinfo {year} {1995})}\BibitemShut
  {NoStop}%
\bibitem [{\citenamefont {Andersson}\ \emph {et~al.}(2000)\citenamefont
  {Andersson}, \citenamefont {Sandstr{\"o}m}, \citenamefont {Kiyan},
  \citenamefont {Hanstorp},\ and\ \citenamefont {Pegg}}]{AnderssonAPS00}%
  \BibitemOpen
  \bibfield  {author} {\bibinfo {author} {\bibfnamefont {K.~T.}\ \bibnamefont
  {Andersson}}, \bibinfo {author} {\bibfnamefont {J.}~\bibnamefont
  {Sandstr{\"o}m}}, \bibinfo {author} {\bibfnamefont {I.~Y.}\ \bibnamefont
  {Kiyan}}, \bibinfo {author} {\bibfnamefont {D.}~\bibnamefont {Hanstorp}}, \
  and\ \bibinfo {author} {\bibfnamefont {D.~J.}\ \bibnamefont {Pegg}},\ }\href
  {\doibase 10.1103/PhysRevA.62.022503} {\bibfield  {journal} {\bibinfo
  {journal} {Phys. Rev. A}\ }\textbf {\bibinfo {volume} {62}},\ \bibinfo
  {pages} {022503} (\bibinfo {year} {2000})}\BibitemShut {NoStop}%
\bibitem [{\citenamefont {Gregoire}\ \emph {et~al.}(2015)\citenamefont
  {Gregoire}, \citenamefont {Hromada}, \citenamefont {Holmgren}, \citenamefont
  {Trubko},\ and\ \citenamefont {Cronin}}]{GregoirePRA15}%
  \BibitemOpen
  \bibfield  {author} {\bibinfo {author} {\bibfnamefont {M.~D.}\ \bibnamefont
  {Gregoire}}, \bibinfo {author} {\bibfnamefont {I.}~\bibnamefont {Hromada}},
  \bibinfo {author} {\bibfnamefont {W.~F.}\ \bibnamefont {Holmgren}}, \bibinfo
  {author} {\bibfnamefont {R.}~\bibnamefont {Trubko}}, \ and\ \bibinfo {author}
  {\bibfnamefont {A.~D.}\ \bibnamefont {Cronin}},\ }\href {\doibase
  10.1103/PhysRevA.92.052513} {\bibfield  {journal} {\bibinfo  {journal} {Phys.
  Rev. A}\ }\textbf {\bibinfo {volume} {92}},\ \bibinfo {pages} {052513}
  (\bibinfo {year} {2015})}\BibitemShut {NoStop}%
\bibitem [{\citenamefont {Frey}\ \emph {et~al.}(1978)\citenamefont {Frey},
  \citenamefont {Breyer},\ and\ \citenamefont {Holop}}]{FreyJPB78}%
  \BibitemOpen
  \bibfield  {author} {\bibinfo {author} {\bibfnamefont {P.}~\bibnamefont
  {Frey}}, \bibinfo {author} {\bibfnamefont {F.}~\bibnamefont {Breyer}}, \ and\
  \bibinfo {author} {\bibfnamefont {H.}~\bibnamefont {Holop}},\ }\href
  {\doibase 10.1088/0022-3700/11/19/005} {\bibfield  {journal} {\bibinfo
  {journal} {J. Phys. B: At. Mol. Phys.}\ }\textbf {\bibinfo {volume} {11}},\
  \bibinfo {pages} {L589} (\bibinfo {year} {1978})}\BibitemShut {NoStop}%
\bibitem [{\citenamefont {Landau}\ \emph {et~al.}(2001)\citenamefont {Landau},
  \citenamefont {Eliav}, \citenamefont {Ishikawa},\ and\ \citenamefont
  {Kaldor}}]{LandauJCP01}%
  \BibitemOpen
  \bibfield  {author} {\bibinfo {author} {\bibfnamefont {A.}~\bibnamefont
  {Landau}}, \bibinfo {author} {\bibfnamefont {E.}~\bibnamefont {Eliav}},
  \bibinfo {author} {\bibfnamefont {Y.}~\bibnamefont {Ishikawa}}, \ and\
  \bibinfo {author} {\bibfnamefont {U.}~\bibnamefont {Kaldor}},\ }\href
  {\doibase 10.1063/1.1386413} {\bibfield  {journal} {\bibinfo  {journal} {J.
  Chem. Phys.}\ }\textbf {\bibinfo {volume} {115}},\ \bibinfo {pages} {2389}
  (\bibinfo {year} {2001})}\BibitemShut {NoStop}%
\bibitem [{\citenamefont {Derevianko}\ \emph {et~al.}(1999)\citenamefont
  {Derevianko}, \citenamefont {Johnson}, \citenamefont {Safronova},\ and\
  \citenamefont {Babb}}]{DereviankoPRL99}%
  \BibitemOpen
  \bibfield  {author} {\bibinfo {author} {\bibfnamefont {A.}~\bibnamefont
  {Derevianko}}, \bibinfo {author} {\bibfnamefont {W.}~\bibnamefont {Johnson}},
  \bibinfo {author} {\bibfnamefont {M.}~\bibnamefont {Safronova}}, \ and\
  \bibinfo {author} {\bibfnamefont {J.}~\bibnamefont {Babb}},\ }\href {\doibase
  10.1103/PhysRevLett.82.3589} {\bibfield  {journal} {\bibinfo  {journal}
  {Phys. Rev. Lett.}\ }\textbf {\bibinfo {volume} {82}},\ \bibinfo {pages}
  {3589} (\bibinfo {year} {1999})}\BibitemShut {NoStop}%
\bibitem [{\citenamefont {Mitroy}\ and\ \citenamefont
  {Bromley}(2003)}]{MitroyPRA03}%
  \BibitemOpen
  \bibfield  {author} {\bibinfo {author} {\bibfnamefont {J.}~\bibnamefont
  {Mitroy}}\ and\ \bibinfo {author} {\bibfnamefont {M.~W.}\ \bibnamefont
  {Bromley}},\ }\href {\doibase 10.1103/PhysRevA.68.052714} {\bibfield
  {journal} {\bibinfo  {journal} {Phys. Rev. A}\ }\textbf {\bibinfo {volume}
  {68}},\ \bibinfo {pages} {052714} (\bibinfo {year} {2003})}\BibitemShut
  {NoStop}%
\bibitem [{\citenamefont {Jiang}\ \emph {et~al.}(2015)\citenamefont {Jiang},
  \citenamefont {Mitroy}, \citenamefont {Cheng},\ and\ \citenamefont
  {Bromley}}]{JiangADNDT15}%
  \BibitemOpen
  \bibfield  {author} {\bibinfo {author} {\bibfnamefont {J.}~\bibnamefont
  {Jiang}}, \bibinfo {author} {\bibfnamefont {J.}~\bibnamefont {Mitroy}},
  \bibinfo {author} {\bibfnamefont {Y.}~\bibnamefont {Cheng}}, \ and\ \bibinfo
  {author} {\bibfnamefont {M.}~\bibnamefont {Bromley}},\ }\href {\doibase
  10.1016/j.adt.2014.10.001} {\bibfield  {journal} {\bibinfo  {journal} {At.
  Data Nucl. Data Tables}\ }\textbf {\bibinfo {volume} {101}},\ \bibinfo
  {pages} {158} (\bibinfo {year} {2015})}\BibitemShut {NoStop}%
\bibitem [{\citenamefont {Porsev}\ and\ \citenamefont
  {Derevianko}(2006)}]{PorsevJOTP06}%
  \BibitemOpen
  \bibfield  {author} {\bibinfo {author} {\bibfnamefont {S.}~\bibnamefont
  {Porsev}}\ and\ \bibinfo {author} {\bibfnamefont {A.}~\bibnamefont
  {Derevianko}},\ }\href {\doibase 10.1134/S1063776106020014} {\bibfield
  {journal} {\bibinfo  {journal} {J. Exp. Theor. Phys.}\ }\textbf {\bibinfo
  {volume} {102}},\ \bibinfo {pages} {195} (\bibinfo {year}
  {2006})}\BibitemShut {NoStop}%
\bibitem [{\citenamefont {Petrunin}\ \emph {et~al.}(1996)\citenamefont
  {Petrunin}, \citenamefont {Andersen}, \citenamefont {Balling},\ and\
  \citenamefont {Andersen}}]{PetruninPRL96}%
  \BibitemOpen
  \bibfield  {author} {\bibinfo {author} {\bibfnamefont {V.}~\bibnamefont
  {Petrunin}}, \bibinfo {author} {\bibfnamefont {H.}~\bibnamefont {Andersen}},
  \bibinfo {author} {\bibfnamefont {P.}~\bibnamefont {Balling}}, \ and\
  \bibinfo {author} {\bibfnamefont {T.}~\bibnamefont {Andersen}},\ }\href
  {\doibase 10.1103/PhysRevLett.76.744} {\bibfield  {journal} {\bibinfo
  {journal} {Phys. Rev. Lett.}\ }\textbf {\bibinfo {volume} {76}},\ \bibinfo
  {pages} {744} (\bibinfo {year} {1996})}\BibitemShut {NoStop}%
\bibitem [{\citenamefont {Andersen}\ \emph {et~al.}(1997)\citenamefont
  {Andersen}, \citenamefont {Petrunin}, \citenamefont {Kristensen},\ and\
  \citenamefont {Andersen}}]{AndersenPRA97}%
  \BibitemOpen
  \bibfield  {author} {\bibinfo {author} {\bibfnamefont {H.}~\bibnamefont
  {Andersen}}, \bibinfo {author} {\bibfnamefont {V.}~\bibnamefont {Petrunin}},
  \bibinfo {author} {\bibfnamefont {P.}~\bibnamefont {Kristensen}}, \ and\
  \bibinfo {author} {\bibfnamefont {T.}~\bibnamefont {Andersen}},\ }\href
  {\doibase 10.1103/PhysRevA.55.3247} {\bibfield  {journal} {\bibinfo
  {journal} {Phys. Rev. A}\ }\textbf {\bibinfo {volume} {55}},\ \bibinfo
  {pages} {3247} (\bibinfo {year} {1997})}\BibitemShut {NoStop}%
\bibitem [{\citenamefont {Petrunin}\ \emph {et~al.}(1995)\citenamefont
  {Petrunin}, \citenamefont {Voldstad}, \citenamefont {Balling}, \citenamefont
  {Kristensen}, \citenamefont {Andersen},\ and\ \citenamefont
  {Haugen}}]{PetruninPRL95}%
  \BibitemOpen
  \bibfield  {author} {\bibinfo {author} {\bibfnamefont {V.}~\bibnamefont
  {Petrunin}}, \bibinfo {author} {\bibfnamefont {J.}~\bibnamefont {Voldstad}},
  \bibinfo {author} {\bibfnamefont {P.}~\bibnamefont {Balling}}, \bibinfo
  {author} {\bibfnamefont {P.}~\bibnamefont {Kristensen}}, \bibinfo {author}
  {\bibfnamefont {T.}~\bibnamefont {Andersen}}, \ and\ \bibinfo {author}
  {\bibfnamefont {H.}~\bibnamefont {Haugen}},\ }\href {\doibase
  10.1103/PhysRevLett.75.1911} {\bibfield  {journal} {\bibinfo  {journal}
  {Phys. Rev. Lett.}\ }\textbf {\bibinfo {volume} {75}},\ \bibinfo {pages}
  {1911} (\bibinfo {year} {1995})}\BibitemShut {NoStop}%
\bibitem [{\citenamefont {Andersen}(2004)}]{AndersenPR04}%
  \BibitemOpen
  \bibfield  {author} {\bibinfo {author} {\bibfnamefont {T.}~\bibnamefont
  {Andersen}},\ }\href {\doibase 10.1016/j.physrep.2004.01.001} {\bibfield
  {journal} {\bibinfo  {journal} {Phys. Rep.}\ }\textbf {\bibinfo {volume}
  {394}},\ \bibinfo {pages} {157} (\bibinfo {year} {2004})}\BibitemShut
  {NoStop}%
\bibitem [{\citenamefont {Lim}\ and\ \citenamefont
  {Schwerdtfeger}(2004)}]{LimPRA04}%
  \BibitemOpen
  \bibfield  {author} {\bibinfo {author} {\bibfnamefont {I.~S.}\ \bibnamefont
  {Lim}}\ and\ \bibinfo {author} {\bibfnamefont {P.}~\bibnamefont
  {Schwerdtfeger}},\ }\href {\doibase 10.1103/PhysRevA.70.062501} {\bibfield
  {journal} {\bibinfo  {journal} {Phys. Rev. A}\ }\textbf {\bibinfo {volume}
  {70}},\ \bibinfo {pages} {062501} (\bibinfo {year} {2004})}\BibitemShut
  {NoStop}%
\bibitem [{\citenamefont {Fox}\ and\ \citenamefont {Turner}(1966)}]{FoxJCP66}%
  \BibitemOpen
  \bibfield  {author} {\bibinfo {author} {\bibfnamefont {K.}~\bibnamefont
  {Fox}}\ and\ \bibinfo {author} {\bibfnamefont {J.~E.}\ \bibnamefont
  {Turner}},\ }\href {\doibase 10.1063/1.1727729} {\bibfield  {journal}
  {\bibinfo  {journal} {J. Chem. Phys.}\ }\textbf {\bibinfo {volume} {45}},\
  \bibinfo {pages} {1142} (\bibinfo {year} {1966})}\BibitemShut {NoStop}%
\bibitem [{\citenamefont {Lévy-Leblond}(1967)}]{LevyLeblondPR67}%
  \BibitemOpen
  \bibfield  {author} {\bibinfo {author} {\bibfnamefont {J.-M.}\ \bibnamefont
  {Lévy-Leblond}},\ }\href {\doibase 10.1103/PhysRev.153.1} {\bibfield
  {journal} {\bibinfo  {journal} {Phys. Rev.}\ }\textbf {\bibinfo {volume}
  {153}},\ \bibinfo {pages} {1} (\bibinfo {year} {1967})}\BibitemShut {NoStop}%
\bibitem [{\citenamefont {Brown}\ and\ \citenamefont
  {Roberts}(1967)}]{BrownJCP67}%
  \BibitemOpen
  \bibfield  {author} {\bibinfo {author} {\bibfnamefont {W.~B.}\ \bibnamefont
  {Brown}}\ and\ \bibinfo {author} {\bibfnamefont {R.~E.}\ \bibnamefont
  {Roberts}},\ }\href {\doibase 10.1063/1.1840977} {\bibfield  {journal}
  {\bibinfo  {journal} {J. Chem. Phys.}\ }\textbf {\bibinfo {volume} {46}},\
  \bibinfo {pages} {2006} (\bibinfo {year} {1967})}\BibitemShut {NoStop}%
\bibitem [{\citenamefont {Ard}\ \emph {et~al.}(2009)\citenamefont {Ard},
  \citenamefont {Garrett}, \citenamefont {Compton}, \citenamefont {Adamowicz},\
  and\ \citenamefont {Stepanian}}]{ArdCPL09}%
  \BibitemOpen
  \bibfield  {author} {\bibinfo {author} {\bibfnamefont {S.}~\bibnamefont
  {Ard}}, \bibinfo {author} {\bibfnamefont {W.}~\bibnamefont {Garrett}},
  \bibinfo {author} {\bibfnamefont {R.}~\bibnamefont {Compton}}, \bibinfo
  {author} {\bibfnamefont {L.}~\bibnamefont {Adamowicz}}, \ and\ \bibinfo
  {author} {\bibfnamefont {S.}~\bibnamefont {Stepanian}},\ }\href {\doibase
  j.cplett.2009.04.007} {\bibfield  {journal} {\bibinfo  {journal} {Chem. Phys.
  Lett.}\ }\textbf {\bibinfo {volume} {473}},\ \bibinfo {pages} {223} (\bibinfo
  {year} {2009})}\BibitemShut {NoStop}%
\bibitem [{\citenamefont {Crawford}\ and\ \citenamefont
  {Garrett}(1977)}]{CrawfordJCP77}%
  \BibitemOpen
  \bibfield  {author} {\bibinfo {author} {\bibfnamefont {O.~H.}\ \bibnamefont
  {Crawford}}\ and\ \bibinfo {author} {\bibfnamefont {W.~R.}\ \bibnamefont
  {Garrett}},\ }\href {\doibase 10.1063/1.433797} {\bibfield  {journal}
  {\bibinfo  {journal} {J. Chem. Phys.}\ }\textbf {\bibinfo {volume} {66}},\
  \bibinfo {pages} {4968} (\bibinfo {year} {1977})}\BibitemShut {NoStop}%
\bibitem [{\citenamefont {Garrett}(1970)}]{GarrettCPL70}%
  \BibitemOpen
  \bibfield  {author} {\bibinfo {author} {\bibfnamefont {W.}~\bibnamefont
  {Garrett}},\ }\href {\doibase 10.1016/0009-2614(70)80045-8} {\bibfield
  {journal} {\bibinfo  {journal} {Chem. Phys. Lett.}\ }\textbf {\bibinfo
  {volume} {5}},\ \bibinfo {pages} {393} (\bibinfo {year} {1970})}\BibitemShut
  {NoStop}%
\bibitem [{\citenamefont {Garrett}(1971)}]{GarrettPRA71}%
  \BibitemOpen
  \bibfield  {author} {\bibinfo {author} {\bibfnamefont {W.~R.}\ \bibnamefont
  {Garrett}},\ }\href {\doibase 10.1103/PhysRevA.3.961} {\bibfield  {journal}
  {\bibinfo  {journal} {Phys. Rev. A}\ }\textbf {\bibinfo {volume} {3}},\
  \bibinfo {pages} {961} (\bibinfo {year} {1971})}\BibitemShut {NoStop}%
\bibitem [{\citenamefont {Qian}\ \emph {et~al.}(2019)\citenamefont {Qian},
  \citenamefont {Zhu},\ and\ \citenamefont {Wang}}]{QianJPCL19}%
  \BibitemOpen
  \bibfield  {author} {\bibinfo {author} {\bibfnamefont {C.-H.}\ \bibnamefont
  {Qian}}, \bibinfo {author} {\bibfnamefont {G.-Z.}\ \bibnamefont {Zhu}}, \
  and\ \bibinfo {author} {\bibfnamefont {L.-S.}\ \bibnamefont {Wang}},\ }\href
  {\doibase 10.1021/acs.jpclett.9b02679} {\bibfield  {journal} {\bibinfo
  {journal} {J. Phys. Chem. Lett.}\ }\textbf {\bibinfo {volume} {10}},\
  \bibinfo {pages} {6472} (\bibinfo {year} {2019})}\BibitemShut {NoStop}%
\bibitem [{\citenamefont {Ulusoy}\ and\ \citenamefont
  {Wilson}(2019)}]{UlusoyMPTC19}%
  \BibitemOpen
  \bibfield  {author} {\bibinfo {author} {\bibfnamefont {I.~S.}\ \bibnamefont
  {Ulusoy}}\ and\ \bibinfo {author} {\bibfnamefont {A.~K.}\ \bibnamefont
  {Wilson}},\ }in\ \href {\doibase 10.1016/B978-0-12-813651-5.00002-4} {\emph
  {\bibinfo {booktitle} {Math. Phys. Theor. Chem.}}},\ \bibinfo {series and
  number} {Developments in Physical \& Theoretical Chemistry},\ \bibinfo
  {editor} {edited by\ \bibinfo {editor} {\bibfnamefont {S.}~\bibnamefont
  {Blinder}}\ and\ \bibinfo {editor} {\bibfnamefont {J.}~\bibnamefont
  {House}}}\ (\bibinfo  {publisher} {Elsevier},\ \bibinfo {year} {2019})\
  p.~\bibinfo {pages} {31}\BibitemShut {NoStop}%
\bibitem [{\citenamefont {Tao}(2001)}]{TaoIRPC01}%
  \BibitemOpen
  \bibfield  {author} {\bibinfo {author} {\bibfnamefont {F.-M.}\ \bibnamefont
  {Tao}},\ }\href {\doibase 10.1080/01442350110071957} {\bibfield  {journal}
  {\bibinfo  {journal} {Int. Rev. Phys. Chem.}\ }\textbf {\bibinfo {volume}
  {20}},\ \bibinfo {pages} {617} (\bibinfo {year} {2001})}\BibitemShut
  {NoStop}%
\bibitem [{\citenamefont {Quemener}\ and\ \citenamefont
  {Julienne}(2012)}]{QuemenerCR12}%
  \BibitemOpen
  \bibfield  {author} {\bibinfo {author} {\bibfnamefont {G.}~\bibnamefont
  {Quemener}}\ and\ \bibinfo {author} {\bibfnamefont {P.~S.}\ \bibnamefont
  {Julienne}},\ }\href {\doibase 10.1021/cr300092g} {\bibfield  {journal}
  {\bibinfo  {journal} {Chem. Rev.}\ }\textbf {\bibinfo {volume} {112}},\
  \bibinfo {pages} {4949} (\bibinfo {year} {2012})}\BibitemShut {NoStop}%
\bibitem [{\citenamefont {Lemeshko}\ \emph {et~al.}(2013)\citenamefont
  {Lemeshko}, \citenamefont {Krems}, \citenamefont {Doyle},\ and\ \citenamefont
  {Kais}}]{LemeshkoMP13}%
  \BibitemOpen
  \bibfield  {author} {\bibinfo {author} {\bibfnamefont {M.}~\bibnamefont
  {Lemeshko}}, \bibinfo {author} {\bibfnamefont {R.~V.}\ \bibnamefont {Krems}},
  \bibinfo {author} {\bibfnamefont {J.~M.}\ \bibnamefont {Doyle}}, \ and\
  \bibinfo {author} {\bibfnamefont {S.}~\bibnamefont {Kais}},\ }\href {\doibase
  10.1080/00268976.2013.813595} {\bibfield  {journal} {\bibinfo  {journal}
  {Mol. Phys.}\ }\textbf {\bibinfo {volume} {111}},\ \bibinfo {pages} {1648}
  (\bibinfo {year} {2013})}\BibitemShut {NoStop}%
\end{thebibliography}%
\end{document}